\documentclass[twocolumn]{aastex63}
\usepackage{graphicx}
\usepackage{soul}  
\usepackage{amsmath}
\usepackage{multirow}
\usepackage{lineno}

\newcommand{\as}{^{\prime \prime}}
\newcommand{\kms}{\rm km~s^{-1}}
\newcommand{\nthp}{\rm N_{\rm 2}H^{\rm +}}
\newcommand{\nhtd}{\rm NH_{\rm 2}D}
\newcommand{\hcop}{\rm HCO^{\rm +}}
\newcommand{\htcop}{\rm H^{13}CO^{\rm +}}
\newcommand{\ceo}{\rm C^{\rm 18}O}
\newcommand{\tco}{\rm ^{\rm 13}CO}
\newcommand{\nht}{N_{\rm H_{2}}}
\newcommand{\mlin}{M_{\rm line}}

\newcommand{\vpeak}{V_{\rm peak}}

\newcolumntype{H}{>{\setbox0=\hbox\bgroup}c<{\egroup}@{}}
\newcolumntype{Z}{>{\setbox0=\hbox\bgroup}c<{\egroup}@{\hspace*{-\tabcolsep}}}

\received{April 6, 2021}
\revised{May 25, 2021}
\accepted{June 4, 2021}

\shorttitle{TRAO FUNS II. IC~5146}
\shortauthors{Chung et al.}

\begin{document}

\title{TRAO Survey of the Nearby Filamentary Molecular Clouds, the Universal Nursery of Stars (TRAO FUNS). II. Filaments and Dense Cores in IC~5146}

\author{Eun Jung Chung} \affiliation{Department of Astronomy and Space Science, Chungnam National University, 99 Daehak-ro, Yuseong-gu, Daejeon 34134, Republic of Korea} \affiliation{Korea Astronomy and Space Science Institute, 776 Daedeokdae-ro, Yuseong-gu, Daejeon 34055, Republic of Korea}

\author{Chang Won Lee} \affiliation{Korea Astronomy and Space Science Institute, 776 Daedeokdae-ro, Yuseong-gu, Daejeon 34055, Republic of Korea} \affiliation{University of Science and Technology, Korea (UST), 217 Gajeong-ro, Yuseong-gu, Daejeon 34113, Republic of Korea}

\author{Shinyoung Kim} \affiliation{Korea Astronomy and Space Science Institute, 776 Daedeokdae-ro, Yuseong-gu, Daejeon 34055, Republic of Korea} \affiliation{University of Science and Technology, Korea (UST), 217 Gajeong-ro, Yuseong-gu, Daejeon 34113, Republic of Korea}

\author{Maheswar Gopinathan} \affiliation{Indian Institute of Astrophysics, Kormangala (IIA), Bangalore 560034, India}

\author{Mario Tafalla} \affiliation{Observatorio Astron$\acute{o}$mico Nacional (IGN), Alfonso XII 3, 28014 Madrid, Spain}

\author{Paola Caselli} \affiliation{Max-Planck-Institut f$\ddot{u}$r Extraterrestrische Physik, Gie$\beta$enbachstrasse 1, 85748 Garching bei M\"unchen, Germany}

\author{Philip C. Myers} \affiliation{Harvard-Smithsonian Center for Astrophysics, 60 Garden Street, Cambridge, MA 02138, USA}

\author{Tie Liu} \affil{Key Laboratory for Research in Galaxies and Cosmology, Shanghai Astronomical Observatory, Chinese Academy of Sciences, 80 Nandan Road, Shanghai 200030, China}

\author{Hyunju Yoo} \affiliation{Korea Astronomy and Space Science Institute, 776 Daedeokdae-ro, Yuseong-gu, Daejeon 34055, Republic of Korea}

\author{Kyoung Hee Kim} \affiliation{Korea Astronomy and Space Science Institute, 776 Daedeokdae-ro, Yuseong-gu, Daejeon 34055, Republic of Korea}

\author{Mi-Ryang Kim} \affiliation{Korea Astronomy and Space Science Institute, 776 Daedeokdae-ro, Yuseong-gu, Daejeon 34055, Republic of Korea}

\author{Archana Soam} \affil{SOFIA Science Centre, USRA, NASA Ames Research Centre, MS-12, N232, Moffett Field, CA 94035, USA}

\author{Jungyeon Cho} \affiliation{Department of Astronomy and Space Science, Chungnam National University, 99 Daehak-ro, Yuseong-gu, Daejeon 34134, Republic of Korea}

\author{Woojin Kwon} \affil{Department of Earth Science Education, Seoul National University (SNU), 1 Gwanak-ro, Gwanak-gu, Seoul 08826, Republic of Korea} \affiliation{SNU Astronomy Research Center, Seoul National University, 1 Gwanak-ro, Gwanak-gu, Seoul 08826, Republic of Korea}

\author{Changhoon Lee} \affil{Korea Astronomy and Space Science Institute, 776 Daedeokdae-ro, Yuseong-gu, Daejeon 34055, Republic of Korea}

\author{Hyunwoo Kang} \affil{Korea Astronomy and Space Science Institute, 776 Daedeokdae-ro, Yuseong-gu, Daejeon 34055, Republic of Korea}

\begin{abstract}
We present the results on the physical properties of filaments and dense cores in IC~5146, as a part of the TRAO FUNS project. We carried out on-the-fly mapping observations using the Taeduk Radio Astronomy Observatory (TRAO) 14~m telescope covering about 1 square degree of the area of IC 5146 using various molecular lines. We identified 14 filaments (24 in total, including sub-filaments) from the $\ceo~(1-0)$ data cube and 22 dense cores from the $\nthp~(1-0)$ data. We examined the filaments' gravitational criticality, turbulence properties, accretion rate from filaments to dense cores, and relative evolutionary stages of cores. Most filaments in IC 5146 are gravitationally supercritical within the uncertainty, and most dense cores are formed in them. We found that dense cores in the hubs show a systemic velocity shift of $\sim$0.3~$\kms$ between the $\nthp$ and $\ceo$ gas. Besides, these cores are subsonic or transonic, while the surrounding filament gas is transonic or supersonic, indicating that the cores in the hubs are likely formed by the dissipation of turbulence in the colliding turbulent filaments and the merging is still ongoing. We estimated a mass accretion rate of 15$- 35~M_{\odot}~\rm Myr^{-1}$ from the filaments to the dense cores, and the required timescales to collect the current core mass are consistent with the lifetime of the dense cores. The structures of filaments and dense cores in the hub can form from a collision of turbulent converging flows, and mass flow along the filaments to the dense cores may play an important role in forming dense cores.
\end{abstract}

\keywords{ISM: clouds --- ISM: kinematics and dynamics --- ISM: structure --- stars: formation}

\section{Introduction} 

It has been noticed that many star-forming regions are associated with elongated filamentary structures of parsec scales in observations at optical, infrared, and submillimeter wavelengths \citep[e.g.,][]{lynds1962,lada2003,goldsmith2008}. The {\em Herschel} Space Observatory has made a significant progress in the study of these filamentary molecular clouds and early star-forming conditions, showing that filamentary structures pervade the molecular clouds, from the low- to high-mass star-forming clouds, and the non-star-forming clouds \citep[e.g.,][]{andre2010,arzoumanian2019}. Over 75\% of the prestellar cores are found to reside in filaments with supercritical mass per unit length in the Aquila molecular cloud complex \citep{konyves2015}. 
A similar result was reported for L1641 molecular clouds in the Orion~A, where $\sim$71\% of the prestellar cores are found to be located on the filaments and they are usually more massive than the prestellar cores off the filaments \citep{polychroni2013}.

Although the {\em Herschel} observations have made a great progress with its extraordinarily high spatial resolution, it does not provide any velocity information on the structures it identifies, which poses a serious limitation. Possible overlaps of filaments with different velocities into the plane of the sky are expected, and submillimeter radio observations, mainly with carbon monoxide and its isotopologues, were carried out to obtain crucial dynamical information on the filaments \citep[e.g.,][]{nagahama1998,hacar2013,panopoulou2014}. It was found that some filaments have multiple velocity substructures \citep[e.g.,][]{hacar2013,hacar2016,hacar2017,hacar2018,henshaw2013,maureira2017,clarke2018,dhabal2018}. One of the interesting things that is revealed by the observations of molecular lines is the mass flow along filaments towards the star clusters in hub-filament structures \citep[HFSs;][]{myers2009}. Velocity gradients along the filaments are observed and mass flow along these filaments are considered to be responsible for the formation of the star clusters in the hubs \citep[e.g.,][]{kirk2013,peretto2014,imara2017,baug2018,yuan2018,trevino2019}. Hence, the filaments and their related structures are supposed to be a prerequisite stage of star formation and closely linked to the formation of prestellar cores and stars. However, a detailed understanding of how they form is still unclear. 

Numerical simulations of the interstellar medium (ISM) show that filaments can be generated by gravitational instability of a sheet, global gravitational collapse of a cloud, and collisions of large-scale turbulence \citep[e.g.,][]{nagai1998,padoan2001,hennebelle2013,gomez2014}. In addition, recent magnetohydrodynamics (MHD) and hydrodynamical turbulent simulations show that clumps in a magnetic field tend to have more elongated filamentary shapes \citep[see][and references therein]{hennebelle2019}. For the formation of dense cores in filaments, it is suggested that cores form in filaments by collapse and fragmentation due to the gravitational instability \citep{inutsuka1997}. With the observational results obtained from {\em Herschel}, a two-step formation scenario has been proposed in which the filaments firstly form by the dissipation of large-scale turbulence, and then the cores are generated in the gravitationally supercritical filaments via fragmentation \citep[e.g.,][]{andre2014,arzoumanian2019}. Another model proposed for the formation of the filaments and the dense cores is by the collision of turbulent flows that form dense cores by the compression and dissipation of turbulence at the stagnation point between two flows \citep[e.g.,][]{ballesteros1999,padoan2002}. 

$^{\as}$TRAO FUNS,$^{\as}$ which is an acronym of the $^{\as}$TRAO survey of the nearby Filamentary molecular clouds, the Universal Nursery of Stars,$^{\as}$ is a project to survey the clouds belonging to the Gould Belt in molecular lines using the Taeduk Radio Astronomy Observatory (TRAO)\footnote{\url{http://radio.kasi.re.kr/trao/main_trao.php}} 14m antenna. We have made observations of nine molecular clouds under the TRAO FUNS project in eight molecular lines. These clouds are found to have diverse physical conditions and range from being quiescent non-star-forming clouds like Polaris to low-mass star-forming clouds such as L1478 of California cloud, and to active high-mass star-forming clouds such as the Orion~B molecular complex. This project mainly aims to obtain the kinematic and chemical structures of the filaments and the associated dense cores to understand the processes involved in their formation.

\begin{figure*} \epsscale{1.17} 
\plotone{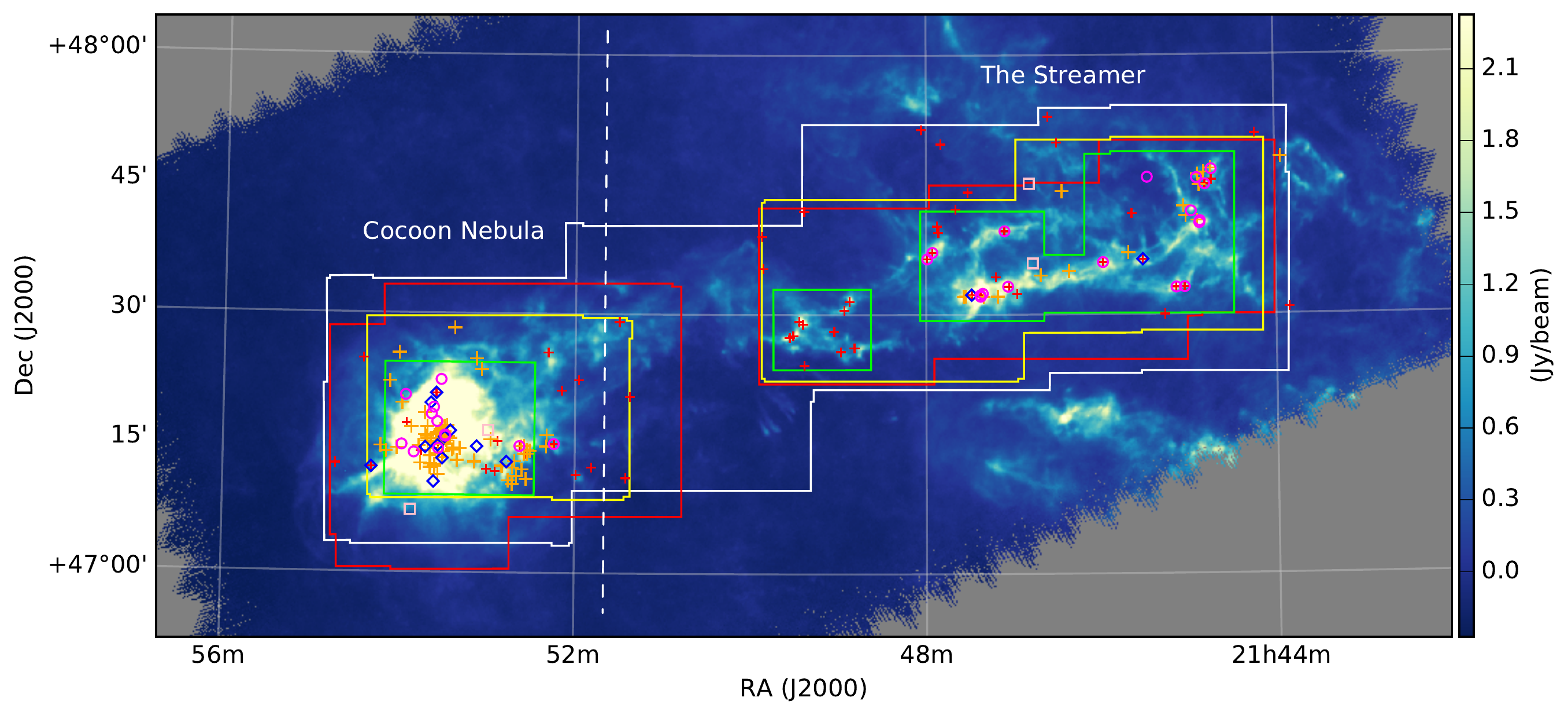}
\caption{Our survey areas of IC~5146 marked over its {\em Herschel} 250 $\mu$m image. The areas observed in different molecular lines are marked with different colors (white for $\tco$ and $\ceo~(1-0)$, red for $\nthp$ and $\hcop ~(1-0)$, yellow for SO~($3_{2}-2_{1}$) and CS~($2-1$), and green for $\nhtd$ ($1_{11}-1_{01}$) and $\htcop~(1-0)$). YSOs identified by {\em Spitzer} are presented as blue diamonds (Flat spectrum), magenta circles (Class I), orange crosses (Class II), and pink squares (Class III) \citep{harvey2008}, and 70~$\mu$m point sources from Herschel/PACS Point Source Catalogue \citep[HPPSC;][]{poglitsch2010} are presented as red crosses. The white dashed line shows R.A.(J2000) of $21^{\rm h}50^{\rm m}31^{\rm s}.5$ which separates the Cocoon Nebula and the Streamer. \label{fig:obsreg}}
\end{figure*}

This is the second paper presenting the results obtained from the TRAO FUNS project, especially on IC~5146 located in the constellation Cygnus. IC~5146 is a nearby star-forming molecular cloud consisting of a reflection nebula in the east, the Cocoon Nebula, and a dark cloud with multiple filaments in the west, the Streamer \citep{lada1994}. {\em Spitzer Space Telescope} observations of the IC~5146 region made using the InfraRed Array Camera (IRAC) and the Multiband Imaging Photometer for {\em Spitzer} (MIPS) resulted in the identification of $\sim$200 candidate young stellar objects \citep[YSOs;][]{harvey2008}. Based on the {\em Herschel} data, \citet{arzoumanian2011} identified 71 YSOs and 45 candidate bound prestellar cores in IC~5146. Observations at 450 and 850~$\mu$m with the James Clerk Maxwell Telescope (JCMT) have been carried out in this cloud, and 15 Class 0/I, 4 Flat, 14 Class II, and 6 Class III YSOs have been found \citep{johnstone2017}. The complex filamentary structure of IC~5146 was revealed in the {\em Herschel} Gould Belt Survey \citep{arzoumanian2011}. The main results from the {\em Herschel} observations are that the filaments in IC~5146 have a narrow characteristic width of $\sim$0.1~pc and the most bound prestellar cores are located on the gravitationally supercritical filaments, similar to the results obtained for other regions by the {\em Herschel} observations \citep[e.g.,][]{andre2014}.

The distance to IC~5146 is still uncertain, and ranges from 460 to 1400~pc \citep[e.g.,][]{harvey2008,nunes2016}. \citet{nunes2016} measured the distance to the embedded clusters in the Streamer and quoted a distance of $\sim$1.2$\pm 0.1$~kpc. Recently, using the newly released {\em Gaia} data \citep[{\em Gaia} DR2,][]{bailer2018}, \citet{dzip2018} estimated the distances to the star-forming clouds in the Gould belt, and estimated a distance of 813$\pm$106~pc for the Cocoon Nebula. Meanwhile, \citet{wang2020} estimated the distance to IC~5146 including the Streamer using {\em Gaia} DR2, and argued that the Streamer is nearer than the Cocoon Nebula. In this work for the analysis, we adopted the most recent distance estimations of 800$\pm$100~pc and 600$\pm$100~pc for the Cocoon and the Streamer, respectively, made using the {\em Gaia} measurements by \citet{wang2020}. Recently the third {\em Gaia} Data Release \citep[{\em Gaia} EDR3;][]{bailer-jones2021} 
was also made, and the distances of these two regions from the {\em Gaia} EDR3 are found to be consistent (within the uncertainties) with the distances adopted here.

The paper is organized in the following manner. Section~\ref{sec:obsdr} describes the observations and data reduction. Section~\ref{sec:results} shows the spatial distributions of the detected molecular lines. Section~\ref{sec:fil} and \ref{sec:dc} explain the identification and the estimations of physical quantities of filaments and dense cores, respectively. Section~\ref{sec:disc} discusses the formation mechanisms of filaments and dense cores in IC~5146 with their physical and chemical properties. Section~\ref{sec:sum} gives summary and conclusions. \\

\section{Observations and Data Reduction} \label{sec:obsdr}

\subsection{Observations}

IC~5146 was observed with the TRAO 14~m telescope. TRAO provides spatial resolution of $\sim$49$^{\as}$ at 110~GHz, corresponding to 0.14~pc at the distance of 600~pc (0.19~pc at 800~pc). TRAO has the frontend of SEcond QUabbin Optical Image Array focal plane array receiver (SEQUOIA-TRAO) consisting of 16 horns, configured in a 4$\times$4 array with a spatial separation of 89$^{\as}$. The backend system used is the fast Fourier transform (FFT) spectrometer, which has 4096$\times$2 channels at 15~kHz resolution ($\sim$0.04~$\kms$ at 110~GHz) and covers a total bandwidth of 62.5~MHz, corresponding to $\sim$170~$\kms$ at 110~GHz. Simultaneous observations of two molecular lines at the frequencies of 85 and 100~GHz or 100 and 115~GHz are allowed. The beam efficiency of the telescope is 0.48 at 98~GHz, and 0.46 at 110~GHz \citep{jeong2019}.

\begin{deluxetable*}{lccCccccc}
\tablecaption{Observing Information \label{tab:lines}} 
\tablecolumns{5} \tablewidth{0pt} 
\tablehead{ 
\colhead{Molecule} & 
\colhead{$\nu_{\rm ref}$ \tablenotemark{a}} & 
\colhead{$\theta_{\rm FWHM}$ \tablenotemark{b}} & 
\colhead{Area \tablenotemark{c}} & 
\colhead{$\theta_{\rm pixel}$ \tablenotemark{d}} & 
\colhead{$\delta v$ \tablenotemark{e}} & 
\colhead{rms \tablenotemark{f}} &
\colhead{rms of mom0 \tablenotemark{g}}&
\colhead{$n_{\rm crit}$ \tablenotemark{h}} \\
\colhead{} & 
\colhead{(GHz)} & 
\colhead{($\as$)} & 
\colhead{(arcmin$^{2}$)} & 
\colhead{(arcsec)} & 
\colhead{($\kms$)} & 
\colhead{(K)} & 
\colhead{(K~$\kms$)} & 
\colhead{(cm$^{-3}$)} }
\startdata
$\ceo \rm~(1-0)$ & 109.782173 $^{1}$ & 49 & \multirow{2}{*}{3360} & \multirow{2}{*}{20} & \multirow{2}{*}{0.1} & \multirow{2}{*}{0.096} & 0.090 & 1.9$\times 10^{3}$ \\ 
$\tco ~(1-0)$ & 110.201353 $^2$ & 49 &  &  &  &  & 0.103 & 1.9$\times 10^{3}$ \\ \hline 
$\nthp ~(1-0)$ & 93.176258 $^1$ & 52 & \multirow{2}{*}{2160} & \multirow{2}{*}{20} & \multirow{2}{*}{0.06} & \multirow{2}{*}{0.066} & 0.071 & 5.7$\times 10^{5}$ \\ 
$\hcop ~(1-0)$ & 89.188525 $^3$ & 57 &  &  &  &  & 0.050 & 1.6$\times 10^{5}$  \\ \hline 
CS ($2-1$) & 97.980953 $^1$ & 52 & \multirow{2}{*}{1780} & \multirow{2}{*}{20} & 0.06 & 0.069 & 0.056 & 3.3$\times 10^{5}$  \\ 
SO $(3_{2}-2_{1})$ & 99.299870 $^4$ & 52 &  &  & 0.1 & 0.056 & 0.052 & 3.5$\times 10^{5}$ \\ \hline 
$\nhtd ~(1_{11}-1_{01})$ & 85.926278 $^3$ & 57 & \multirow{2}{*}{864} & \multirow{2}{*}{20} & 0.1 & 0.058 & 0.078 & 3.8$\times 10^{6}$ \\
H$\tco^{+} ~(1-0)$ & 86.754288 $^3$ & 57 &  &  & 0.06 & 0.068 & 0.067 & 1.5$\times 10^{5}$ \\
\enddata 
\tablenotemark{}{} \\
\tablenotemark{a}{~Rest frequency of each molecular line. The references are: $^1$Lee et al. (2001), $^2$Lovas (2004), $^3$Cologne Database for Molecular Spectroscopy (CDMS: M\"uller et al. 2001, https://cdms.ph1.uni-koeln.de/cdms/portal/), and $^4$Submillimeter, Millimeter, and Microwave Spectral Line Catalog provided by Jet Propulsion Lab oratory (Pickett et al. 1998).}\\ 
\tablenotemark{b}{~FWHM of the telescope beam (Jeong et al. 2019).}\\ 
\tablenotemark{c}{~Total observed area.} \\
\tablenotemark{d,e}{~The pixel size and channel width of the final data cube.} \\
\tablenotemark{f}{~Noise level in T$_{\rm A}^{\ast}$ of the final data cube.} \\
\tablenotemark{g}{~Noise level of integrated intensity moment 0 map.} \\
\tablenotemark{h}{~Critical density, which is estimated from $A / \gamma$ where $A$ is the Einstein $A$ coefficient and $\gamma$ is the collisional rate coefficient at 10~K from LAMDA (Sch\"oier et al. 2005). 
$n_{\rm crit}$ of SO~$(3_{2}-2_{1})$ is calculated with $\gamma$ at 60~K, which is the lowest temperature provided by LAMDA.} 

\end{deluxetable*}

We choose eight molecular lines to investigate the physical properties of filaments and dense cores. $\ceo~(1-0)$ and $\nthp~(1-0)$ molecular lines are chosen as tracers of relatively less dense filaments and denser cores, respectively. $\hcop~(1-0)$ and CS~$(2-1)$ molecular lines were chosen because they are known to be good tracers of infall motions in prestellar cores \citep[e.g.,][]{lee2001}. SO and $\nhtd$ are known to be respectively the most and least sensitive molecules to the gas freeze-out \citep[e.g.,][]{tafalla2006}. Hence, we selected SO~($3_{2}-2_{1}$) and $\nhtd$~($1_{11}-1_{01}$) lines to investigate the chemical evolution of dense cores. We made the simultaneous observations in two molecular lines with sets of $\ceo~(1-0)$ and $\tco~(1-0)$, $\nthp~(1-0)$ and $\hcop~(1-0)$, CS~(2-1) and SO~($3_{2}-2_{1}$), and $\nhtd$~($1_{11}-1_{01}$) and $\htcop~(1-0)$, respectively. 

\begin{figure*} \epsscale{1.2}
\plotone{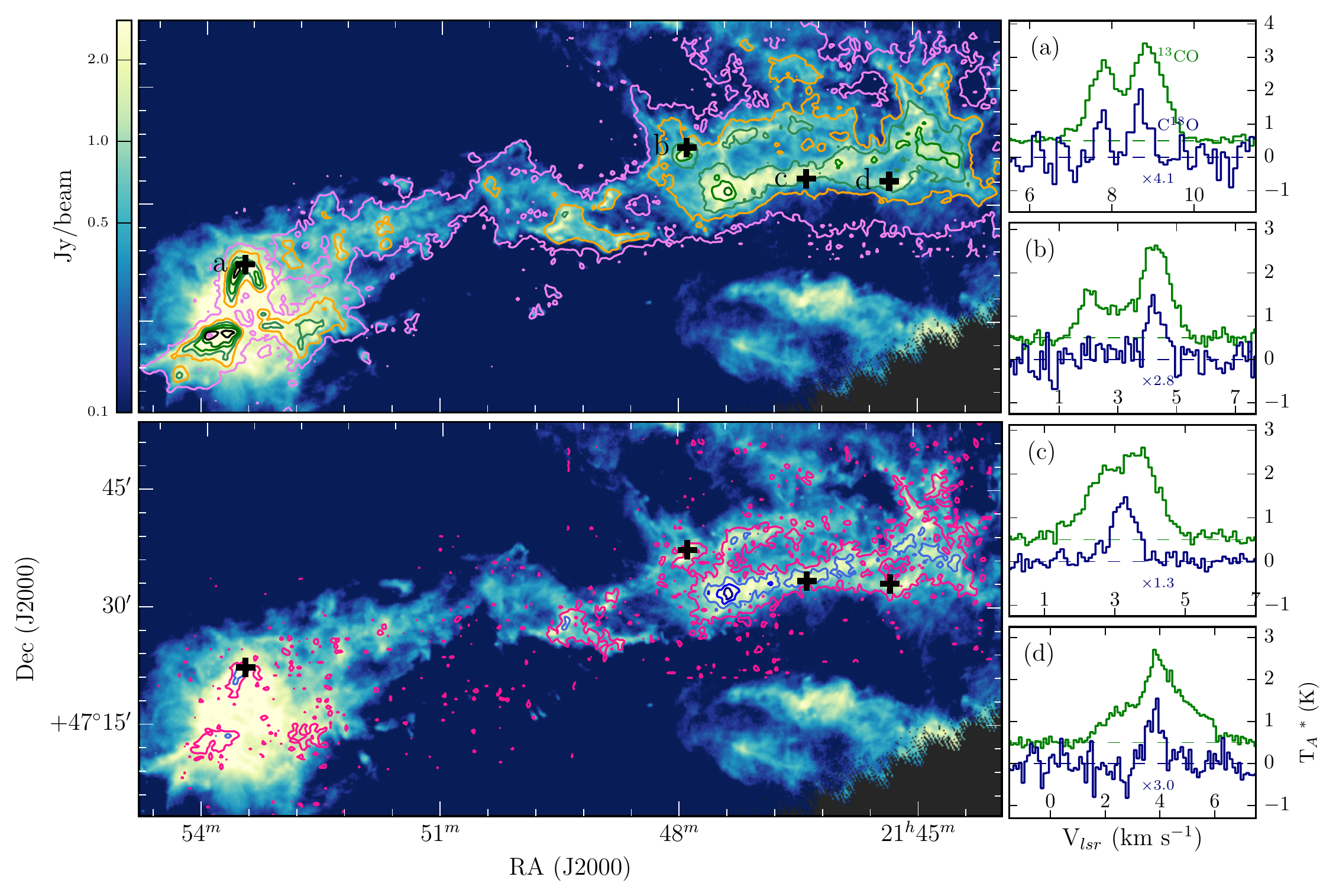}
\caption{Integrated intensity maps of $\tco~(1-0)$ (top left) and $\ceo~(1-0)$ (bottom left) with contours on the $Herschel ~250 ~ \mu \rm m$ image and spectra of the selected positions (right). The contour levels of $\tco$ are 5, 20, 35, $\cdots$, 95$\times \sigma$ and those of $\ceo$ are 3, 8, 13, and 18$\times \sigma$. The positions of the presented spectra in the right are indicated with crosses. In the right spectra panels, $\ceo$ line intensity is scaled up (the factor is written under each $\ceo$ spectrum). \label{fig:13c18omom0}}
\end{figure*}

The observations were carried out from 2018 May to 2019 November. These data are obtained in on-the-fly mapping mode. To cover the whole IC~5146 region including the Cocoon Nebula and the Streamer in the $\ceo~(1-0)$ and $\tco~(1-0)$ molecular line set, we divided it into four subregions having box shapes (referred as 'tiles' hereafter) with a size of $28^{\prime} \times 30^{\prime}$. Five tiles were made for the $\nthp~(1-0)$ and $\hcop~(1-0)$ line set, of which three were of $20^{\prime} \times 20^{\prime}$ size and two were of $20^{\prime} \times 24^{\prime}$ size. For the third molecular line set of CS~(2-1) and SO~($3_{2}-2_{1}$), three tiles (one $28^{\prime} \times 22^{\prime}$ and two $29^{\prime} \times 20^{\prime}$) are made. For the last molecular line set of $\nhtd$~($1_{11}-1_{01}$) and $\htcop~(1-0)$, five tiles with various sizes from 6$^{\prime} \times 6^{\prime}$ to 17$^{\prime} \times 18^{\prime}$ were made to cover the dense core regions detected in $\nthp~(1-0)$. The scanning rate was 55$\as$ per second and the recording time is 0.2~s. We chose scan steps of 5$\as$ to 11$\as$ along the scan direction and separations of 5$\as$ to 33$\as$ between the rows to increase the observation efficiency according to the sizes of tiles. The scan steps used are smaller than the recommended Nyquist spacing to avoid undersampling. We made several maps alternately along the RA and Dec directions to obtain the uniform target sensitivity. Figure~\ref{fig:obsreg} shows the observed regions for each set of molecular lines over the {\em Herschel} 250~$\mu$m continuum image. \\

\subsection{Data Reduction}

The data were reduced as follows. First, the raw on-the-fly data for each map of each tile were read and converted into a {\sc Class}\footnote{\url{http://www.iram.fr/IRAMFR/GILDAS}} format map after the subtraction of baseline (with first order) in {\sc Otftool}. A resulting cell size of 10$^{\as}$ was chosen and noise-weighting was applied. Further reductions and inspection of the data were done using the {\ttfamily otfpro} {\sc Class} script. The baseline subtractions were done iteratively. Both ends of the raw spectra were cut off and baselines of the spectra were subtracted with a second-order polynomial. After that, the spectra were resampled with a channel width of 0.06~$\kms$, and both ends were cut off again, giving the spectra a velocity range of 60~$\kms$; the baselines of the spectra were again subtracted but with a first-order polynomial. Finally, the maps were merged into a final fits cube with 20$^{\as}$ cell size and 0.1~$\kms$ velocity channel width for $\ceo~(1-0)$, $\tco~(1-0)$, SO~($3_{2}-2_{1}$), and $\nhtd$~($1_{11}-1_{01}$) data, and 20$^{\as}$ cell size and 0.06~$\kms$ velocity channel width for the $\nthp~(1-0)$, $\hcop~(1-0)$, CS~(2-1), and $\htcop~(1-0)$ data. The basic information on the observations and the final data is given in Table~\ref{tab:lines}. The final rms levels achieved were $\lesssim$0.1$\rm ~K[T_{\rm A}^{\ast}$] for the $\ceo$ and $\tco$ lines and $\sim$0.07$\rm ~K[T_{\rm A}^{\ast}$] for other molecular lines. \\

\section{Results} \label{sec:results}

\subsection{$\tco$ and $\ceo$ emissions} 

Figure~\ref{fig:13c18omom0} shows the integrated intensity maps of $\tco~(1-0)$ and $\ceo~(1-0)$ lines, which are found to nicely delineate the large-scale structure of IC~5146. The line intensity maps are integrated over the velocity range from -0.8 to 10.7 $\kms$ for $\tco~(1-0)$, and 0.7 to 9.4 $\kms$ for $\ceo~(1-0)$. The distribution of $\tco~(1-0)$ emission is well matched with that of the {\em Herschel} 250~$\mu$m emission, while $\ceo~(1-0)$ lines are only detected in relatively compact regions of high continuum flux. In the eastern region of IC~5146, the ball shape of the Cocoon Nebula is shown in the 250~$\mu$m emission as well as H$\alpha$ image \citep[e.g.,][]{arzoumanian2011}. However, the CO isotopologues have a different distribution from those of the dust and ionized gas. They have the shape of three-leafed clover instead of a roundish structure. In the Streamer, hub-filament structures and filament shapes can be seen in the $\tco$ and $\ceo~(1-0)$ maps. The molecular gases of $\tco$ and $\ceo$ have relatively higher LSR velocity toward the southeast than toward the northwest (see the spectra in Figure~\ref{fig:13c18omom0} and Figure~\ref{af:pvds} in the Appendix). 

\begin{figure*} \epsscale{1.}
\plotone{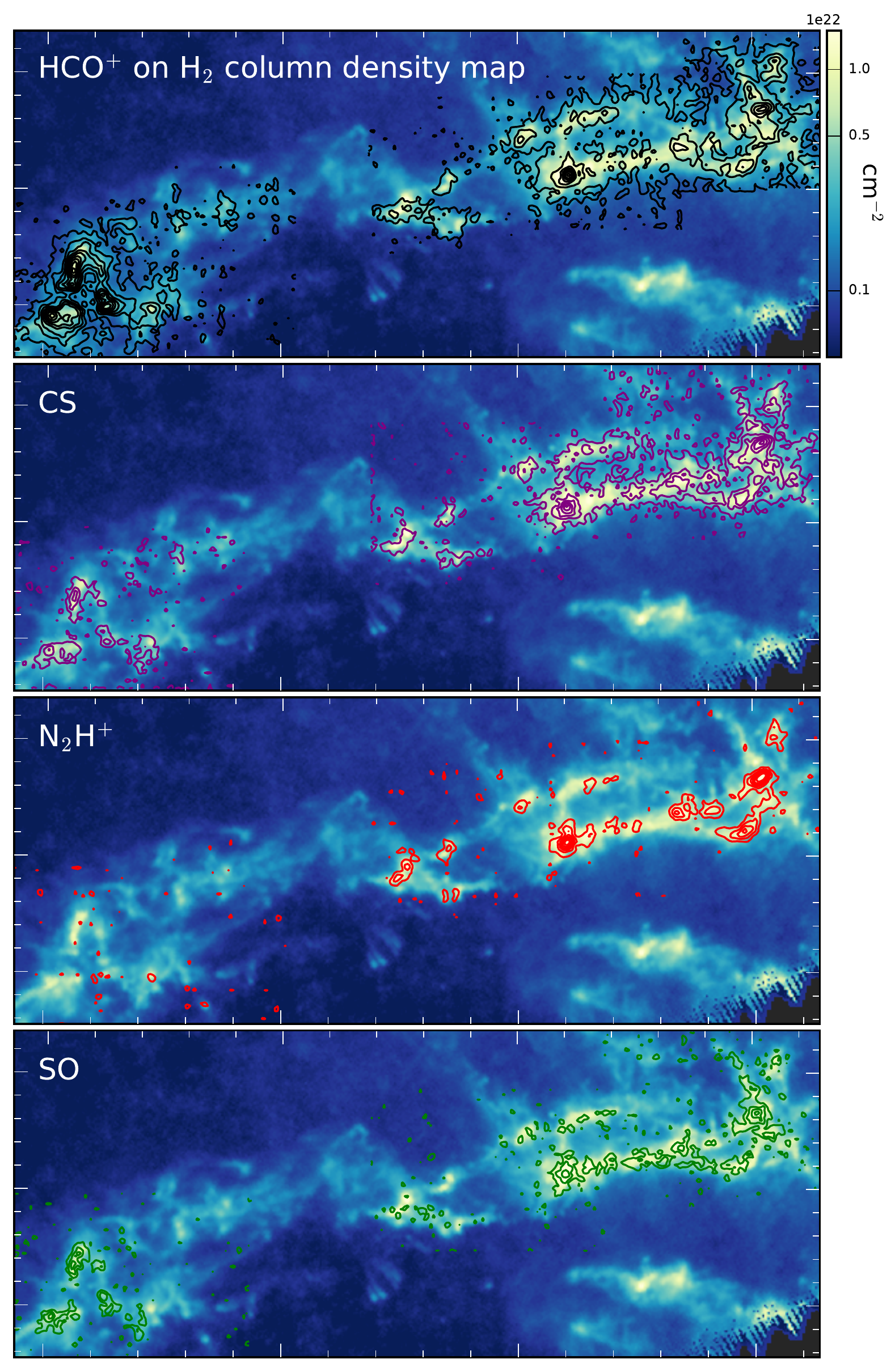}
\caption{Integrated intensity contour maps of $\hcop~(1-0)$, CS~(2-1), $\nthp~(1-0)$, and SO~($3_{2}-2_{1}$) on an H$_{2}$ column density map \citep{arzoumanian2011}. The contours are drawn every 5$\sigma$ from the 3$\sigma$ level for $\hcop$, $\nthp$, and CS emissions, and every 3$\sigma$ for SO emission. 1$\sigma$ values for the integrated intensity maps for each line are given in Table~\ref{tab:lines}. \label{fig:othermom0}}
\end{figure*}

In Figure~\ref{fig:13c18omom0}, the right panels present the $\tco$ and $\ceo~(1-0)$ spectra at four positions indicated in the maps with crosses, showing that overlapping of the multiple velocity components seems common. Shown here are the various shapes of $\tco$ and $\ceo~(1-0)$ spectra found towards IC~5146. In position (a), $\tco$ and $\ceo~(1-0)$ lines trace the same multiple velocity components at 7.6 and 8.8~$\kms$. These two components are shown to be clearly separated in the $\tco~(1-0)$ spectrum as well as in the $\ceo~(1-0)$ spectrum, though the former shows much larger line width than the latter. However, in positions (b) and (c), the $\tco~(1-0)$ spectra show different peaks when compared with the $\ceo~(1-0)$ spectra. In (b), the $\ceo~(1-0)$ line has only one peak at $\sim$4.5~$\kms$, but the $\tco~(1-0)$ line shows an additional peak at $\sim$2$~\kms$, suggesting that $\tco$ covers less dense and larger bulk structures where $\ceo$ is not detected. However, toward brighter regions, $\tco$ can be self-absorbed as shown in position (c). Looking at the $\tco~(1-0)$ spectrum alone, it appears that there are two different velocity components. However, it can be seen that the dip between the two peaks of $\tco~(1-0)$ exactly matches with the peak position of the $\ceo~(1-0)$ spectrum which has a relatively smaller optical depth than $\tco~(1-0)$, indicating that the double components of the $\tco~(1-0)$ line are caused by the self-absorption of the $\tco~(1-0)$ spectrum due to its high optical depth. The $\tco~(1-0)$ spectrum at position (d) shows a larger line width ($\Delta V_{\rm FWHM} \sim 4~\kms$) and appears to be composed of multiple Gaussian components while the $\ceo~(1-0)$ spectrum shows only a single component. The $\tco~(1-0)$ line is useful to trace the less dense filament material and covers a large area, but has limitations due to its large optical depth in the area of high column density. After visual inspections of all the spectra, we chose the $\ceo~(1-0)$ line to trace the velocity structure of the dense filament material, which is thought to be more closely related to the star formation \citep[e.g.,][]{nishimura2015}.

\subsection{$\hcop$, {\rm CS}, $\nthp$, and {\rm SO} emission}

We present the distributions of $\hcop~(1-0)$, CS~(2-1), $\nthp~(1-0)$, and SO~$(3_{2}-2_{1})$ emission in Figure~\ref{fig:othermom0}. The $\hcop~(1-0)$, CS~(2-1), and SO~$(3_{2}-2_{1})$ intensities are integrated over the velocity ranges from 0.62 to 9.98, -1.66 to 9.2, and 0.86 to 9.2~$\kms$, respectively. The $\nthp~(1-0)$ intensity is integrated from -5.9 to 13.6~$\kms$ including all hyperfine components. 

\begin{figure*} \epsscale{1.17}
\plotone{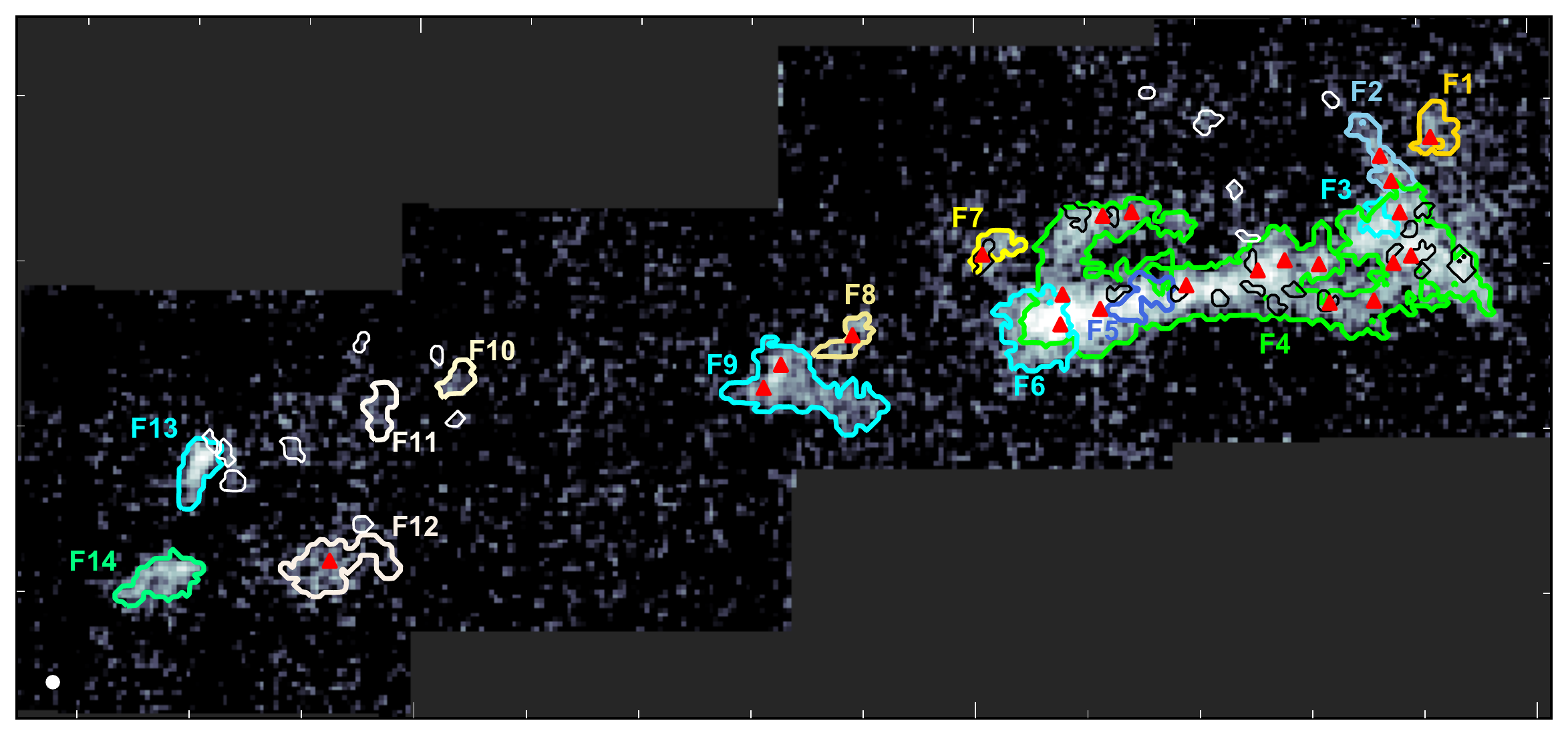}
\caption{Identified filaments on the $\ceo~(1-0)$ integrated intensity image. Filaments larger than 6$\times \theta_{\rm beam}$ are marked as F1 to F14. The others, having sizes of 1 to 4$\times \theta_{\rm beam}$ and aspect ratio less than 3, are shown with thin white and black lines without their names. The red triangles represent the locations of 22 dense cores identified with $\nthp~(1-0)$ data (Section~\ref{sec:dcid}). The 49$^{\as}$ FWHM beam of the TRAO telescope at 110~GHz is shown by white circle at the bottom left corner. \label{fig:filid}}
\end{figure*}

The distribution of $\hcop~(1-0)$ emission matches well with the region of overall high H$_{2}$ column density derived from the 70$-500~\mu$m {\em Herschel} data \citep{arzoumanian2011}. In the Cocoon Nebula, $\hcop~(1-0)$ emission shows a shell-like structure. The abundance of $\hcop$ is maintained by the balance between its formation from the reactions between H$_{3}^{+}$ and CO and its destruction from dissociative recombination with electrons. Besides, it is reported that in the far-ultraviolet irradiated environments $\hcop$ can more easily recombine with free electrons, leading to a decrease in its abundance \citep[e.g.,][]{pety2017}. Hence, the central hole in $\hcop~(1-0)$ emission may be due to the dominance of its destruction processes through the electronic recombination in an environment that contains abundant free electrons produced by the central B-type star. In the Streamer, the $\hcop~(1-0)$ emission appears to be well matched with the H$_{2}$ column density distribution as well as with $\ceo~(1-0)$. 

CS~(2-1) emission (shown with contours in the second panel from the top of Figure~\ref{fig:othermom0}) is in good agreement with the $\ceo~(1-0)$ emission (the bottom panel in Figure~\ref{fig:13c18omom0}) in the Cocoon Nebula as well as in the Streamer. 

For the $\nthp$ and SO species (presented with contours in the third and bottom panels of Figure~\ref{fig:othermom0}, respectively), the emission shows a clumpy distribution in the bright $\ceo~(1-0)$ and/or dense H$_{2}$ regions. It is noticeable that $\nthp~(1-0)$ is hardly detected in the Cocoon Nebula at the rms level of $\sim$0.07~K[T$_{\rm A}^{*}$] while other species are detected above the 3$\sigma$ level. The lack of $\nthp~(1-0)$ emission in the Cocoon Nebula may have something to do with the presence of the central B-type star, which may increase the dust temperature enough for CO to be evaporated rather than depleted. CO freezes out at low temperature ($< 20$~K) but starts to evaporate back into the gas phase at $\sim$25~K \citep[e.g.,][]{hocuk2014}. Indeed, the mean dust temperature of the Cocoon Nebula is 23~K and goes up to $\sim$30~K. The abundant CO can play a role in destroying $\nthp$, resulting in making the $\nthp$ abundance lower \citep{caselli1999,caselli2002ii,bergin2002}. In the Streamer, some distinct features can be noticed based on the distribution of various molecular species. One representative example is the $\nthp~(1-0)$ bright region in the very northwestern part (F1 region in Figure~\ref{fig:filid}), where there is weak $\ceo~(1-0)$, relatively bright CS~(2-1), but no SO~$(3_{2}-2_{1})$ emission. The spatial differentiation between molecules may be due to the different evolutionary stages as well as distinct physical conditions. The chemical differentiation is studied in more detail in Section~\ref{sec:corechem}. \\

\section{Filament Properties} \label{sec:fil}

\subsection{Filament Identification} \label{sec:filid}

We identified velocity coherent filaments using the three-dimensional information of the $\ceo~(1-0)$ data cube. There are several algorithms available that can be used to identify the structures of clumps and filaments \citep[e.g.,][]{rosolowsky2008, sousbie2011a, sousbie2011b, menshchikov2012, hacar2013, koch2015, ossenkopf2019}. Consequently, it is important to make a comparison between the results obtained by them. However, this is definitely beyond the scope of the paper. Thus, in this paper, we simply introduce the various algorithms  developed for the filament identification and describe in detail the one that we used in this work.

\begin{figure*} \epsscale{1.2}
\plotone{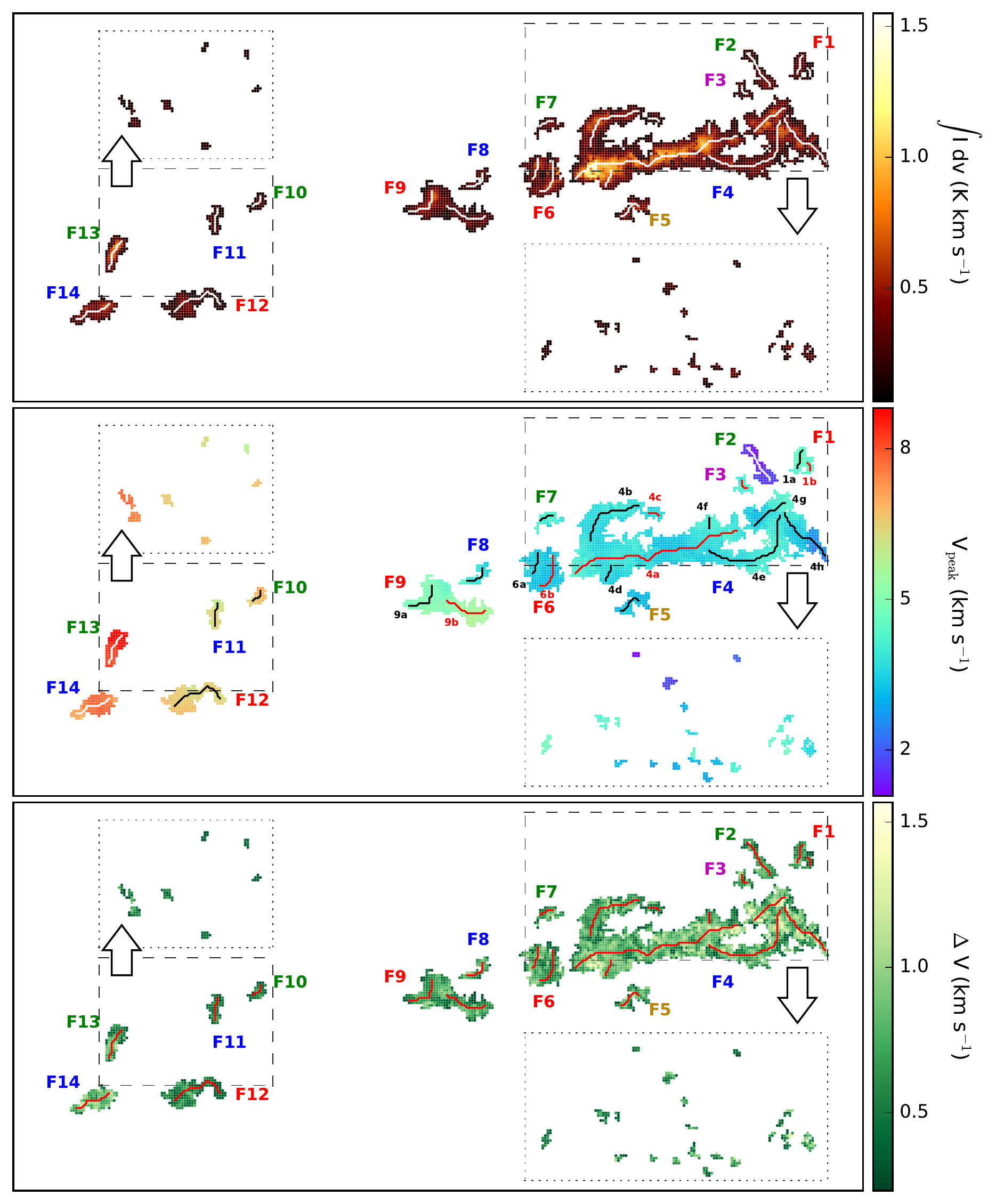}
\caption{Integrated intensity map (top), velocity field map (middle), and line width map (bottom) of each filament. Peak velocity ($V_{\rm peak}$) and line width ($\Delta$V) are quantities derived by the Gaussian fitting method. Locations of the filament's ridges (skeletons) are drawn on top of the maps and the notations of sub-filaments are given in the middle panel. A small offset is given to the original position of each filament to avoid spatial overlaps and distinguish them from each other. The small clumps in the left and right dotted square boxes are shifted by (0, $+18^{\prime}$) and (0, $-29^{\prime}$) from their original positions (dashed square boxes), respectively. \label{fig:fila0vpvw}}
\end{figure*}

The algorithms of {\tt astrodendro} \citep{rosolowsky2008}, {\sc DisPerSE} \citep{sousbie2011b}, {\tt FIVE} \citep{hacar2013}, and {\sc filfinder} \citep{koch2015} have been used for finding filamentary structures in the astronomical data. The {\sc filfinder} uses two-dimensional image data and is not suitable for the identification of the filaments in a three-dimensional data cube. The other algorithms, namely {\sc DisPerSE}, {\tt astrodendro}, and {\tt FIVE}, can be used with the three-dimensional data cube, but the methodology used to identify the filaments in each of them is quite different. The {\sc DisPerSE} finds critical points where the intensity gradient equals zero in a map and examines the persistence, which is the absolute difference between the pair of critical points. If the persistence is larger than the persistence threshold given as a free parameter, it connects the critical points to make arcs of integral shapes, producing the skeletons of the filaments as their ridges. However, with only the information on the ridges of the filaments, this algorithm is not able to give any detailed information on the physical quantities of the whole velocity coherent three-dimensional structure. Meanwhile, {\tt astrodendro}, a {\sc python} package utilizing the {\sc dendrograms} is designed to find the hierarchical structures in the molecular line data cube and provide three-dimensional isosurface hierarchical structures. The {\tt astrodendro} finds structures, called leaves, from local maxima and appends the surrounding regions with lower flux densities, increasing the volume of the structures. When they meet neighboring structures, they continue to merge into larger structures, the so-called branches and trunks. This process is found to be useful to identify the isosurface hierarchical structures and thus the filamentary structures in the molecular clouds with a simple hierarchical structure like L1478 \citep{chung2019}. However, in a complex molecular cloud where multiple velocity components are mixed up in a complicated way with their significant intensity variance, e.g., F4 of IC~5146 (see Figures~\ref{fig:filid} and \ref{fig:fila0vpvw}), the use of isosurface structures is found to be not effective for identifying velocity coherent filamentary structures in detail. 

Above all, we aim to find velocity coherent structures in the position-position-velocity (PPV) space regardless of the intensity. The {\tt FIVE} algorithm is found to be a well designed algorithm for identifying coherent structures in the PPV cube by applying the friends-of-friends (FoF) technique to the central velocity information of the molecular line \citep{hacar2013}. Its concept is very simple, and it is straightforward to find velocity coherent structures in the PPV space, making the {\tt FIVE} algorithm more suited to our aim.

Hence we take the concept of the {\tt FIVE} algorithm to find continuously connected structures in the PPV space and to segregate structures having different velocity components that appear to be connected owing to overlap in the line of sight. We first decomposed the multiple velocity components of $\ceo~(1-0)$ spectra using the tool {\ttfamily FUNStools.Decompose}\footnote{\url{https://github.com/radioshiny/funstools}} (S. Kim et al. 2021, in preparation). We then identified filaments with an FoF-like algorithm, \texttt{FindingFilaments} (\texttt{FF}). 

\texttt{FUNStools.Decompose} is a tool that automatically decomposes multiple Gaussian components from the $\ceo~(1-0)$ data cube. The algorithm primarily decides the number of components and their velocity positions in the smoothed spectrum using the first, second, and third derivatives based on a conceptual idea that the velocity component in a filament is continuous with the surroundings. Then, the fitting results are given as the initial guess, and a Gaussian fitting is performed again. 

The parameters that we used for $\ceo~(1-0)$ data of the IC~5146 are 3$\sigma$ level for the lower limit of intensity, smoothing parameters of two pixels ($40^{\as} \sim 1 \theta_{\rm beam}$) and three velocity channels (0.3~$\kms$) in the initial fitting stage and one pixel and one velocity channel in the final fitting stage. We gave low and high limits of fitted velocity dispersion of 0.1 and 2.0~$\kms$ to avoid creating pseudostructures such as spikes or baseline-like structure with a wide velocity line width. We found $\sim$3700 velocity components for 3236 pixels, and pixels having double and triple velocity components are $\sim$15\% and $<$1\%, respectively. The velocity structures appear to be somewhat complicated, especially in the western region, and the results of decomposition are significantly uncertain in the pixels with low signal-to-noise ratio (SNR).

\texttt{FindingFilaments (FF)} algorithm is an algorithm designed for identifying filamentary structures that uses a similar concept to the FoF. FoF is the algorithm originally used to find a group of galaxies in an external galaxy survey, and to create groups of adjacent galaxies within a certain range in three-dimensional space of R.A., decl., and redshift \citep[e.g.,][]{huchra1982}. The \texttt{FF} algorithm is similar to the FoF in that it collects adjacent points in three-dimensional space of R.A., decl., and velocity, but differs from the FoF in that it treats overlapping components in the line-of-sight direction, i.e., components at the same R.A. and decl. but having different velocities, as different ones. The data set used in the \texttt{FF} algorithm is the pixel number, the emission amplitude, the central velocity, and the velocity dispersion for the Gaussian components that are decomposed from the spectra.

The \texttt{FF} algorithm works in an iterative manner based on the steps mentioned below. First, it selects a decomposed Gaussian seed component with the maximum amplitude and gives the structure a number. Second, it selects the other Gaussian components in the neighboring pixels whose pixel distance is less than $\sqrt 2$ from the seed component, and checks the velocity differences of the seed and other components in the neighboring pixels. At this stage, if the velocity difference between the seed and other components in the neighboring pixel is less than the velocity dispersion of the seed ($\sigma_{v}^{\rm seed}$), the neighboring component is assigned as a friend of the seed and given the same structure number. If more than one velocity component in a neighboring pixel is within the range of velocity dispersion from the velocity of the seed, then only the closest one becomes the friend of the seed. Third, after every neighboring component has been checked, the assigned friends of the seed become the seeds of the structure for the next turn, and the same second and third steps are repeated until there are no more friends to assign. 

We do not use any criterion of intensity gradient or initial assumption of filament direction but use the distance in the PPV space. Hence, the result changes only with the given criterion of velocity difference between the seed and the neighboring component for which we use $\sigma_{v}^{\rm seed}$ in our study. However, it turned out that the result does not change with an even smaller criterion value  ($\sim 0.1~\kms$) than $\sigma_{v}^{\rm seed}$. Hence, the \texttt{FF} algorithm applying to the decomposed Gaussian components firmly identifies the velocity coherent structures in the PPV space.

We found 44 structures that are larger than one beam size ($\theta_{\rm FWHM} \sim 49^{\as}$). Of these, 30 structures have sizes of less than $\lesssim 4 \times \theta_{\rm beam}$ and aspect ratios ($d_{\rm max}/d_{\rm min}$ where $d_{\rm max}$ and $d_{\rm min}$ are the largest and the smallest diameters of the structure, respectively) less than 3, i.e., they can be considered roundish clumps rather than elongated filaments. The other 14 structures have sizes larger than $6 \times \theta_{\rm beam}$, and elongated shapes. Hence, we mainly analyzed the largest 14 structures, and assigned numbers to the filaments as F1 to F14 from west to east. For the other 30 smaller structures ($\lesssim 4 \times \theta_{\rm beam}$), we assigned numbers as Clump1 to Clump30 (CL1 to CL30) again starting from west to east in our analysis. 

The distribution of the filaments and the clumps identified is presented in Figure~\ref{fig:filid}. The 14 filaments are identified with their numbers, but the small clumps are presented as unnumbered thin white and black lines. Filament4 (F4) is the largest filament of IC~5146 that is continuously connected in the plane of the sky and also in velocity space. Four filaments (F2, F3, F5, and F6) and a large number of small clumps are found to be overlap F4 in the plane of the sky.  

To estimate the length of the filaments, the skeleton is determined with {\sc filfinder} \citep{koch2015} using the integrated intensity map of each filament. {\sc filfinder} uses the medial axis transform method, which gives the skeleton of the set of central pixels of inscribed circles having a maximum radius. We already identified filaments with the three-dimensional data cube, and the resulting filaments are continuous and coherent in the PPV three-dimensional space; the skeletons are used to calculate the length and the width of each filament. Hence we used the major skeleton only and excluded small minor skeletons in each filament. However, some filaments show multiple structures in the integrated intensity map of the filaments (F1, F4, F6, and F9, see Figure~\ref{fig:fila0vpvw}(a)). Hence, we used multiple skeletons to estimate physical quantities of each sub-filament (e.g., F1a, F1b). \\

\subsection{Physical Properties of the Identified Filaments} \label{sec:filpp}

Physical properties of the 14 filaments, such as H$_{2}$ column density, H$_{2}$ mass, length, width, mass per unit length, and nonthermal velocity dispersion, are derived. The results are listed in Table~\ref{tab:ppfila}. For the 30 clumps, H$_{2}$ column densities, H$_{2}$ masses, effective radii, and virial masses are calculated and given in Table~\ref{tab:ppclumps} in the Appendix. \\

\begin{deluxetable*}{l r@{~to~}l c c c r@{$\pm$}l r@{$\pm$}l r@{$\pm$}l r@{$\pm$}l c H}
\tablecaption{Physical Properties of Filaments \label{tab:ppfila}}
\tablewidth{0pt}
\tablehead{
\colhead{Fil. ID} &
\multicolumn{2}{c}{$V_{\rm peak}$ Range} &
\colhead{$\bar \sigma_{\rm tot}$} &
\colhead{$L$} &
\colhead{$W$} & 
\multicolumn{2}{c}{$\bar N_{\rm H_{2}}$} &
\multicolumn{2}{c}{$M$} & 
\multicolumn{2}{c}{$\mlin$} & 
\multicolumn{2}{c}{$\mlin^{\rm crit}$} & 
\colhead{Cores} &
\colhead{} \\ 
\colhead{} &
\multicolumn{2}{c}{($\kms$)} &
\colhead{($\kms$)} &
\colhead{(pc)} &
\colhead{(pc)} & 
\multicolumn{2}{c}{($10^{20}~\rm cm^{-2}$)} &
\multicolumn{2}{c}{($M_{\odot}$)} & 
\multicolumn{2}{c}{($M_{\odot}~\rm pc^{-1}$)} & 
\multicolumn{2}{c}{($M_{\odot}~\rm pc^{-1}$)} & 
\colhead{} &
\colhead{} \\
\colhead{(1)} & \multicolumn{2}{c}{(2)} & \colhead{(3)} & \colhead{(4)} & \colhead{(5)} & \multicolumn{2}{c}{(6)} & \multicolumn{2}{c}{(7)} & \multicolumn{2}{c}{(8)} & \multicolumn{2}{c}{(9)} & \colhead{(10)} & \colhead{} 
}
\startdata
1a & 4.2 & 4.6 & 0.34 & 0.46 & 0.18 & 39 & 10 & 13.5 & 0.6 & 30 & 9 & 54 & 13 & 1 & 2 \\ 
1b & 4.3 & 4.5 & 0.28 & 0.20 & 0.19 & 22 & 7 & 3.8 & 0.5 & 19 & 14 & 36 & 12 &  &  \\ 
2 & 1.2 & 1.9 & 0.34 & 0.97 & 0.26 & 37 & 12 & 23.2 & 0.9 & 24 & 4 & 52 & 14 & 2 &  \\ 
3 & 3.7 & 4.5 & 0.35 & 0.26 & 0.22 & 28 & 9 & 7.1 & 0.5 & 27 & 15 & 56 & 24 &  &  \\  \vspace{2mm}
4a & 3.0 & 4.4 & 0.40 & 4.23 & 0.46 & 77 & 45 & 398.9 & 2.6 & 94 & 3 & 73 & 28 & 7 & 14$^{\dagger}$ \\ 
4b & 3.4 & 4.3 & 0.35 & 1.68 & 0.42 & 49 & 20 & 95.1 & 1.5 & 57 & 5 & 58 & 20 & 2 &  \\ 
4c & 3.2 & 3.7 & 0.30 & 0.26 & 0.19 & 23 & 8 & 7.4 & 0.6 & 29 & 16 & 43 & 17 &  &  \\ 
4d & 3.3 & 4.2 & 0.42 & 0.40 & 0.40 & 88 & 34 & 45.9 & 0.9 & 116 & 41 & 83 & 23 &  &  \\ 
4e & 3.2 & 4.5 & 0.38 & 2.57 & 0.48 & 45 & 17 & 111.3 & 1.7 & 43 & 2 & 69 & 28 & 2 &  \\ \vspace{2mm}
4f & 3.6 & 4.3 & 0.37 & 0.23 & 0.41 & 53 & 21 & 19.2 & 0.7 & 82 & 50 & 63 & 23 &  &  \\ 
4g & 2.6 & 4.5 & 0.42 & 0.92 & 0.42 & 53 & 23 & 60.1 & 1.1 & 66 & 10 & 81 & 39 & 1 &  \\ 
4h & 2.2 & 4.3 & 0.39 & 1.51 & 0.39 & 46 & 20 & 63.6 & 1.3 & 42 & 4 & 71 & 33 & 2 &  \\ 
5 & 2.8 & 3.4 & 0.35 & 0.59 & 0.32 & 45 & 20 & 24.8 & 0.9 & 42 & 10 & 58 & 22 &  &  \\ 
6a & 3.0 & 3.7 & 0.37 & 0.51 & 0.35 & 48 & 22 & 32.2 & 0.9 & 63 & 17 & 62 & 24 &  &  \\ \vspace{2mm}
6b & 2.8 & 3.6 & 0.34 & 0.89 & 0.34 & 55 & 23 & 51.0 & 1.2 & 58 & 9 & 54 & 21 &  &  \\ 
7 & 3.3 & 4.7 & 0.35 & 0.34 & 0.33 & 35 & 11 & 12.2 & 0.7 & 36 & 15 & 57 & 22 &  & 2 \\ 
8 & 3.4 & 4.4 & 0.36 & 0.57 & 0.20 & 32 & 9 & 11.8 & 0.7 & 21 & 5 & 59 & 16 & 1 & 2 \\ 
9a & 4.4 & 5.4 & 0.34 & 0.92 & 0.37 & 48 & 22 & 50.0 & 1.1 & 54 & 8 & 52 & 16 & 2 & 4 \\ 
9b & 4.4 & 5.8 & 0.34 & 1.02 & 0.42 & 36 & 11 & 43.6 & 1.2 & 43 & 6 & 52 & 19 &  & 2 \\ \vspace{2mm}
10 & 6.5 & 7.0 & 0.29 & 0.45 & 0.19 & 25 & 8 & 11.7 & 0.9 & 26 & 11 & 39 & 9 &  &  \\ 
11 & 5.9 & 6.5 & 0.29 & 0.73 & 0.25 & 27 & 8 & 18.7 & 1.1 & 26 & 7 & 40 & 8 &  & 1 \\ 
12 & 6.1 & 6.8 & 0.30 & 1.86 & 0.67 & 30 & 11 & 67.7 & 2.0 & 36 & 4 & 43 & 12 & 1 & 2 \\ 
13 & 8.0 & 8.8 & 0.37 & 1.01 & 0.26 & 61 & 33 & 69.7 & 1.7 & 69 & 13 & 65 & 24 &  & 2 \\ 
14 & 6.7 & 8.0 & 0.43 & 1.25 & 0.40 & 50 & 17 & 77.1 & 1.8 & 62 & 10 & 87 & 26 &  & 1 \\  
\enddata 
\tablecomments{Columns: (1) Filament ID. F1, F4, F6, and F9, which are continuously connected in the 3D space but show substructures in their integrated intensity map, are divided into several sub-filaments (see text). (2) The largest and smallest $\vpeak$ in $\kms$. (3) Averaged total velocity dispersion of the molecule of mean mass ($\mu=2.8$) in $\kms$ (Fuller \& Myers 1992). 
(4) Length of filament measured from the easternmost (or northernmost) point to the westernmost (or southernmost) point of the skeleton in parsec. (5) Filament width in parsecs, i.e., FWHM of radial profile of H$_{2}$ column density. (6) Averaged $\nht$ and its dispersion in units of $10^{20}~ \rm cm^{-2}$. (7) H$_{2}$ mass of filament in $M_{\odot}$. (8) Mass per unit length of filament in $M_{\odot} ~ \rm pc^{-1}$. (9) Effective critical mass per unit length of filament derived with the mean total velocity dispersion in $M_{\odot}~ \rm pc^{-1}$ (see Section~\ref{sec:mlin} for details). The given uncertainties of $M$, $M_{\rm line}$, and $M_{\rm line}^{\rm crit}$ are estimated from the observational rms error. 
(10) Number of dense cores identified with $\nthp$ data (\S~\ref{sec:dcid}). Among the 22 dense cores found, one is linked with Clump22 in R.A., decl., and velocity space. } 

\end{deluxetable*}

\subsubsection{H$_{2}$ column density and mass}

H$_{2}$ column densities ($N_{\rm H_{2}}$) in the filaments were estimated from $\ceo~(1-0)$ data using the formula derived under the assumption of the local thermodynamic equilibrium (see Appendix~\ref{as:h2cd} for more details), and ranged between $\sim$6$\times 10^{20}$ and 2$\times$10$^{22}~ \rm cm^{-2}$. H$_{2}$ masses of the filaments are in the range from $\sim$4 to 400~$M_{\odot}$. The masses of the 30 clumps which are given in Table~\ref{tab:ppclumps} range from $\sim$1 to 7~$M_{\odot}$. The given uncertainty of mass is assigned by propagating the observational rms error. \\

\subsubsection{Length, width, and mass per unit length}

The lengths of the filaments are estimated along the skeleton without any correction for inclination. The widths of the filaments are measured from the radial profile of H$_{2}$ column density along the distance from the skeleton, where the FWHM of a Gaussian fit gives the width of the filament.

The lengths of the main filaments are about 0.3--4.2~pc and the widths are about 0.2--0.7~pc assuming the distances of 800~pc and 600~pc for the Cocoon (F10 to F14) and the Streamer (F1 to F9), respectively. The width is significantly larger than the filament width of 0.1~pc derived from the {\em Herschel} continuum data \citep[e.g.,][]{arzoumanian2019}. These large widths of the filaments in our study are probably due to the limited spatial resolution (0.14$-$0.2 pc) of our observations and/or relatively flat distribution of CO, which may be due to the possible depletion of CO molecules in the region of the filament with high column density and/or the lower dynamic range of the distribution of CO emission in comparison with the continuum emission. 


The mean masses per unit length ($\mlin$) of the filaments were calculated by simply dividing the mass by the length obtained above, and given in Table~\ref{tab:ppfila} with the errors propagated from the observational uncertainties. $\mlin$ is found to be between $\sim$20 and 120~$M_{\odot}~\rm pc^{-1}$. We derived the virial mass and the virial parameter, $\alpha_{\rm vir} = M_{\rm vir}/M$, for 30 clumps with a low aspect ratio ($\lesssim$~3) using Equation A5, instead of deriving the mass per unit length. Their estimated values are found in the ranges $\sim$0.5$- 5~M_{\odot}$ and $\sim$0.4$-$1.2, respectively. \\

\begin{figure*}
\includegraphics[width=1\textwidth,height=0.5\textwidth]{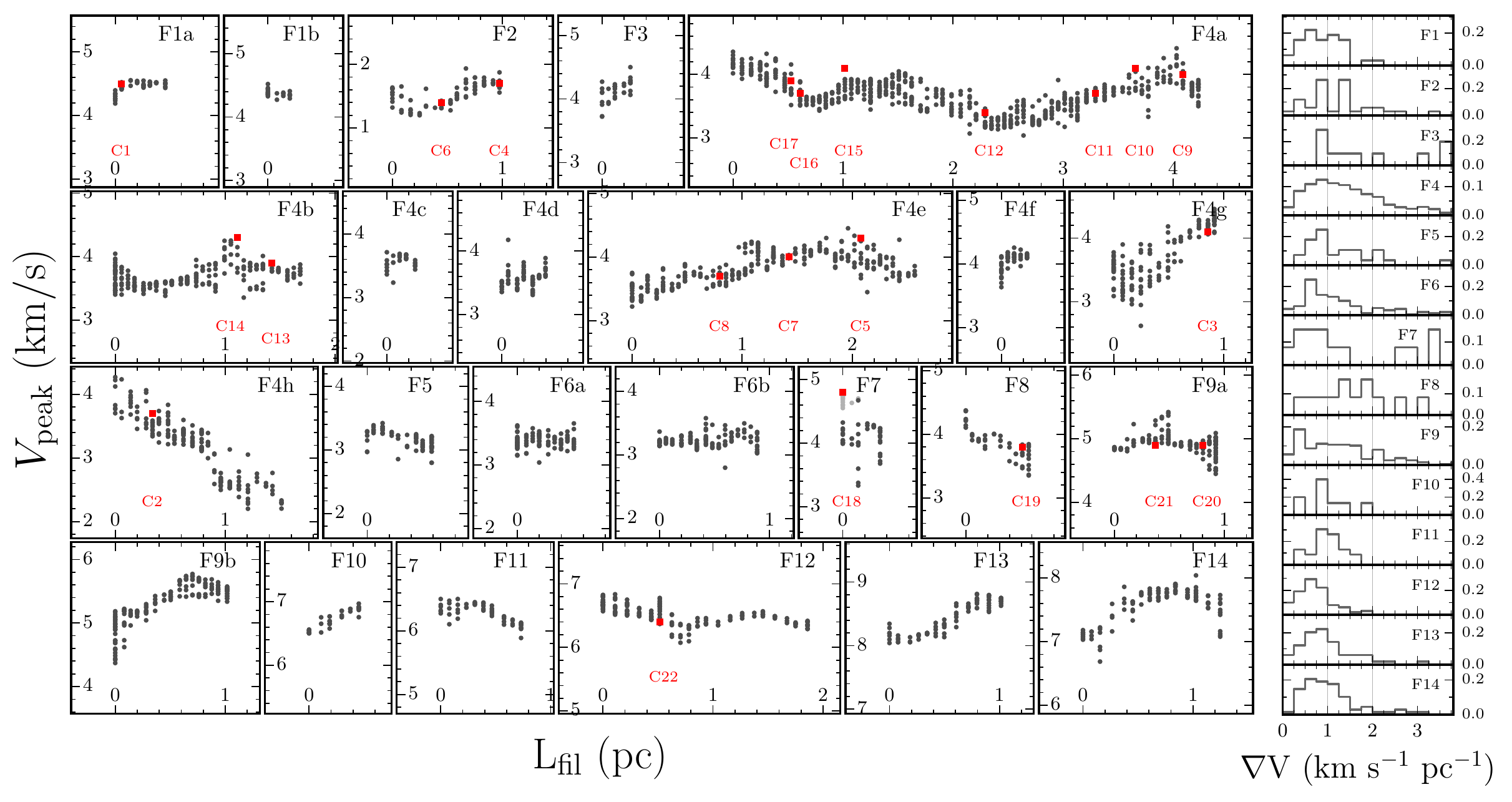}
\caption{Velocity structures along the filaments. Left: $L_{\rm fil}$ is the distance from the easternmost point along the skeleton of each filament. The $L_{\rm fil}$ for $V_{\rm peak}$ that is not on the skeleton uses $L_{\rm fil}$ of the nearest skeleton. $V_{\rm peak}$ from $\ceo~(1-0)$ is presented with solid black dots. $V_{\rm peak}$ of each core denoted with red squares is derived from the averaged $\nthp~(1-0)$ spectrum (see Figure~\ref{fig:corespec}). The gray dots around C18 are $V_{\rm peak}$ of Clump22 from $\ceo~(1-0)$, showing that C18 shares the PPV space with CL22, not with F7. Right: normalized histograms of velocity gradient ($\nabla V$) in $\kms~\rm pc^{-1}$ of each filament. \\ \label{fig:lfilv}}
\end{figure*}

\begin{figure*}
\includegraphics[width=1\textwidth]{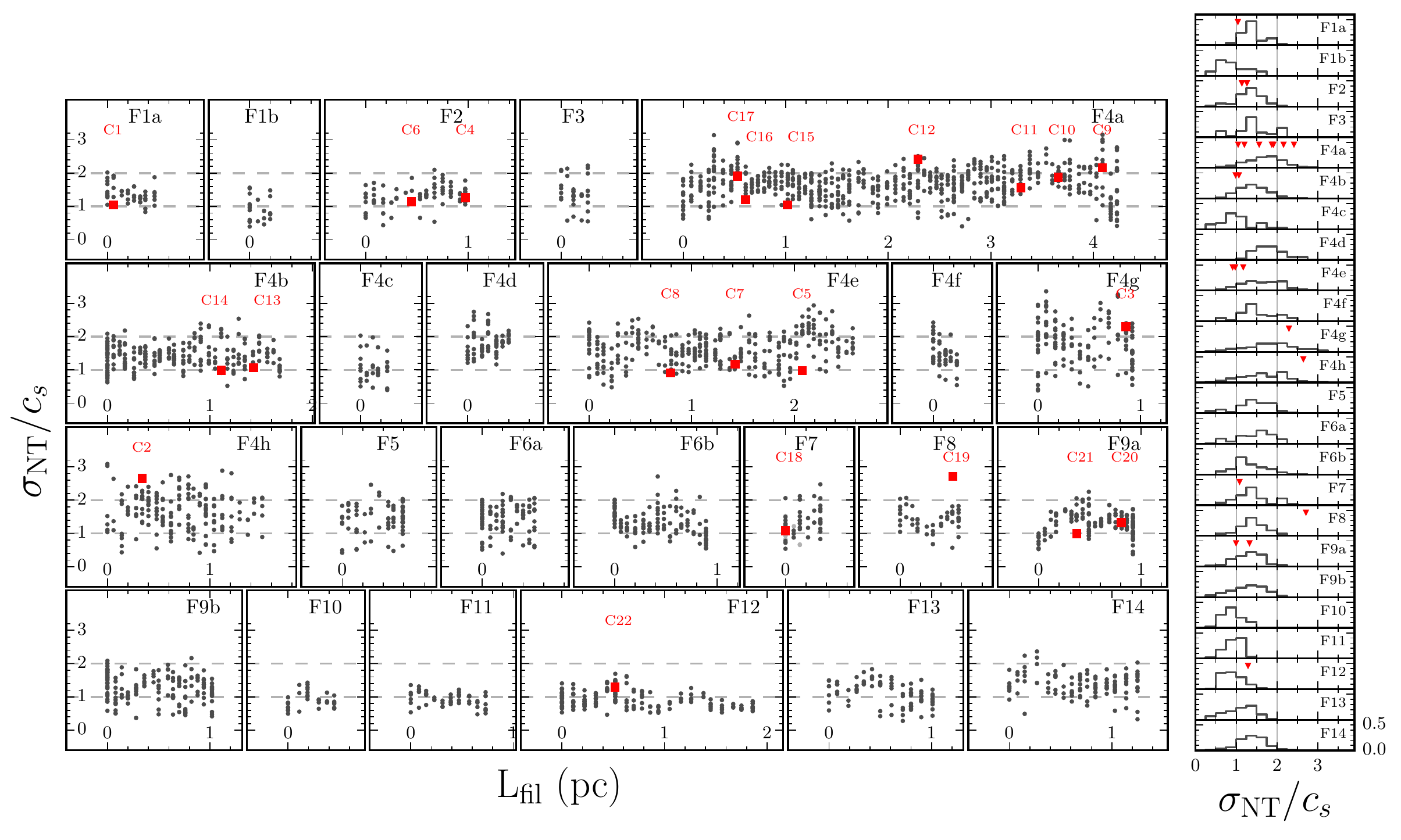}
\caption{Velocity dispersions in all identified filaments and dense cores. Left: Nonthermal velocity dispersions normalized by the local sound speed ($\sigma_{\rm NT} / c_{\rm s}$) are presented as a function of the position along each filament. The $\sigma_{\rm NT} / c_{\rm s}$ derived from $\ceo~(1-0)$ is denoted with solid black dots. $\sigma_{\rm NT} / c_{\rm s}$ derived from the averaged $\nthp~(1-0)$ spectrum of each dense core (see Figures~\ref{fig:corespec}) is shown as red squares and the core numbers are given. Right: normalized histograms of $\sigma_{\rm NT} / c_{\rm s}$ derived from $\ceo~(1-0)$ spectra. The red inverted triangles indicate $\sigma_{\rm NT} / c_{\rm s}$ of dense cores. \\
\label{fig:lfildv}}
\end{figure*}

\subsubsection{Global velocity field} \label{sec:vfield}

Global velocity fields of the filaments are presented in the middle panel of Figure~\ref{fig:fila0vpvw} and in Figure~\ref{fig:lfilv}. It is shown that the filaments have coherent and continuously changing velocity fields. Overall IC~5146 has filaments with LSR velocities from 1.2 to 8.8 $\kms$ and the velocity in each filament changes by $\sim 1 - 2~\kms$.

Figure~\ref{fig:lfilv} presents in detail the velocity fields of the filaments along the skeletons. $L_{\rm Fil}$ is the distance of a position in the filament measured along the skeleton from its easternmost point. There are several positions that are not on the skeleton. The $L_{\rm Fil}$ for those were assigned as the $L_{\rm Fil}$ of the locations on the skeleton nearest to such positions. The velocities monotonically increase or decrease along the skeletons in the majority of the filaments. However, oscillatory behavior as seen in the Taurus L1495/B213 complex and California L1478 \citep[e.g.,][]{tafalla2015,chung2019} can be also seen in the filament F4a. It was suggested that the increment of the velocity from the center to both the edges in F4a can be a signature of edge-driven collapse and fragmentations \citep{wang2019}. 

We measured velocity gradients employing the {\it gradient} numpy {\sc Python} code.\footnote{\url{https://numpy.org/doc/stable/reference/generated/numpy.gradient.html}} We used the Gaussian-decomposed velocity field of each filament, and the {\it gradient} algorithm computes the gradient at every position using second-order central differences. The right panel of Figure~\ref{fig:lfilv} shows the normalized histogram of the derived velocity gradient. This is an upper limit, because we have not corrected the inclination. Most of the filaments have velocity gradients between 0 and 2~$\kms ~\rm pc^{-1}$, and about 46\% of the velocity gradients are larger than 1~$\kms ~\rm pc^{-1}$. F4, which is the largest filament in IC~5146, appears to have distinguishably larger velocity gradients than the other filaments. Excluding F4, only 36\% of the filaments have larger $\nabla V$ than 1 $\kms ~\rm pc^{-1}$. Especially in the filaments F12, F13, and F14 of the Cocoon Nebula region, the proportion of positions having $\nabla V \leq 1~ \kms ~\rm pc^{-1}$ reaches 70\%. In F4, the proportion of positions with $\nabla V > 1~ \kms ~\rm pc^{-1}$ is 51\%, and this portion shows a significantly higher gradient when compared to the values from other filaments. \\

\subsubsection{Nonthermal velocity dispersion}

Nonthermal velocity dispersion ($\sigma_{\rm NT}$) is calculated with the following equation:
\begin{equation} 
	\sigma_{\rm NT} = \sqrt{\sigma_{\rm obs}^{2} - \frac{k_{\rm B} T}{m_{\rm obs}}} , 
\end{equation} 
where $\sigma_{\rm obs}$ is the total velocity dispersion estimated from the FWHM of the observed molecular line, $k_{\rm B}$ is the Boltzmann constant, $T$ is the gas temperature, and $m_{\rm obs}$ is the mass of the observed molecule of $\ceo$. Dust temperature from the {\em Herschel} data \citep{andre2010,arzoumanian2011} is used for the gas temperature after the data have been convolved to the pixel grid of TRAO and resampled to the pixel-grid of TRAO data. The uncertainty of $\sigma_{\rm NT} / c_{\rm s}$ is measured from that of the velocity dispersion and that of the dust temperature caused by the different resolution of {\em Herschel} and TRAO, and $\sigma_{\rm NT} / c_{\rm s}$ varies within about $\pm$5~\%.

Figure~\ref{fig:lfildv} shows the distribution of nonthermal velocity dispersion along the filament skeleton and the normalized histogram of $\sigma_{\rm NT} / c_{\rm s}$ of the filament. Filaments in IC~5146 appear to be mostly transonic $(1 < \sigma_{\rm NT}/c_{\rm s} < 2)$. However, some filaments (e.g., F4) show supersonic motions within a somewhat limited area. We found some correlation between the presence of dense cores in the filaments and the nonthermal velocity dispersion in the filaments. Most of all, dense cores tend to exist in the filaments with large nonthermal velocity dispersion mostly in the transonic regime. Furthermore, all the filaments that are dominant with subsonic motions ($\sigma_{\rm NT}/c_{\rm s} \sim 1$) are found to have no dense core except for one, filament F12. \\ 

\subsection{Comparison of $\ceo$ filaments with the Herschel continuum filaments} \label{sec:herskel}

\citet{arzoumanian2011} investigated the filament properties of IC~5146 using {\em Herschel} dust continuum data. They produced dust temperature ($T_{\rm dust}$) and H$_{2}$ column density ($N_{\rm H_{2}}$) maps of IC~5146 based on the 70 to 500 $\mu$m data, and applied the {\sc DisPerSE} algorithm \citep{sousbie2011a} to the curvelet image to find 27 filaments. In this section, we examine and compare the filaments of IC~5146 identified using $\ceo~(1-0)$ emission with those identified by \citet{arzoumanian2011}. We will refer to these two types of filaments as $\ceo$ filaments and {\em Herschel} continuum filaments, respectively.

\begin{figure*} \epsscale{1.17}
\plotone{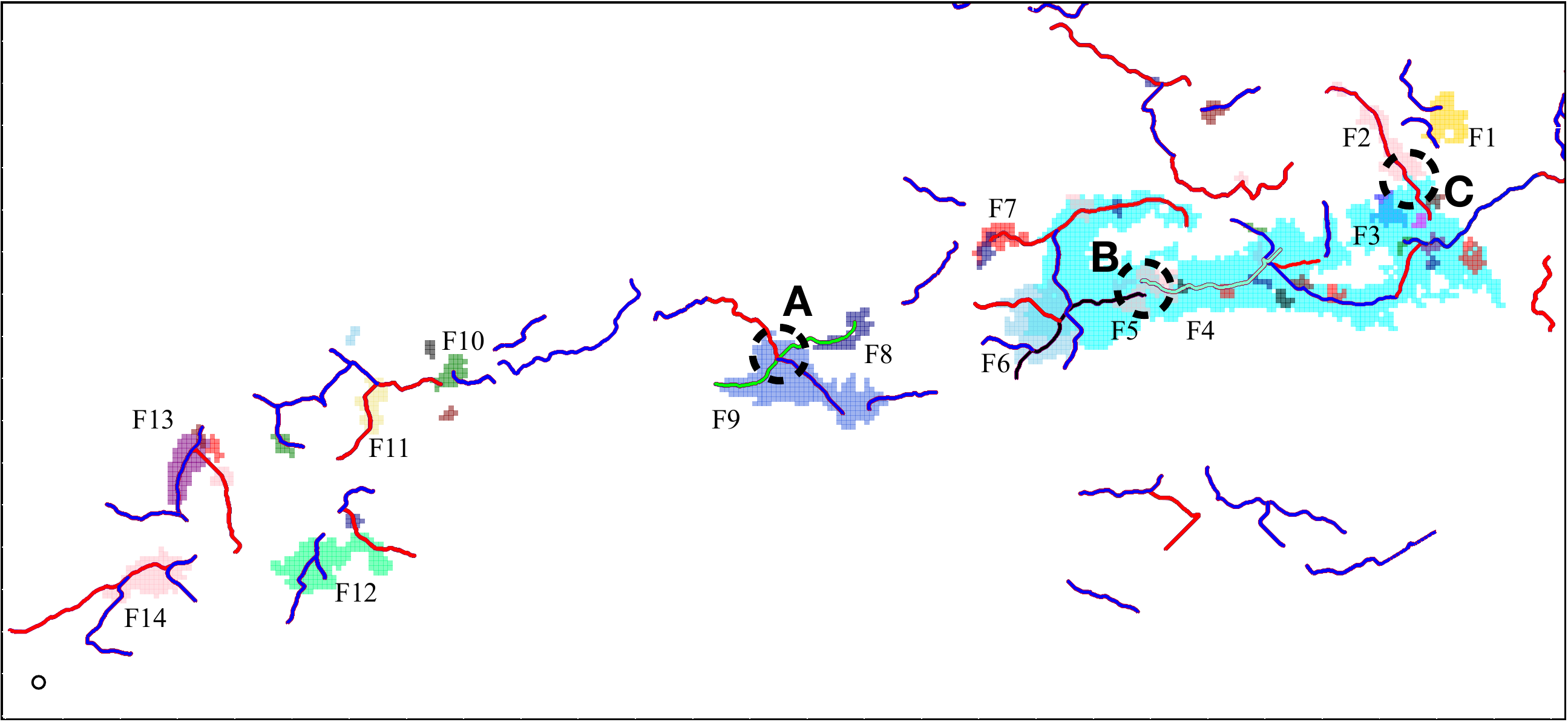}
\caption{Comparison between $\ceo$ filaments and {\em Herschel} continuum filaments. $\ceo$ filament outlines are drawn and tagged with their numbers. The skeletons of the {\em Herschel} continuum filaments are shown as colored lines. The dashed circles of A, B, and C show the representative regions where the $\ceo$ filaments and {\em Herschel} continuum filaments are differently identified. See the text. \label{fig:hskel}}
\end{figure*}

Figure~\ref{fig:hskel} compares the skeletons of the {\em Herschel} continuum filaments and the outlines of the $\ceo$ filaments. The main features of the filaments found with these two different tracers appear to be similar in the bright regions, but are different at the less bright region and the regions where there are multiple $\ceo$ filaments with different velocity components overlapped in the line-of-sight direction. 

Due to the superior sensitivity, the filaments identified using the {\em Herschel} continuum data are also found in areas where $\ceo~(1-0)$ emission is absent. These {\em Herschel} continuum filaments are well connected to the filaments where the $\ceo~(1-0)$ line is bright, so that one $\ceo$ filament consists of several {\em Herschel} continuum filaments. For example, in the case of F9 in region~A, three different {\em Herschel} continuum filaments depicted in bright green, red, and blue meet, but only one $\ceo$ filament, F9, is identified. In contrast, the filaments identified using the dust continuum emission from the {\em Herschel} can be separated into two or more structures in the analyses done with the $\ceo~(1-0)$ emission since the {\em Herschel} dust emission is detected in relatively less bright regions than $\ceo~(1-0)$. For example, the {\em Herschel} continuum filament shown in bright green in region~A includes the $\ceo$ filaments F8 and F9. The nearest points of F8 and F9 $\ceo$ filaments have similar velocities of $\sim$4.4~$\kms$, but they are slightly off (about one FWHM beam size) in the plane of the sky. 

Identification of the continuum filaments cannot be consistent with that of the $\ceo$ filaments if there are multiple filaments with different velocities along the line of sight. The {\em Herschel} continuum filament illustrated with a red line in region C is overlaid on the $\ceo$ filaments F2 and F4. As indicated by white arrows in the $p-v$ diagram (Figure~\ref{af:pvds} in the Appendix), F2 and F4 have totally different velocities in the $\ceo~(1-0)$ observations even though they appear to be continuously connected in the continuum emission. Another inconsistency in the identification of the filaments in line and continuum emission can be seen around the $\ceo$ filament F5 in region B. F5 overlaps with F4 but has a different velocity to it. In the dust continuum emission, the overlapping region appears brighter and the ridges found in the {\em Herschel} continuum filament are separated into two (skeletons with black and lime colors). \\

\section{Dense Core Properties} \label{sec:dc}

\subsection{Dense Core Identification} \label{sec:dcid}

The $\nthp$ line is usually optically thin \citep[except toward the central regions of pre-stellar cores; e.g.,][]{keto2010}, and is an appropriate tracer of dense cores in nearby star-forming regions. We identified dense cores by applying the \textsc{FellWalker} clump-finding algorithm \citep{berry2015} to the $\nthp~(1-0)$ integrated intensity images. In running this algorithm, an object with a peak intensity higher than 3$\sigma$ and the size larger than 2.5$\times$the beam size of 52$^{\as}$ is identified as a real dense core. Pixels with intensities $> 0.5 \sigma$ are allowed to be associated with a peak. Two neighboring peaks are considered as separate if the difference between the peak values and the minimum value (dip value) between the peaks is larger than 1$\sigma$, the two peaks are considered as separate. 

\begin{figure*} \epsscale{1.17}
\plotone{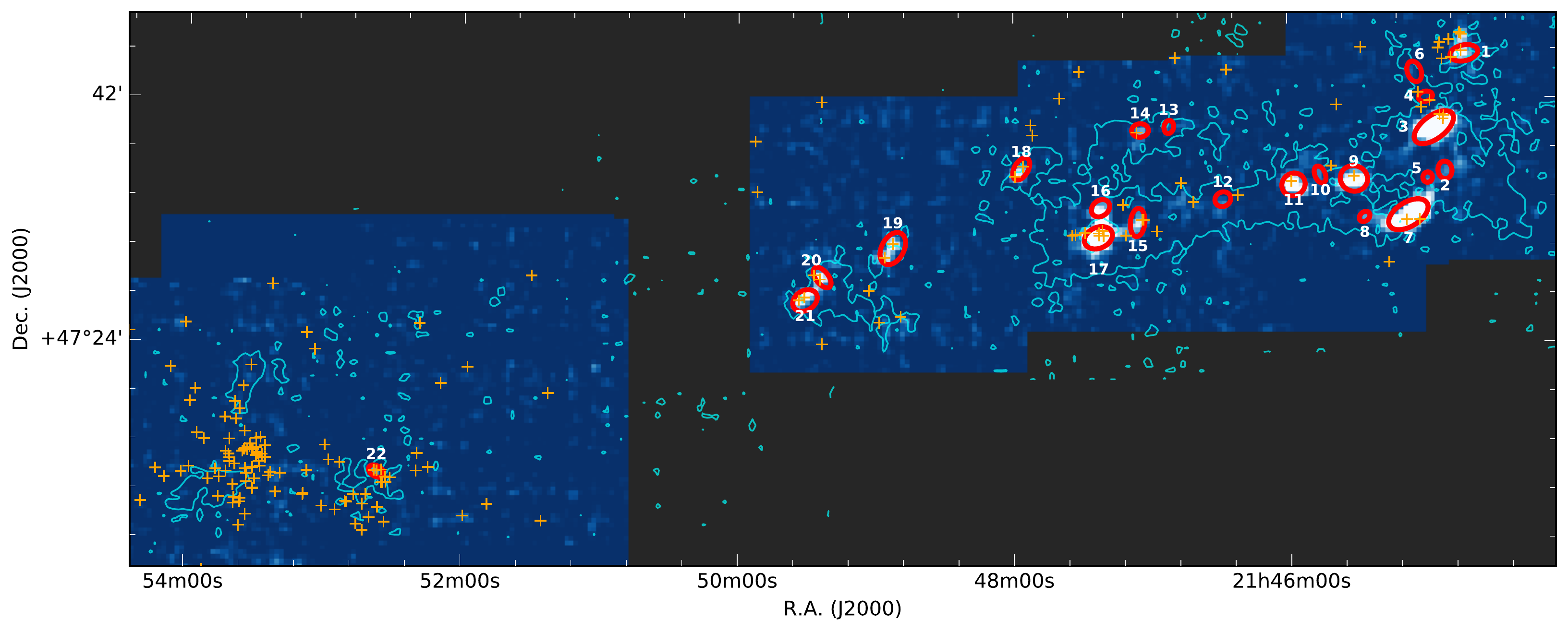}
\caption{Dense cores identified with {\sc FellWalker} are marked with red ellipses and the core ID numbers are labeled in white. The background color image is the integrated intensity map of $\nthp~(1-0)$. The $\ceo~(1-0)$ 3$\sigma$ level contour is given with aqua color. Positions of YSO candidates from {\em Spitzer} \citep{harvey2008} and 70~$\mu$m point sources from {\em Herschel}/PACS Point Source Catalogue \citep[HPPSC;][]{poglitsch2010} are presented as orange crosses. \label{fig:fellcore}}
\end{figure*}

\begin{figure*} \epsscale{1.15}
\plotone{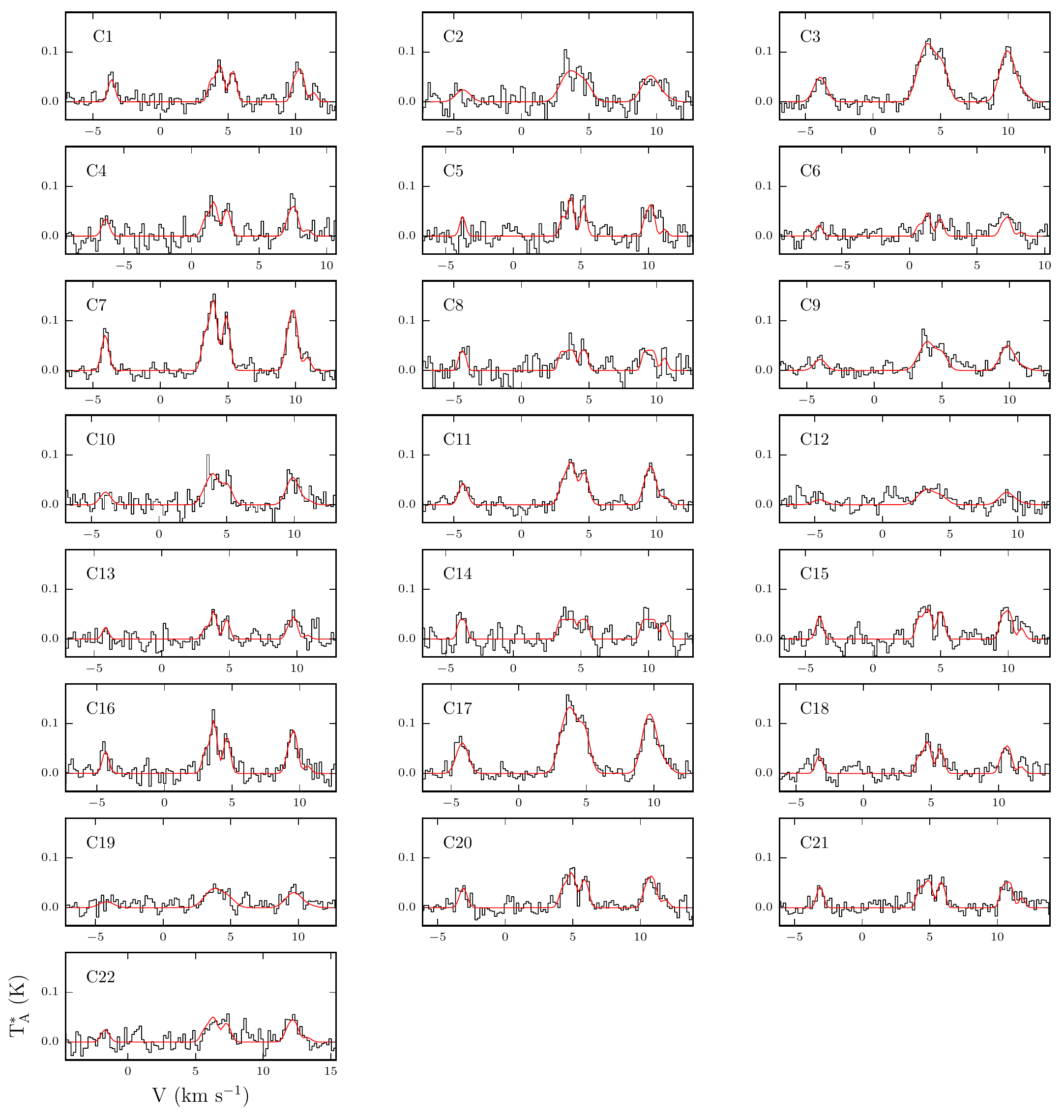}
\caption{The averaged $\nthp~(1-0)$ spectra for dense cores. Red profiles overlaid on the spectra are the results of the hyperfine fit with seven Gaussian components for $\nthp$ lines. \label{fig:corespec}}
\end{figure*}

In total, we found 22 dense cores: one core in the Cocoon Nebula and 21 cores in the Streamer. Information about the identified dense cores is in Table~\ref{tab:ppcore}. The position, size, and position angle (PA) of the dense cores are the results that the applied \textsc{FellWalker} algorithm gives. $V_{\rm peak}$ and $\Delta V$($\nthp$) are estimated by simultaneously fitting seven Gaussian functions for the seven hyperfine components of $\nthp~(1-0)$ using the line parameters given by \citet{caselli1995}. Filament ID that shares the PPV space with the dense core is provided. It is noticeable that C18 is not related with the filament F7, but with the clump CL22, while every other dense core shares the PPV space with various filaments and not with the clumps. There are 14 starred cores that are well matched with the positions of the YSO candidates from {\em Spitzer} \citep{harvey2008} and 70~$\mu$m point sources from {\em Herschel}/PACS Point Source Catalogue \citep[HPPSC;][]{poglitsch2010} and eight starless cores where no YSOs are found. The positions and averaged spectra of the dense cores are presented in Figures~\ref{fig:fellcore} and \ref{fig:corespec}. \\

\subsection{Mass and Virial Mass} \label{sec:coremass}

We derive masses of dense cores with the integrated intensity of $\nthp~(1-0)$ and the virial masses as described in Appendix~\ref{as:coremass} and tabulate the quantities in Table~\ref{tab:ppcore}. The virial parameters, $\alpha_{\rm vir} = M_{\rm vir}/M$, of the dense cores range between $\sim$0.3 and 2.6. A significant number of dense cores (10 among the 22) are virialized ($\alpha \leq 1$), and three dense cores have $1 < \alpha < 2$ and are likely close to gravitationally bound considering the large uncertainty in $\alpha$ (a factor of $\sim$2). We found three most massive dense cores C3, C7, and C17. Their masses are estimated to be $\gtrsim 15~M_{\odot}$ and their virial parameters are found to be quite small ($\alpha_{\rm vir} \leq 0.5$), giving a hint that there must be active star formation. In fact, they are found to contain multiple YSOs. \\  

\begin{deluxetable*}{lcccrrcc r @{$\pm$} l r @{$\pm$} l r @{$\pm$} l cc}
\tablecaption{Information about the Identified Dense Cores \label{tab:ppcore}}
\tablehead{
\colhead{Core ID} &
\multicolumn{2}{c}{Position} &
\colhead{} &
\multicolumn{2}{c}{Size\tablenotemark{$^{\rm a}$}} & 
\colhead{PA} &
\colhead{$\vpeak^{\nthp}$} &
\multicolumn{2}{c}{$\Delta V^{\nthp}$} &
\multicolumn{2}{c}{$M_{\rm obs}$\tablenotemark{$^{\rm b}$}} &
\multicolumn{2}{c}{${\alpha_{\rm vir}}$\tablenotemark{$^{\rm b}$}} &
\colhead{Fil. ID\tablenotemark{$^{\rm c}$}} &
\colhead{YSOs\tablenotemark{$^{\rm d}$}}\tabularnewline 
\cline{2-3} \cline{5-6} 
\colhead{} &
\colhead{R.A.} &
\colhead{Decl.} &
\colhead{} &
\colhead{Major} & 
\colhead{Minor} &
\colhead{} &
\colhead{} &
\colhead{} &
\colhead{} &
\colhead{} &
\colhead{} &
\colhead{} \tabularnewline
\colhead{} &
\colhead{(hh:mm:ss)} &
\colhead{(dd:mm:ss)} & 
\colhead{} &
\colhead{(pc)} & 
\colhead{(pc)} & 
\colhead{(deg)} &
\colhead{($\kms$)} &
\multicolumn{2}{c}{($\kms$)} &
\multicolumn{2}{c}{($M_{\odot}$)} &
\multicolumn{2}{c}{} &
\colhead{} &
\colhead{}}
\startdata
C1 & 21:44:42.5 & +47:45:18.5 & & 0.33 & 0.14 & 101 & 4.50 & 0.49 & 0.05 & 4.2 & 3.0 & 0.5 & 0.4 & F1a & 3 \\ 
C2 & 21:44:51.4 & +47:36:42.3 & & 0.17 & 0.10 & 8 & 3.80 & 1.20 & 0.20 & 1.8 & 1.2 & \multicolumn{2}{c}{$~~\cdots$} & F4h &  \\ 
C3 & 21:44:56.1 & +47:39:51.7 & & 0.57 & 0.26 & 126 & 4.20 & 1.04 & 0.04 & 23.7 & 16.7 & 0.5 & 0.3 & F4g & 4 \\ 
C4 & 21:44:59.6 & +47:42:09.0 & & 0.11 & $\leq$0.15 & 104 & 1.70 & 0.59 & 0.05 & 0.5 & 0.4 & $\leq$2.7 & 2.3 & F2 & 2 \\ \vspace{2mm}
C5 & 21:44:59.1 & +47:36:11.2 & & $\leq$0.15 & $\leq$0.15 & 172 & 4.29 & 0.47 & 0.06 & 0.8 & 0.6 & $\leq$1.5 & 1.4 & F4e &  \\ 
C6 & 21:45:04.3 & +47:43:59.5 & & 0.23 & 0.09 & 20 & 1.40 & 0.54 & 0.06 & 1.0 & 0.7 & 1.7 & 1.5 & F2 &  \\ 
C7 & 21:45:07.7 & +47:33:28.2 & & 0.54 & 0.27 & 121 & 3.98 & 0.57 & 0.02 & 16.5 & 11.6 & 0.3 & 0.2 & F4e & 2 \\ 
C8 & 21:45:26.8 & +47:33:20.8 & & 0.02 & $\leq$0.15 & 131 & 3.70 & 0.38 & 0.07 & 0.9 & 0.6 & \multicolumn{2}{c}{$~~\cdots$} & F4e &  \\ 
C9 & 21:45:31.2 & +47:36:09.5 & & 0.32 & 0.29 & 71 & 4.00 & 0.89 & 0.06 & 7.3 & 5.2 & 0.9 & 0.7 & F4a & 2 \\ \vspace{2mm}
C10 & 21:45:46.0 & +47:36:28.2 & & 0.18 & $\leq$0.15 & 22 & 4.00 & 0.87 & 0.17 & 1.4 & 1.0 & \multicolumn{2}{c}{$~~\cdots$} & F4a &  \\ 
C11 & 21:45:57.7 & +47:35:45.4 & & 0.27 & 0.24 & 126 & 3.80 & 0.68 & 0.05 & 6.4 & 4.5 & 0.5 & 0.4 & F4a & 1 \\ 
C12 & 21:46:28.6 & +47:34:42.8 & & 0.14 & 0.11 & 116 & 3.40 & 1.13 & 0.41 & 1.0 & 0.7 & \multicolumn{2}{c}{$~~\cdots$} & F4a &  \\ 
C13 & 21:46:52.0 & +47:40:03.7 & & 0.09 & $\leq$0.15 & 169 & 3.90 & 0.49 & 0.08 & 0.6 & 0.5 & \multicolumn{2}{c}{$~~\cdots$} & F4b &  \\ 
C14 & 21:47:04.5 & +47:39:49.1 & & 0.16 & 0.08 & 95 & 4.30 & 0.47 & 0.10 & 1.7 & 1.2 & \multicolumn{2}{c}{$~~\cdots$} & F4b & 1 \\ \vspace{2mm} 
C15 & 21:47:05.8 & +47:33:02.3 & & 0.33 & 0.11 & 174 & 4.10 & 0.49 & 0.04 & 3.8 & 2.7 & 0.5 & 0.4 & F4a & 2 \\ 
C16 & 21:47:22.0 & +47:34:06.4 & & 0.21 & 0.13 & 128 & 3.67 & 0.54 & 0.03 & 3.8 & 2.7 & 0.4 & 0.3 & F4a &  \\ 
C17 & 21:47:23.0 & +47:31:55.9 & & 0.35 & 0.22 & 113 & 3.90 & 0.94 & 0.06 & 17.0 & 12.0 & 0.4 & 0.3 & F4a & 8 \\ 
C18 & 21:47:56.5 & +47:37:01.3 & & 0.27 & 0.09 & 150 & 5.00 & 0.52 & 0.04 & 2.9 & 2.1 & 0.6 & 0.5 & CL22$^{\rm e}$ & 2 \\ 
C19 & 21:48:52.6 & +47:31:10.5 & & 0.42 & 0.24 & 151 & 3.90 & 1.04 & 0.10 & 5.4 & 3.8 & 1.6 & 1.4 & F8 & 2 \\ \vspace{2mm}
C20 & 21:49:23.4 & +47:29:02.0 & & 0.27 & 0.06 & 37 & 5.00 & 0.59 & 0.06 & 3.0 & 2.1 & 0.6 & 0.5 & F9a & 2 \\ 
C21 & 21:49:30.9 & +47:27:20.4 & & 0.30 & 0.20 & 122 & 4.90 & 0.45 & 0.40 & 4.4 & 3.1 & \multicolumn{2}{c}{$~~\cdots$} & F9a & 2 \\ 
C22 & 21:52:36.7 & +47:14:33.4 & & 0.20 & $\leq$0.20 & 60 & 6.40 & 0.64 & 0.12 & 1.6 & 1.2 & \multicolumn{2}{c}{$~~\cdots$} & F12 & 7 \\ 
\enddata
\tablenotemark{}{} \\
\tablenotemark{$^{\rm a}$}{The core size is corrected for the beam smearing effect with the equation of $s_{\rm corr}=\sqrt{{s_{\rm obs}}^{2} - \theta^{2}}$ where $s_{\rm corr}$ is the corrected size, $s_{\rm obs}$ is the observed size, and $\theta$ is the standard deviation of the Gaussian beam profile (Berry 2015). 
For the cores with smaller observed size than $\theta$, the beam size is given as the upper limit of the core size.}\\ 
\tablenotemark{$^{\rm b}$}{The uncertainty of $M_{\rm obs}$ is estimated from the observational rms error and that of $\alpha_{\rm vir}$ is propagated from those observational uncertainties.}\\
\tablenotemark{$^{\rm c}$}{Filament ID given in Table~\ref{tab:ppfila}.}\\
\tablenotemark{$^{\rm d}$}{Number of YSO candidates from Spitzer (Harvey et al. 2008) 
and HPPSC (Poglitsch et al. 2010).} \\
\tablenotemark{$^{\rm e}$}{Dense core C18 is located in the overlapped region of filament F7 and clump CL22, but its velocity is the same as that of CL22.}

\end{deluxetable*}
\vspace{5mm}

\subsection{Chemical and Dynamical Properties of Starless Cores} \label{sec:corechem}

SO and CO molecules are known to easily deplete in the cold dense cores while the $\nthp$ molecule can survive well at the very evolved stage of the dense core \citep[e.g.,][]{caselli1999,tafalla2006}. In addition, $\nhtd$, deuterated ammonia, can survive in the gas phase in the interior of the highly evolved prestellar cores, and comparison of the distributions of these molecules around the dense cores would be very useful to infer how the cores are chemically evolved \citep[e.g.,][]{crapsi2007}. Here we explain how the distribution of these molecules is different from core to core, especially for starless cores (C2, C5, C6, C8, C10, C12, C13, and C16).

Figures~\ref{af:corechem1} -- \ref{af:corechem5} in the Appendix present the integrated intensity maps of $\nthp~(1-0)$, SO~$(3_{2}-2_{1})$, and $\ceo~(1-0)$ of dense cores. The $\nthp~(1-0)$ emission in our sample of starless cores found in IC~5146 is mostly weak but centrally concentrated, and its peak positions are approximately coincident with those of the 250~$\mu$m continuum emission. The SO~$(3_{2}-2_{1})$ line seems also to trace the dense region of the cores. But its emission is as weak as $\nthp~(1-0)$ in most starless cores. On the other hand, the $\ceo~(1-0)$ emission appears much brighter than the other two molecular lines, tracing wide regions of the clouds. Looking at the distribution of the $\ceo~(1-0)$ emission toward the $\nthp$ starless cores, we do not see any significant hint of CO freeze-out in those cores. The differences among the spatial distributions of our tracers are probably from a combination of the critical densities and the chemical properties of the tracers. The $\ceo~(1-0)$ line with relatively low critical density ($\rm \sim 1.9 \times 10^3~cm^{-3}$) would be easily detected over the wide area of the less dense clouds, while tracers such as SO and $\nthp$ with relatively higher critical densities ($\rm \gtrsim 10^5~cm^{-3}$) are detected only in the very dense core regions only (see $n_{\rm crit}$ in Table~\ref{tab:lines}). However, all those distributions would be modified with their chemical properties such as the freeze-out or enhancement in the cold dense regions of the cores. In the case of starless cores in IC~5146, weak emission in $\nthp$ and SO, and no significant depletion in CO, give a hint that most of starless cores may not be highly evolved.

$\nhtd~(1_{11}-1_{01})$ is not detected at the rms level of 0.06~K[T$_{\rm A}^{\ast}$] with 0.1~$\kms$ channel resolution. 
We estimated a 3$\sigma$ upper limit of the $\nhtd$ abundance using the column density obtained from Equation (4) in \citet{wienen2021} as $\sim 5\times 10^{-10}$. This upper limit abundance can be compared with the abundance of the $\nhtd$ as a function of time from the {\tt NAUTILUS} chemical model by \citet{majumdar2017}, giving an upper limit for the age of the starless cores in our study as a few million years, which is consistent with the range of the statistical time scale of starless dense cores (e.g., Lee \& Myers 1999). This indicates that our dense cores may have a wide range of ages. However, many of dense cores found in IC~5146 may not be fully evolved to have detectable $\nhtd$ emission. This is also consistent with the idea that most of dense starless cores are chemically young from their characteristic distribution of $\nthp$, SO, and $\ceo$. On the other hand there are several other starless cores showing a signature of highly evolved status such as gas infalling motions as mentioned below. Thus we cannot rule out another possibility that some of the starless cores are highly evolved, but $\nhtd$ may be present onlt in the limited small area of the central region of the core and thus be beam-diluted within our large beam of the TRAO telescope due to the relatively large distance to IC~5146. In fact, the high critical density of the $\nhtd~(1_{11}-1_{01})$ line (Table~\ref{tab:lines}) suggests that even in the case of dynamically evolved cores, the small emitting region with volume density around $\rm 10^6~cm^{-3}$ could be heavily diluted with the present observations.

Many dense cores have been found to show inward motions, which are considered as one of the essential conditions for star formation \citep[e.g.,][]{lee1999,evans2015,yen2019,kim2021}. Inward motions can be traced with optically thick (e.g., $\hcop$, CS) and thin lines (e.g., $\nthp$) by detecting an asymmetric profile of double peaks in the optically thick line where the blue peak is brighter than the red peak, and a single peaked profile in the optically thin line.

To examine any infall motions around the cores, the $\hcop~(1-0)$ and CS~(2-1) spectra are presented in Figures~\ref{af:corechem1} -- \ref{af:corechem5} in the Appendix. Cores C2, C6, and C16 are infall candidates. Dense cores C2 and C16 show a blue asymmetric infall signature around the core center. C6 does not show a clear central dip but a brighter blue peak and a red shoulder. CS~(2-1) spectra of the infall candidates do not show as significant blue asymmetries as $\hcop~(1-0)$ spectra, but they mostly present brighter blue peaks and red shoulders. \\ 

\section{Discussion} \label{sec:disc}

\subsection{Nonthermal motion and systemic velocity of filaments and dense cores}

A colliding model has been proposed where filaments can form by the collision of turbulent flows, and the dissipation of turbulence creates the dense cores \citep[e.g.][]{ballesteros1999,padoan2002}. It is expected that if the cores form from the collision between turbulent filaments, the initially turbulent filaments would remain supersonic while the cores formed in them would be in subsonic motion. Hence, diagnosing the kinematic properties of the filaments and dense cores would examine this collision mechanism for the formation of filaments and dense cores. 

\begin{figure*} \epsscale{1.17} \plotone{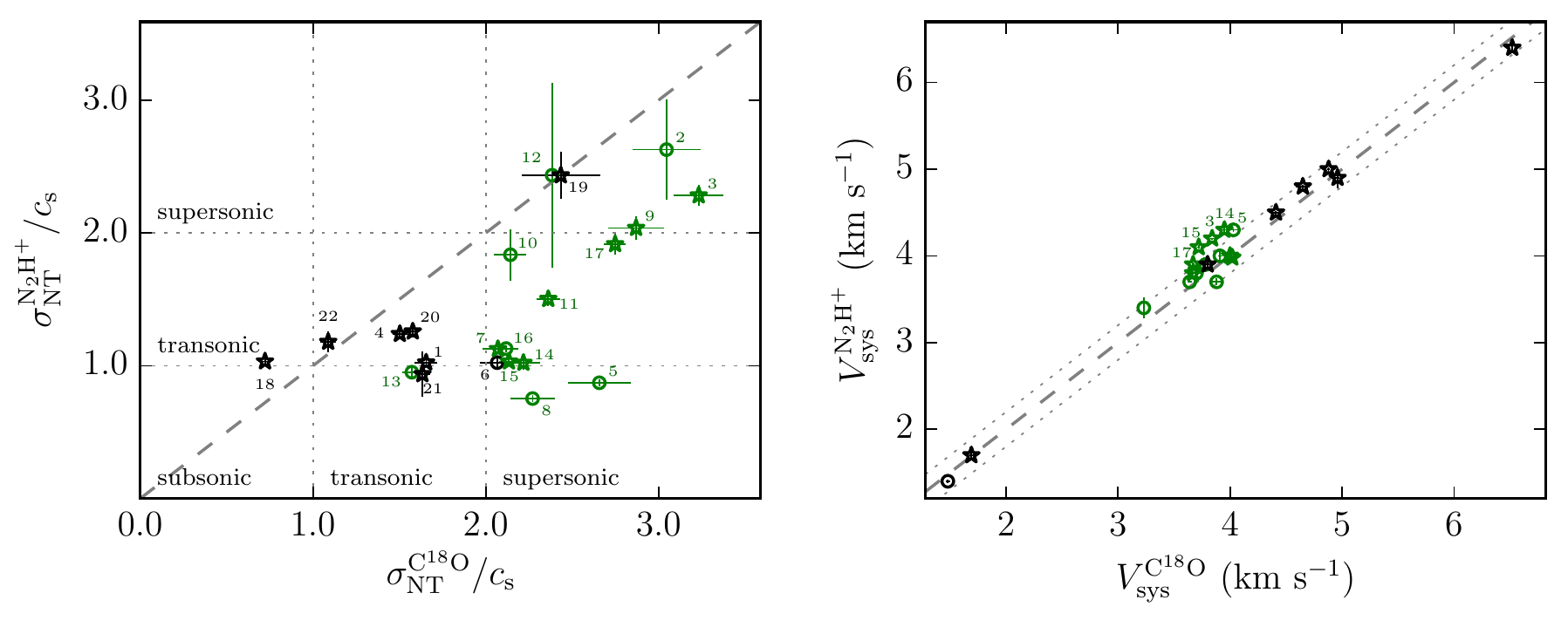} 
\caption{Left: Nonthermal velocity dispersions of the dense cores ($\sigma_{\rm NT}^{\nthp} / c_{\rm s}$) and their surrounding filaments ($\sigma_{\rm NT}^{\ceo} / c_{\rm s}$). The gray dashed line connects the positions where the nonthermal velocity dispersions for the dense cores and filaments are identical. The star symbols are to indicate the samples where YSOs are believed to exist by \citet{harvey2008} and \citet{poglitsch2010}. The green symbols are for the samples belonging to filament F4. Right: peak velocity of the dense cores traced by $\nthp~(1-0)$ and the surrounding materials of filaments traced by $\ceo~(1-0)$. The gray dashed and dotted lines indicate the line of equality and that line displaced by the sound speed (about 0.2~$\kms$ at 13~K). \vspace{2mm} \label{fig:vpeakdcs}}\end{figure*}

Figure~\ref{fig:vpeakdcs} displays the nonthermal velocity dispersions and systemic velocities for the $\ceo$ and $\nthp$ gas to compare the line width and systemic velocity properties between the filaments and the dense cores. The velocity dispersion and systemic velocity of the $\ceo$ gas are derived from the Gaussian fitting of the $\ceo~(1-0)$ spectrum averaged over the area of the dense core, and thus would represent the kinematic properties of the filament material in the line-of-sight direction of the dense core. The velocity dispersion and systemic velocity of $\nthp~(1-0)$ were obtained in the same manner as those of $\ceo~(1-0)$.

In the left panel of Figure~\ref{fig:vpeakdcs}, the nonthermal velocity dispersions of $\ceo$ and $\nthp~(1-0)$ are compared. All the dense cores have smaller $\sigma_{\rm NT}^{\nthp} / c_{\rm s}$ than $\sigma_{\rm NT}^{\ceo} / c_{\rm s}$ except the dense cores C12, C18, C19, and C22, where $\sigma_{\rm NT}^{\nthp} / c_{\rm s}$ is almost the same as $\sigma_{\rm NT}^{\ceo} / c_{\rm s}$ within their uncertainties. The ratios of $\sigma_{\rm NT}^{\nthp} / c_{\rm s}$ to $\sigma_{\rm NT}^{\ceo} / c_{\rm s}$ for the most of the dense cores range between $\sim0.4$ and  $\sim0.7$.

Similar kinetic features between dense cores and the surrounding filaments have also been found in a previous study of L1478 in the California cloud by \citet{chung2019}. In that study, the dense cores are divided into two groups according to their nonthermal velocity dispersions. One group contains cores where the nonthermal velocity dispersions are similar to those of their surrounding filaments, and the other group contains cores where the nonthermal velocity dispersions are smaller than those of their surrounding filaments. The cores in the former group are found to be in a single-shaped filament but the cores in the latter group are located in the hub-filament structures.

Likewise, this study also shows that all the dense cores (except C1, C6, and C21) having $\sigma_{\rm NT}^{\nthp} / c_{\rm s}$ smaller than $\sigma_{\rm NT}^{\ceo} / c_{\rm s}$ by about the sound speed ($c_{\rm s}$) are located in the filament F4 showing hub-filaments structures in the easternmost (near C17) and also westernmost (near C3) regions. In addition, the central region of F4, near C11, appears to be a hub where the sub-filaments of F4a, F4e, and F4f meet. C15 is also located at the intersection of F4a and F4d (see Fig.~\ref{af:accRfig} in the Appendix). These nonthermal motions of dense cores, i.e., $\sigma_{\rm NT}^{\nthp} / c_{\rm s}$ smaller than $\sigma_{\rm NT}^{\ceo} / c_{\rm s}$, are consistent with the expectations of the collision scenario. 

One thing we should notice here is that the nonthermal velocity dispersion may arise from turbulence and/or any bulk motions such as gas infall or bipolar outflow. Infall motions are found in three starless cores, C2, C6, and C16 (Section~\ref{sec:corechem}), and CO outflows have been observed in  five starred cores, C3, C7, C11, C17, and C18 \citep{dobashi2001}. However, such motions can be hardly traced with $\nthp$ and $\ceo~(1-0)$ because these lines are usually optically thin and poorly sensitive to those gas motions \citep[e.g.,][]{lee2001,fuente2012,lo2015}. Most of all, the infall speeds found in starless cores are quite low, of the order of 0.1 $\rm km~s^{-1}$, making it hard for them to play a major role in broadening of the lines. Moreover, these lines likely trace the gas at the centers of the outflows in the starred cores, i.e., the outflow motions can be partially traced by these lines only at the position of the driving source \citep[e.g.,][]{su2004,lee2018}. Thus in our case where the line profiles under discussion were obtained from averaging profiles over the area of the dense cores, it is unlikely that infall or outflow motions would significantly affect the broadening of the lines and both $\sigma_{\rm NT}^{\ceo}$ and $\sigma_{\rm NT}^{\nthp}$. Hence, the main origin of the nonthermal velocity dispersions obtained from the $\ceo$ and $\nthp~(1-0)$ lines is thought to be from the turbulence. Therefore, the different nonthermal velocity dispersions of $\ceo$ and $\nthp~(1-0)$ of the cores can be interpreted with the dissipation of  turbulence after the collisions of turbulent filaments. Hence, F4 and most of the dense cores in it, at least the cores in the hubs (C3, C5, C15, and C17), might have been formed as a result of collisions of turbulent flows.

The systemic velocity shift between the $\ceo$ and $\nthp$ gas also supports the collision scenario for the formation of dense cores in the hubs. In the right panel of Figure~\ref{fig:vpeakdcs}, the systemic velocities of the dense cores and the filament material correlate well with each other, and in the majority of them the differences are smaller than or similar to the sound speed ($\sim$0.20~$\kms$ at 13~K which is the mean dust temperatures of the dense cores measured from the {\em Herschel} data). However, C3, C5, C14, C15, and C17 have a larger offset than the sound speed between the systemic velocities of the dense cores and the filamentary gas. The systemic velocity shift between the $\nthp$ and $\ceo$ gas of the five cores is $0.32 \pm 0.06 ~\kms$. This discrepancy is firstly attributed to the uncertain $V_{\rm sys}^{\nthp}$ due to the low S/N. However, the $\nthp~(1-0)$ spectra of C3 and C17 show high S/N, and the velocity difference between the $\ceo$ and $\nthp~(1-0)$ emission can be caused by the merging of filaments with different velocities.

The relative core-to-envelope motions have been studied to investigate the core formation mechanism, but no displacement in the systemic velocities between the dense gas and the surrounding gas has been found in the low-mass star-forming regions \citep[e.g.,][]{kirk2007,hacar2011,punanova2018,chung2019}. However, recently a velocity difference of $\sim$0.3~$\kms$ between $\ceo$ and $\nthp~(1-0)$ lines was observed in the infrared dark clouds (IRDCs) G035.39-00.33 and G034.43+00.24 \citep{henshaw2013,barnes2018}. These clouds are proposed to be formed by the collision of filaments, and the velocity difference indicates that the merging of filaments is still ongoing.

In fact, filaments and clumps overlapped on the line of sight can be easily found around most of the dense cores (see Figure~\ref{fig:fellcore}). C3 and C17, in particular, are located at the hubs where the largest filaments merge. Inspecting the $\ceo~(1-0)$ spectra in the regions, multiple velocity components are presented in the cores (Figure~\ref{af:c3_c18ospec} in the Appendix). The offset value of $V_{\rm sys}^{\ceo}$ and $V_{\rm sys}^{\nthp}$ is similar to the velocity shift between $\ceo$ and $\nthp$ gas in the IRDCs G035.39-00.33 and G034.43+00.24. Hence, the velocity difference of the five cores with respect to the systemic velocities of their parent filaments can imply that the dense cores are forming via the merging of the filaments. \\

\subsection{Are the filaments and clumps in IC~5146 gravitationally bound?} \label{sec:mlin}

\begin{figure*} 
\includegraphics[width=0.98\textwidth,height=0.53\textwidth]{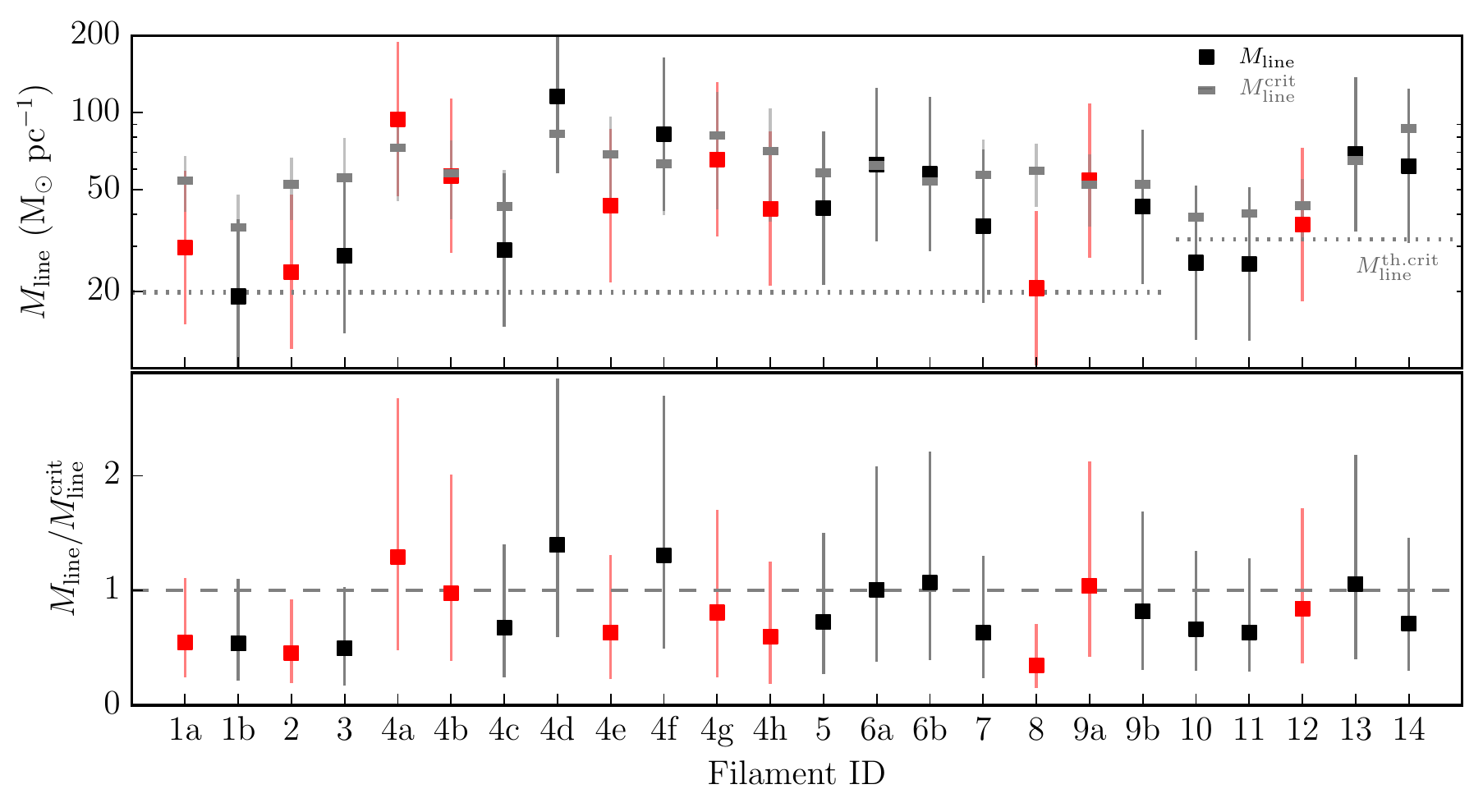} \caption{Criticality of the filaments. Top : the mass per unit length ($\mlin$) is presented with a square as a function of the filament ID. The error bars indicate the uncertainties of a factor of 2 in $\mlin$. Red and black indicate filaments with and without $\nthp$ dense cores. The effective critical mass per unit length ($\mlin^{\rm crit}$) derived with the average $\sigma_{\rm tot}$ is also given by a gray bar. The horizontal gray dotted lines denote the equilibrium values of 20 and 32~$M_{\odot}~ \rm pc^{-1}$ for an isothermal cylinder in pressure equilibrium at 15~K (the mean dust temperature for the Streamer) and 23~K (the mean dust temperature for the Cocoon), respectively. Bottom : ratios of mass per unit length to effective critical mass per unit length. The gray dashed line indicates the line where $\mlin$ and $\mlin^{\rm crit}$ are identical. \vspace{2mm} \label{fig:mline}} \end{figure*}
\begin{figure} \epsscale{1.17} \plotone{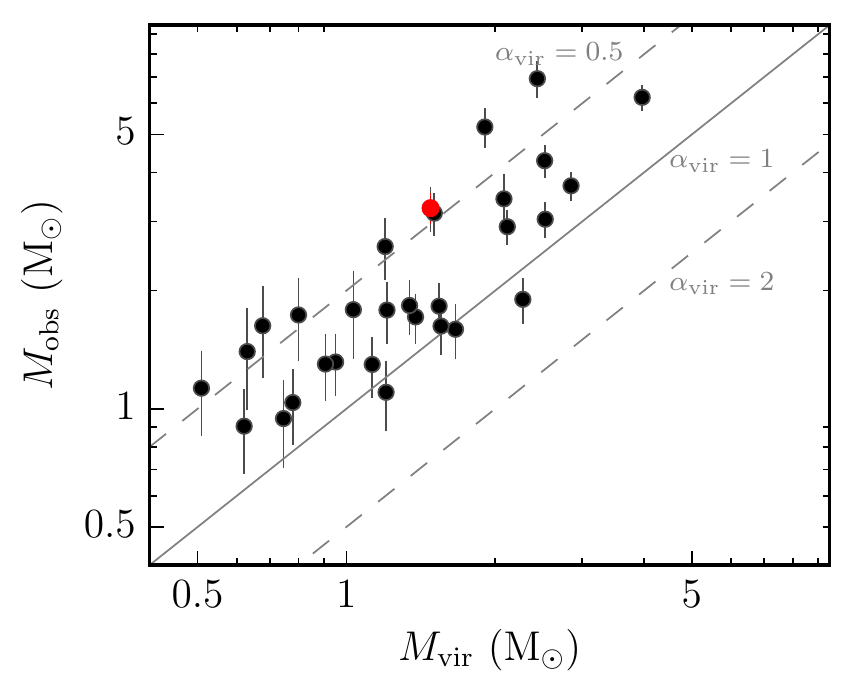} \caption{Observed mass ($M_{\rm obs}$) and virial mass ($M_{\rm vir}$) of 30 clumps. The red one is Clump22, which has a dense core (C18). The virial parameters, $\alpha_{\rm vir}$, of 0.5, 1, and 2 are drawn with gray lines. \vspace{2mm} \label{fig:mclump}} \end{figure}

Under the assumption that a filament is an infinitely long, self-gravitating, isothermal cylinder with only thermal support, the equilibrium mass per unit length is given by
\begin{equation}
	M_{\rm line}^{\rm th.crit} = \frac{2 c_{\rm s}^{\rm 2}}{G},
\end{equation}

\noindent where $c_{\rm s}$ and $G$ are the isothermal sound speed and gravitational constant, respectively \citep{inutsuka1992,inutsuka1997}. 
The median value of the dust temperature in IC~5146 is about 23~K in the Cocoon Nebula and about 15~K in the Streamer \citep{arzoumanian2011}. The corresponding $\mlin^{\rm th.crit}$ is $\sim$32~$M_{\odot}~\rm pc^{-1}$ at 23~K and $\sim$20~$M_{\odot}~\rm pc^{-1}$ at 15~K. On the other hand, the effective critical mass per unit length that includes the nonthermal components of turbulent motions is calculated as 
\begin{equation}
	M_{\rm line}^{\rm crit} = \frac{2 \sigma_{\rm tot}^{\rm 2}}{G} ,
\end{equation}
\noindent where $\sigma_{\rm tot}$ is the average total velocity dispersion of the mean free particle of molecular weight $\mu=2.8$ within a filament \citep[e.g.,][]{arzoumanian2013,peretto2014}. The effective critical mass per unit length of each filament is calculated using the average total velocity dispersion, and these values are tabulated in Table~\ref{tab:ppfila}.

Figure~\ref{fig:mline} gives the mass per unit length of filaments with the the critical mass per unit length, indicating that the majority of those in IC~5146 are supercritical or marginally critical ($\gtrsim \mlin^{\rm crit}$ or $\mlin^{\rm th.crit}$) within uncertainties of a factor of 2. Meanwhile, the filaments F2 and F8, which harbour dense cores, have $\mlin / \mlin^{\rm crit} \lesssim  1$, and are subcritical even with the uncertainties of a factor of 2. One of the possible explanations for the subcritical filaments with dense cores is that the CO depletion via freeze-out onto dust grains as the core evolves results in the underestimation of the filament mass. Filaments with $\nthp$ cores clearly must have CO freeze-out, since otherwise $\nthp$ emission would not have been seen. This implies that  the $\ceo$-based masses are probably an underestimate of the true value. In Figure~\ref{fig:compnh2} in the Appendix, H$_{2}$ column densities derived from our $\ceo~(1-0)$ and from the {\em Herschel} data are compared. F2 and F8 are shown to have a shallower slope of $N_{\rm H_{2}}^{\ceo}$ to $N_{\rm H_{2}}^{\rm Herschel}$ than that in total, especially where $N_{\rm H_{2}}^{\rm Herschel} \gtrsim 5 \times 10^{21} \rm cm^{-2}$. The presence of cores in the filament may reflect fragmentation that occurred in the past history of the filament, but the current criticality of a filament indicates its ability to make cores in the future. If filament velocity dispersions and other properties have changed over time, their criticality may have changed as well. Hence, we speculate that F2 and F8 formed the dense cores in the supercritical, or at least transcritical, conditions, but they appear to be subcritical subsequently due to the depletion of CO as a result of the core evolution inside the filament.

We also test gravitational instability for the 30 clumps that show a smaller aspect ratio than the filaments by deriving their virial parameters. Figure~\ref{fig:mclump} shows masses of the 30 clumps as a function of their virial mass ($M_{\rm vir}$). Most of the clumps are virialized ($M_{\rm obs} > M_{\rm vir}$). The clump CL22, which is found to be consistent in the velocity field as well as in the plane of the sky with the dense core C18, has $\alpha_{\rm vir} \sim 0.5$, indicating that CL22 is virialized and gravitationally contracting. 

We conclude that most of the filaments and clumps in IC~5146 are gravitationally bound systems that can form prestellar cores and stars by contraction and fragmentation. \\ 

\subsection{Velocity field and gradient: filaments as passages of mass flow?} \label{sec:massflow}

Several previous observations of molecular lines toward filaments have found that filaments have velocity gradients \citep[e.g.,][]{hacar2016,barnes2018,chung2019}. One explanation for the velocity gradients in the filaments is the gravitational accretion onto the filaments. The most representative example is the Taurus B211/3 filament, in which striations parallel to the $B$-field but perpendicular to the main filament ridge are observed \citep{palmeirim2013}. The velocity difference between the filament and the striations is $\sim$1~$\kms$, being well matched with what is expected from the gravitational freefall motions. Another example of a velocity gradient, which is now along the filament, is presented by \citet{kirk2013} in the Serpens South filament. The reported velocity gradient in the Serpens South is 1.4$\pm$0.2~$\kms~\rm pc^{-1}$ with an assumption for the inclination angle of 20$^{\circ}$. Their study suggested that the mass flow along the filaments into the central star clusters plays an important role in their ongoing star formation. \citet{trevino2019} investigated Monoceros~R2, a representative hub-filament system forming high-mass stars. They found increasing velocity gradients along the filaments toward the hub, claiming that the gas accelerates near the gravitational potential well of the hub and makes the formation of high-mass stars possible in the hub region. All of these findings indicate that the flow of material along the filaments will possibly directly affect the formation of dense cores and stars.

\begin{deluxetable*}{l c c c c c}  
\tablecaption{Mass Accretion Rates from Filaments to Cores \label{tab:accM}}
\tablewidth{0pt}
\tablehead{
\colhead{} &
\colhead{$\bigtriangledown V_{\parallel \rm , obs}$} &
\colhead{$\dot M_{\parallel \rm , i}~^{\rm a}$} &
\colhead{$\sum \dot M_{\parallel \rm , i}~^{\rm a}$} &
\colhead{$t_{\rm acc} ~^{\rm b}$} & 
\colhead{$t_{\rm ff} ~^{\rm c}$}
\\
\colhead{} & 
\colhead{($\kms~\rm pc^{-1}$)} & 
\colhead{($M_{\odot}~\rm Myr^{-1}$)} & \colhead{($M_{\odot}~\rm Myr^{-1}$)} &
\colhead{(Myr)} & 
\colhead{(Myr)} 
}
\startdata
C3-F4g & 1.5$\pm$0.1 & 26$\pm$10 & \multirow{3}{*}{35$\pm$11} & \multirow{3}{*}{0.7$\pm$0.2} &  \multirow{3}{*}{0.3} \\ 
C3-F4e & 0.5$\pm$0.1 & 4$\pm$1 &  &  &   \\ 
C3-F4h & 0.9$\pm$0.1 & 6$\pm$3 &  &  &   \\ 
\hline 
C11-F4a & 0.6$\pm$0.1 & 12$\pm$5 & \multirow{2}{*}{15$\pm$5} & \multirow{2}{*}{0.4$\pm$0.1} &  \multirow{2}{*}{0.3} \\ 
C11-F4e & 0.3$\pm$0.1 & 3$\pm$1 &  &  &   \\ 
\hline 
C15-F4aE~$^{\rm d}$ & 1.0$\pm$0.1 & 10$\pm$5 & \multirow{3}{*}{19$\pm$6} & \multirow{3}{*}{0.2$\pm$0.1} &  \multirow{3}{*}{0.3} \\ 
C15-F4aW~$^{\rm d}$ & 0.3$\pm$0.1 & 5$\pm$2 &  &  &   \\ 
C15-F4d & 0.2$\pm$0.1 & 5$\pm$2 &  &  &   \\ 
\hline 
C16-F4b & 0.7$\pm$0.1 & 5$\pm$2 & \multirow{2}{*}{26$\pm$14} & \multirow{2}{*}{0.8$\pm$0.4}~$^{\rm e}$ &  0.2 \\ 
C17-F4a & 0.7$\pm$0.1 & 20$\pm$14 &  &  &  0.2 \\ 
\hline \hline
\enddata 
\tablenotemark{}{} \\
\tablenotemark{$\rm a$}{The mass accretion rate and the total mass accretion rates from filaments to the cores are given assuming tan$(\alpha) = 1$ (i.e., the inclination angle $\alpha=45^{\circ}$) for all filaments. The real accretion rate should be divided by tan$(\alpha)$, and the accretion rate can be varied by a factor of 2.7 to 0.4 between the inclination angle of 20$^{\circ}$ and 70$^{\circ}$.}\\
\tablenotemark{$\rm b$}{The accretion time is the time taken to gather the current core mass through the accretion flow from the filaments, i.e., $t_{\rm acc} = \frac{M_{\rm core}}{\sum \dot{M_{\parallel \rm , i}}}$.}\\
\tablenotemark{$\rm c$}{The free fall time, $t_{\rm ff} = \sqrt{3 \pi / 32 \rm G \rho_{0}}$.}\\
\tablenotemark{$\rm d$}{F4a/E and F4a/W refer to the eastern and western filament regions of C15.}\\
\tablenotemark{$\rm e$}{The accretion time is measured for the total mass of C16 and C17.}

\end{deluxetable*}

It is interesting to note that F4, which harbors the largest number of dense cores and YSOs, has the largest velocity gradient ($\bigtriangledown V$) in IC~5146. The mean $\bigtriangledown V$ of F4 is 1.7$\pm$1.2~$\kms \rm ~pc^{-1}$, and the portion of the filament with $\bigtriangledown V$ larger than $2~\kms \rm ~pc^{-1}$ is close to 30\%. However, the other filaments have mean $\bigtriangledown V$ of 1.3$\pm$0.9~$\kms \rm ~pc^{-1}$, and the portion with $\bigtriangledown V$ larger than $2~\kms \rm ~pc^{-1}$ is less than 15\%. Since we derived the velocity gradient without correcting for the inclination ($\alpha$), projection effects should be considered. The observed velocity gradient is identical to the true velocity gradient multiplied by $\rm tan(\alpha)$. If the inclination changes from 20$^{\circ}$ to 70$^{\circ}$, the true velocity gradient can vary by a factor of 2.7 to 0.4 of the observed velocity gradient. Therefore it is possible that the relatively large velocity gradient inferred in F4 may not be true in reality but appears to be so as a result of the projection effect.

\citet{kirk2013} have estimated the accretion mass of the filament to the stellar cluster in Serpens South with the following equation by assuming that the filament would have a simple cylindrical shape with mass $M$, length $L$, radius $r$, and motions of velocity $V_{\parallel}$ along the filament long axis:
\begin{align} \label{eq:accM}
	\dot{M}_{\parallel} &= V_{\parallel} \times (M / \pi r^{2} L) \times (\pi r^{2}) \nonumber \\
	&=  V_{\parallel} \times (M / L).
\end{align}

With the inclination angle $\alpha$ of the filament to the plane of the sky, $L_{\rm obs} = L \rm cos({\alpha})$ and $V_{\parallel \rm , obs} = V_{\parallel} \rm sin({\alpha})$, and $V_{\parallel \rm , obs} = \bigtriangledown V_{\parallel \rm , obs} \times L_{\rm obs}$. Then the mass accretion rate becomes
\begin{equation} \label{eq:accM2}
	\dot{M}_{\parallel} = \bigtriangledown V_{\parallel \rm , obs} \times M \times \rm tan^{-1}({\alpha}).
\end{equation}

Using this equation, we attempted to estimate the mass accretion rate from the filament to the cores of C3, C11, C15, and C17, located in the hubs of HFSs where converging flows can be seen. However, a direct application of equation~\ref{eq:accM2} using the total mass and global velocity gradient of the filament is not straightforward, as there are multiple dense cores in filament F4. For example, a large velocity gradient can be seen along the F4h to the core C3. However, there is a core, C2, in the middle of F4h, and the global velocity gradient, mass, and length of F4h are not appropriate to measure the mass accretion rate from F4h to C3. Therefore, we carefully examined the systemic motions along with the filament's skeleton. We selected a local area of filament at which filament gas material possibly flows to the cores. For C3, the accretion rates are calculated along the F4g, F4e, and F4h. For C11, although it is not large, the velocity gradients are found along the F4a and F4e, and the mass accretion rates are derived for both directions. Accretion rates for C15 are estimated from the areas of F4a and F4d area. C16 and C17 are located quite close to each other, and the mass accretion rate is measured from F4b and F4a, respectively. We have limited the filament area to an area where the velocity gradient is visible but seemingly related to the cores only in the hub-like region. The filament areas that we use are shown by arrows in the top panel of Figure~\ref{af:accRfig} in the Appendix. 

We applied the mean gas density of the local area for $(M / \pi r^{2} L)$ in Equation~\ref{eq:accM} to estimate the local accretion rates for cores. Assuming that the thickness of a filament along the line of sight is equal to its width, the mean gas density of the $i$-th filament area becomes $\bar{\rho_{i}} = \bar N_{\rm H_{2}, i}/W_{i}$ where $\bar N_{\rm H_{2}, i}$ and $W_{i}$ are the mean H$_{2}$ column density and and the width of the $i$-th filament. Then, the mass accretion rate from the $i$-th filament area to the core can be estimated from the equation
\begin{equation}
	\dot{M}_{\parallel , i} = \frac{\pi}{4} \cdot \frac{\bigtriangledown V_{\parallel \rm , obs, \it i} \times \bar{\rho_{i}} \times L_{i} \times W_{i}^{2}}{\rm tan({\alpha})},
\end{equation}
\noindent where $\bigtriangledown V_{\parallel \rm , obs, \it i}$ and $L_{i}$ are the observed velocity gradient and length of the $i$-th filament area. $\bigtriangledown V_{\parallel \rm , obs, \it i}$ is derived from a linear least-squares fit for the systemic velocities along with the filament skeleton. The region where the velocity gradient is estimated with this fit is drawn with a green line in the bottom panel of Figure~\ref{af:accRfig} in the Appendix. The accretion time, $t_{\rm acc} = M_{\rm core} / \sum \dot{M}_{\parallel , i}$, is calculated to compare with the freefall time of $t_{\rm ff} = \sqrt{3 \pi / 32 \rm G \rho_{0}}$. The results are given in Table~\ref{tab:accM}. 

The accretion rates toward dense cores in F4 of IC~5146 were found to be in the range from 15 to 35~$M_{\odot}~\rm Myr^{-1}$, which is similar to those found for the Serpens South filament \citep[28~$M_{\odot}~\rm Myr^{-1}$;][]{kirk2013} and smaller than that of the Monoceros R2 filaments \citep[70~$M_{\odot}~\rm Myr^{-1}$;][]{trevino2019}. C3 is placed at the hub of the western-HFS, and the mass accretion rate of 35~$M_{\odot}~\rm Myr^{-1}$ is comparable to that of Serpens South. The time scales to collect the current core mass via the accretion flows from filaments are found to be 0.2 to 0.8~Myr. Though relatively longer than the freefall time of $\sim$0.3~Myr, this is consistent with the lifetime of YSOs within the uncertainty. It is reported that the global lifetime of the prestellar core phase is $1.2 \pm 0.3$~Myr \citep{konyves2015}. The lifetimes of Class~I and Class~II are known to be around 0.5~Myr and 1~Myr, respectively \citep[see][and references therein]{evans2009}. Hence, the accretion time of the cores is roughly in agreement with the time for the formation of YSOs and prestellar cores in the filaments. In conclusion, it is likely that the accretion flow from the filaments to the cores in IC~5146 plays a significant role in the star forming processes in IC~5146. \\

\subsection{Filaments in the Cocoon Nebula}

The Cocoon Nebula and the Streamer have been investigated together due to their proximity in the plane of the sky \citep[e.g.,][]{lada1994,johnstone2017,wang2017}. However, the distances of the Cocoon and the Streamer are known to be quite different. Our observations of the $\ceo~(1-0)$ line indicate that the velocity range of the Cocoon Nebula ($6 - 9~\kms$) also differs from that of the Streamer ($1-7~\kms$). Besides, the Cocoon and the Streamer have significantly different star formation environments. The Cocoon Nebula has $\sim$100 YSOs, while the Streamer has $\sim$20 YSOs \citep{harvey2008}. Also, there is a massive B-type star BD+46$^{\circ}$ 3474 at the center of the Cocoon Nebula \citep{herbig2008}. Hence, comparing the physical properties of filaments in the Cocoon and in the Streamer would be meaningful. In this section, we compare the physical properties of filaments in the Cocoon Nebula with those in the Streamer, focusing on the more evolved Cocoon Nebula.

In the Cocoon Nebula region, five filaments are identified (F10--F14). Only one dense core is detected in the Cocoon Nebula, while a few tens of dense cores are found in the filaments associated with the Streamer. There is no significant difference found between the filaments identified with the Cocoon Nebula and those with the Streamer in terms of their physical properties such as the H$_{2}$ column density and the mass per unit length ($\mlin$). The H$_{2}$ column density ranges between $\sim$2.5 and $6.1 \times 10^{21}~\rm cm^{-2}$ in the Cocoon filaments and between $\sim$2.2 and $8.8 \times 10^{21}~\rm cm^{-2}$ in the Streamer filaments. The Cocoon filaments have $\mlin$ of $\sim$25$-68~M_{\odot}~\rm pc^{-1}$, and the Streamer filaments have $\mlin$ of $\sim$19$- 115~M_{\odot}~\rm pc^{-1}$. The critical value of mass per unit length at which thermal pressure can support the gravitational contraction is $\sim$32~$M_{\odot}~ \rm pc^{-1}$ at 23~K for the Cocoon's filaments and $\sim$20~$M_{\odot}~ \rm pc^{-1}$ at 15~K for the Streamer's filaments \citep{ostriker1964}. 

F10, F11, and F14 in the Cocoon have smaller $\mlin$ than $\mlin^{\rm crit}$, and thus they are seemingly thermally supported and no dense cores form. The filament F12 where a small $\nthp$ dense core is detected has $\mlin$ of $\sim$36~$M_{\odot}~\rm pc^{-1}$, similar to the critical $\mlin$ within the uncertainty. F13 has the effective critical $\mlin$ of $71 \pm 24~M_{\odot}~\rm pc^{-1}$. The observed mass per unit length of F13 is $\sim$68~$M_{\odot}~\rm pc^{-1}$. Hence, F13 is gravitationally supercritical but is devoid of any dense cores. One possible explanation for the F13 not forming dense cores despite being physically supercritical is that its dynamical state may be controlled by other means of support such as turbulence and/or magnetic field that are not considered here. However, the nonthermal velocity dispersions show that the filaments associated with the Cocoon Nebula have subsonic or transonic turbulent motions, while those associated with the Streamer are transonic or supersonic. Hence, we tentatively conclude that the thermal pressure and magnetic field rather than the turbulence may be relatively more important in filaments of the Cocoon Nebula than in filaments of the Streamer.

The roles of gravity, turbulence, and magnetic field may change along with the formation and evolution of filaments and cores. The models of clouds and star formation suggest that the magnetic field and turbulence may play different roles at different stages of evolution \citep[e.g.,][]{crutcher2012}. Besides, the types of clumps that evolve can be affected by the balance of the three factors. It has recently been suggested that the subtle difference in the relative significance between the gravity, turbulence, and magnetic field can produce different fragmentation from the clump to the core scale \citep{tang2019}. To investigate the precise roles of the gravity, turbulence, and magnetic field in forming stars, more observational constraints, particularly observations of polarization, are required. The filaments in Cocoon are interesting targets in which to investigate the precise roles played by the gravity, turbulence, and magnetic field in their formation and evolution. Further investigations on the relative importance of magnetic field relative to gravity and turbulence will be given in our next paper (E. J. Chung et al. 2021, in prep.). 

\begin{deluxetable*}{lrrrrrrrH}
\tablecaption{Evolution Indicators of Starless Dense Cores \label{tab:coreevol}}
\tablewidth{0pt}
\tablehead{
\colhead{Core ID} &
\colhead{$N$($\nthp$)} & 
\colhead{$\Delta V$($\nthp$)} & 
\colhead{$f_{D} \rm (CO)$} & 
\colhead{$N$(H$_{2}$)} & 
\colhead{Det(SO)} &
\colhead{Infall Asy.} & 
\colhead{Total\tablenotemark{$\ast$}} \\
\colhead{} & 
\colhead{($10^{12}~ \rm cm^{-2}$)} & 
\colhead{($\kms$)} & 
\colhead{} & 
\colhead{($10^{20}~ \rm cm^{-2}$)} & 
\colhead{} & 
\colhead{} & 
\colhead{} & 
\colhead{} \\
\colhead{} &
\colhead{(1)} & 
\colhead{(2)} & 
\colhead{(3)} & 
\colhead{(4)} & 
\colhead{(5)} & 
\colhead{(6)} & 
\colhead{(7)} & 
\colhead{} } 
\startdata 
C2 & 0.97 (1) & 0.49 (1) & 1.1 (1) & 108.1 (1) & Y (0) & Y (1) & 5 & 4h \\ 
C5 & 0.95 (1) & 0.31 (0) & 0.9 (0) & 65.5 (0) & N (1) & N (0) & 2 & 4e \\ 
C6 & 0.68 (0) & 0.47 (0) & 1.1 (1) & 73.9 (1) & Y (0) & Y (1) & 3 & 2 \\ 
C8 & 1.01 (1) & 0.47 (0) & 1.0 (0) & 56.0 (0) & N (1) & N (0) & 2 & 4e \\ 
C10 & 0.82 (0) & 0.45 (0) & 0.5 (0) & 55.7 (0) & N (1) & N (0) & 1 & 4a \\ 
C12 & 0.88 (0) & 1.62 (1) & 0.7 (0) & 86.3 (1) & Y (0) & N (0) & 2 & 4a \\ 
C13 & 0.75 (0) & 0.78 (1) & 1.0 (0) & 64.9 (0) & Y (0) & N (0) & 1 & 4b \\ 
C16 & 2.29 (1) & 0.57 (1) & 1.3 (1) & 166.2 (1) & Y (0) & Y (1) & 5 & 4a \\ 
\enddata
\tablecomments{These physical parameters provide the evolutionary status of starless dense cores (Crapsi et al. 2005). 
Columns numbered (1)--(6) show $\nthp$ column density ($N$($\nthp$)), $\nthp$ line width ($\Delta V$($\nthp$)), CO depletion factor ($f_{D} \rm (CO)$), H$_{2}$ column density at the peak position ($N$(H$_{2}$)), the detection of SO emission (Det(SO)), and infall asymmetry of the $\hcop$ spectrum (Infall Asy.). $\rm Det(SO)$ indicates whether the SO line was detected above its 3$\sigma$ level. Each number in parenthesis is the point given for each parameter of each core if its property is ranked to mean that it is more evolved than the average for that property among the cores. $^\ast$The total is sum of points, and a larger total means that the core is more evolved.} 
\end{deluxetable*}
\vspace{5mm}

\subsection{Formation of Filaments and Dense Cores in the Streamer} \label{sec:formfdc}

This section discusses possible roles of the filamentary structures in the formation and evolution of filaments and dense cores. Most of the filaments in the Streamer are gravitationally supercritical, and the kinematic properties of filaments and dense cores imply that the mechanisms forming filaments and dense cores can differ depending on the environment of each filament. We have found 21 dense cores with the $\nthp~(1-0)$ line in the Streamer, and investigating the relative evolutionary stages of the cores may help in understanding the possible formation history of the Streamer. 

Observations toward the starless cores found that dynamical and chemical properties such as line width, inward motion, and depletion of C-bearing molecules, yet enhancement of deuterated species in the dense cores, can be used as indicators of the evolution of cores. In particular, \citet{crapsi2005} examined dynamical and chemical properties of several tens of starless cores, and found that the most evolved cores have higher $\nthp$, $\rm N_{2}D^{+}$, and H$_{2}$ column densities, a higher ratio of $N\rm (N_{2}D^{+})$ to $N(\nthp)$, higher CO depletion factor ($f_{D}$(CO),  the ratio of the canonical CO abundance to the observed CO abundance), larger $\nthp$ line width with infall signature, and more compact density distribution than the others.

Adopting the indicators provided by Crapsi et al. (2005), 
we investigated the relative evolutionary status of eight starless cores in IC~5146. Table~\ref{tab:coreevol} shows the available quantities of $\nthp$ column density, $\nthp~(1-0)$ line width, CO depletion factor, and H$_{2}$ column density at the peak position of $\nthp~(1-0)$ emission for the eight starless cores. We used the canonical CO abundance ([CO]/[H$_{2}$]) of $9.1 \times 10^{-5}$ from \citet{pineda2010}, and derived the observed CO abundance from the CO column density and H$_{2}$ column density obtained in Section~\ref{sec:filpp}. We included one more factor, i.e., the detection of SO~$(3_{2}-2_{1})$ over 3$\sigma$ as an evolution indicator for the cores in our discussion. We gave one point for each parameter for each of the cores and summed the points to get a total. A larger sum of the points implies that that core is relatively more evolved. Among the eight starless cores, C2 and C16 secured the highest sums, implying that they are the most evolved starless cores, while C10 and C13 appear to be the least evolved cores.

\begin{figure*} \epsscale{1.17} \plotone{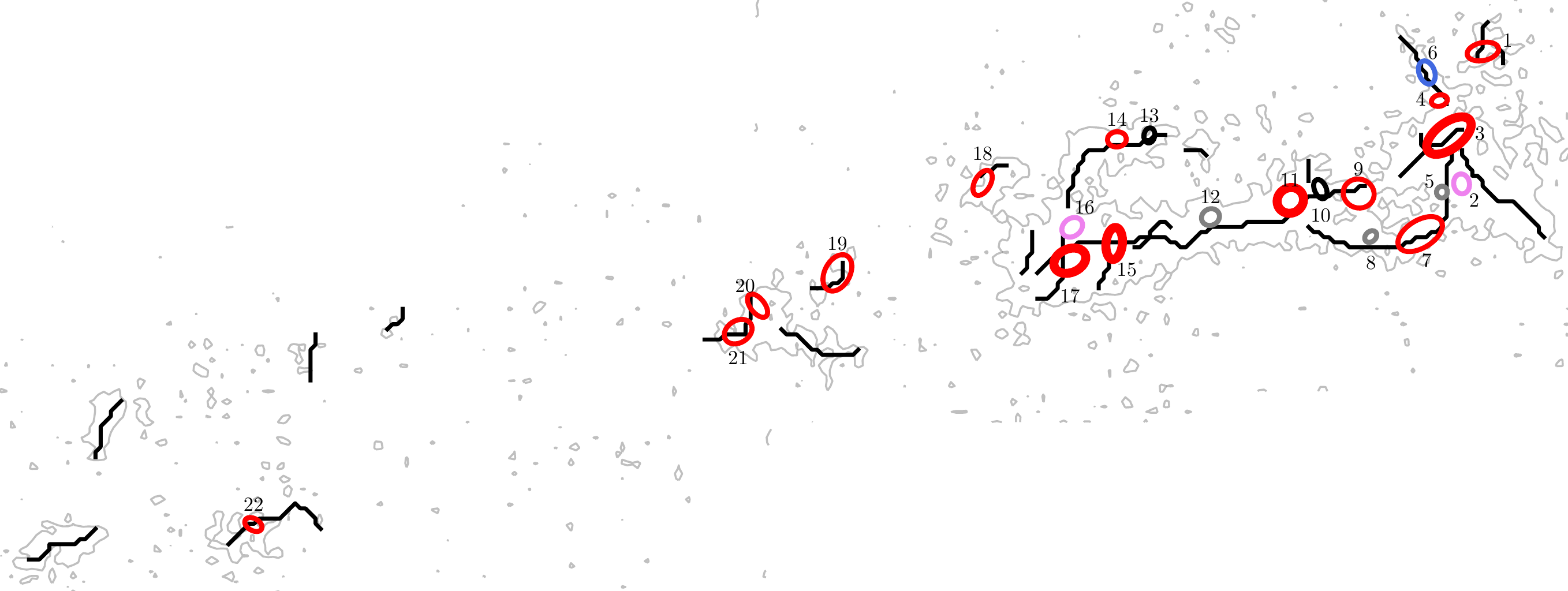} 
\caption{Dense cores in the filaments. The integrated intensity of $\ceo~(1-0)$ emission is drawn with contours at the 3$\sigma$ level, and skeletons of filaments are overlaid with solid lines. The 14 cores with YSOs are drawn in red, and the eight starless cores are color-coded based on the total points for grading their evolutionary status in Table~\ref{tab:coreevol}, i.e., black, gray, blue, purple, and pink colors for the cores with total points from 1 to 5, respectively; a larger total means that the core is more evolved. The thick red ellipses indicate cores with YSOs located in the hubs.  \vspace{2mm} \label{fig:cores_evol_skel}}\end{figure*}

Figure~\ref{fig:cores_evol_skel} shows the distribution of dense cores that are color-coded based on their relative evolutionary stage. The cores with YSOs are located in the supersonic filaments and hubs of HFSs while most of the starless cores are found in single-shaped, transonic filaments. This indicates that the filamentary accretion flow as well as turbulent gas motions in the filament may play an important role in the formation of dense cores and stars.

\citet{seo2019} investigated the kinematics and chemistry of star-forming regions in the Taurus molecular cloud, and proposed three star formation types, i.e., fast, slow, and isolated modes, depending on the filamentary structures involved in the star formation. The fast mode may apply for star formation at the hub where the column density is high. The converging flow from filaments to the hub and ram pressure may promote the mass accretion and the formation of the cores and the stars. The slow mode would work for the formation of stars at a gravitationally supercritical filament as well as the formation of pressure-confined cores at a gravitationally subcritical filament. The isolated mode is the classical mode of star formation where cores and stars form in isolated clumps by gravitational contraction. 

We note that the cores with YSOs and starless cores in IC~5146 can be explained well by these modes. The formation of dense cores and YSOs in C3, C11, C15, and C17 in hubs of IC~5146 seems consistent with the suggested fast mode. On the other hand, the starless core C6 seems to be controlled by a slow mode because it is located in a single-shaped filament (F2). C18, the dense core linked with a clump, but not with a filament, seems to be a case of the isolated mode. 

Meanwhile, sub-filaments not located in hubs of F4 harbor both cores with YSOs and cores without. The starless cores C2, C5, C8, C10, C12, C13, and C16 are embedded between or next to the dense cores with YSOs. This seems to be because when a chain of cores forms in a filament, a core having a deeper gravitational potential well grows first, and then the adjacent less massive core gains mass slowly. \\ 

\section{Summary and Conclusion} \label{sec:sum}

We carried out mapping observations of the clouds and dense cores in IC~5146 in molecular lines with the TRAO 14m antenna to investigate how stars form in relation with filamentary structures in IC~5146. The main results of our observations are as follows.

\begin{enumerate}
\item From $\ceo~(1-0)$ data, 14 filaments (size $\gtrsim$ 6$\times \theta_{\rm FWHM}$) and 30 clumps (size $\lesssim$ 4$\times \theta_{\rm FWHM}$ and aspect ratio $\lesssim 3$) are identified by performing a Gaussian decomposition for the observed spectra and a friends-of-friends algorithm for the decomposed Gaussian components. The basic physical quantities of the filaments such as H$_{2}$ column density, length, width, mass, mass per unit length, and mean velocity gradient are estimated. 
\item From $\nthp~(1-0)$ data, 22 dense cores (21 in the Streamer and 1 in the Cocoon Nebula) are found. Among the 22 dense cores, 14 cores are found to have YSOs while the other cores that are identified for the first time in this study are starless. Their positions, sizes, peak velocities, line widths, masses, and virial parameters are derived.
\item We compared the identified $\ceo$ filaments with the {\em Herschel} continuum filaments found in the dust continuum emission \citep{arzoumanian2011}. 
In some cases, filaments seen as a single entity in the continuum observations are found to consist of multiple filaments of different systemic velocities in our $\ceo~(1-0)$ map. This indicates that our observations of molecular lines are useful for extracting velocity coherent structures from those having multiple velocities that are spatially overlapped and hence would have been treated as a single filament based on the continuum maps alone. 
\item Based on the comparison of nonthermal velocity dispersions derived from $\ceo$ and $\nthp~(1-0)$, we divided the dense cores into two groups: one where the nonthermal velocity dispersions of the filaments and dense cores are nearly the same, and the other where the nonthermal velocity dispersions of the dense cores are smaller than those of the filaments. Among the dense cores in the latter group, three dense cores located in hubs show a different systemic velocity from that of filament material. This agrees with what the collision model of turbulent flows predicts in the formation of the filaments and the dense cores. Hence, we propose that the hubs and dense cores in them may have been formed through the collision of turbulent flows. 
\item Most of the filaments in IC~5146 have a larger mass per unit length ($\mlin$) than the critical $\mlin$ within the uncertainty, and hence they are gravitationally supercritical. Most of the dense cores are found on the supercritical filaments, but four dense cores are found in the subcritical filaments. Only one dense core is found to be in a virialized small clump. 
\item Every filament shows a continuously coherent velocity field, and its velocity gradient is on an average $\lesssim$~1~$\kms~\rm pc^{-1}$. F4, the largest filament with most of the dense cores in IC~5146, shows the largest velocity gradient, up to about $2-3~\kms~\rm pc^{-1}$. We estimate accretion rates of $\sim 15 - 35~ M_{\odot}~\rm Myr^{-1}$ from filaments onto the cores in the filaments where the velocity gradients were measured. The time scales to gain current core masses via the accretion flow are $\sim 0.2-0.8$~Myr, which is consistent with the time scales of the YSOs formed in IC~5146. This suggests that the accretion flows along the filaments may have played a significant role in the formation of stars in IC~5146.
\item The formation processes of dense cores and stars found in the filaments of IC~5146 appear to be well explained based on the three modes suggested by \citet{seo2019} -- the fast, slow, and isolated modes. The cores with YSOs tend to be located in the turbulent hubs, and hence might have been formed in the fast mode, while starless cores located on the transonic single-shaped filaments might be currently forming in a slow mode. One dense core embedded in the clump CL22 is likely to have formed by the isolated mode. \\
\end{enumerate}

\acknowledgments

We appreciate the referee and the editor for the valuable comments and suggestions. This work was supported by the National Research Foundation of Korea(NRF) grant funded by the Korea government(MSIT) (No. NRF-2019R1I1A1A01042480) and the Basic Science Research Program through the National Research Foundation of Korea (NRF) funded by the Ministry of Education, Science and Technology (NRF-2019R1A2C1010851). K. H. Kim is supported by the center for Women In Science, Engineering and Technology (WISET) grant funded by the Ministry of Science and ICT (MSIT) under the program for returners into R\&D (WISET-2019-288; WISET-2020-247; WISET-2021-080). A.S. acknowledge the support from the NSF through grant AST-1715876.\\

\makeatletter
\renewcommand\@biblabel[1]{}
\makeatother


\appendix \label{sec:append}
\addcontentsline{toc}{section}{Appendices}
\renewcommand{\thesubsection}{\Alph{subsection}}

\setcounter{equation}{0}
\renewcommand{\theequation}{A\arabic{equation}}

\begin{figure*} \epsscale{1.17}
\plotone{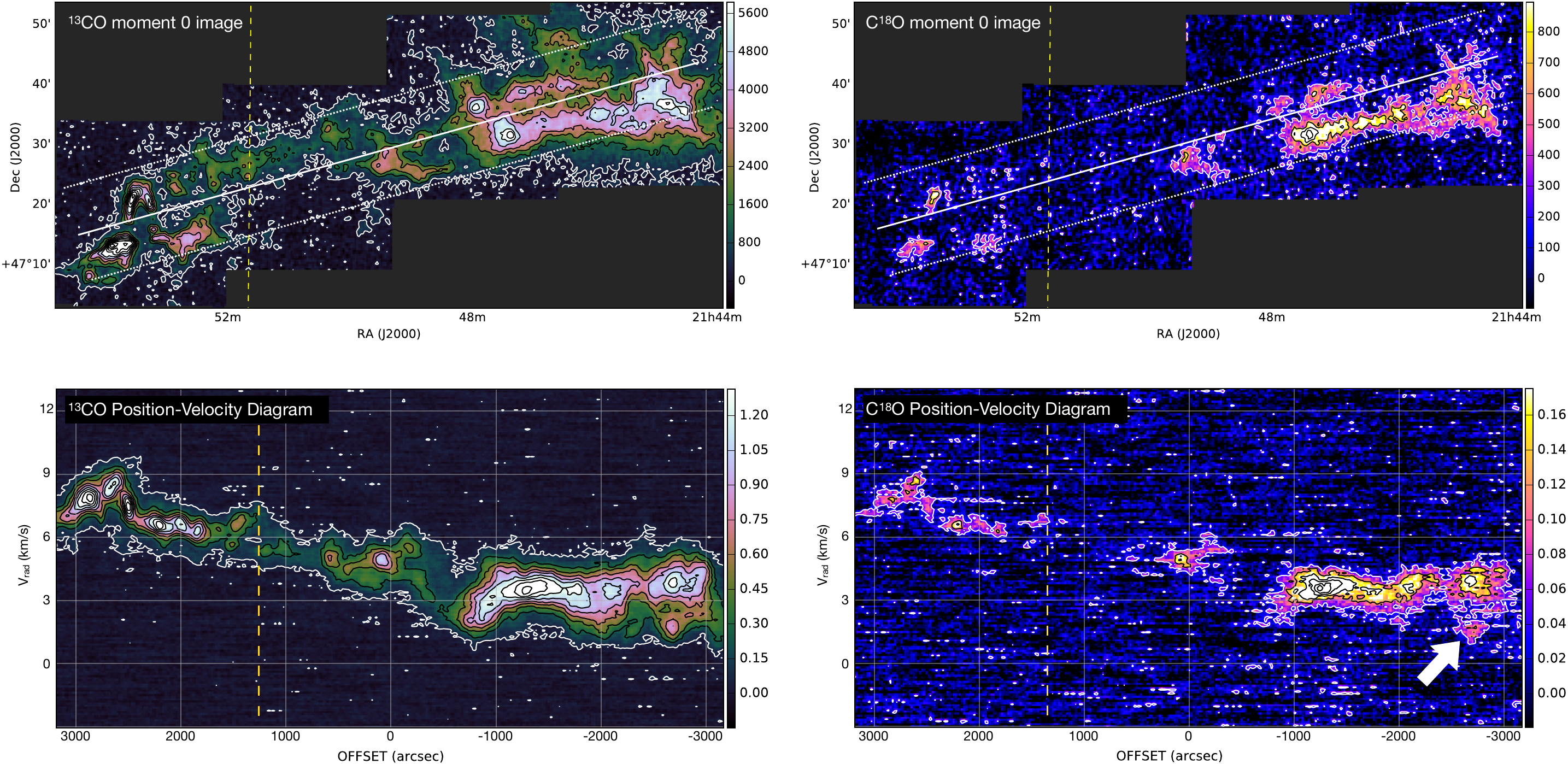}
\caption{The integrated intensity maps (top) and the position-velocity cuts along the white solid line in the integrated intensity maps (bottom) of $\tco$ and $\ceo~(1-0)$ emissions, respectively. The $p-v$ diagrams are averaged within the regions of white dotted lines in the integrated intensity maps. The yellow dashed lines indicate R.A.(J2000) of 21h 50m 31.5s which separates the Cocoon Nebula and the Streamer as in Figure~\ref{fig:obsreg}. The white arrow in the $\ceo ~ \it p-v$ diagram indicates the F2 emission which has quite different velocity from the F4's velocity. \label{af:pvds}}
\end{figure*}

\begin{center} \begin{figure*} 
\includegraphics[width=0.9\textwidth,height=0.96\textwidth]{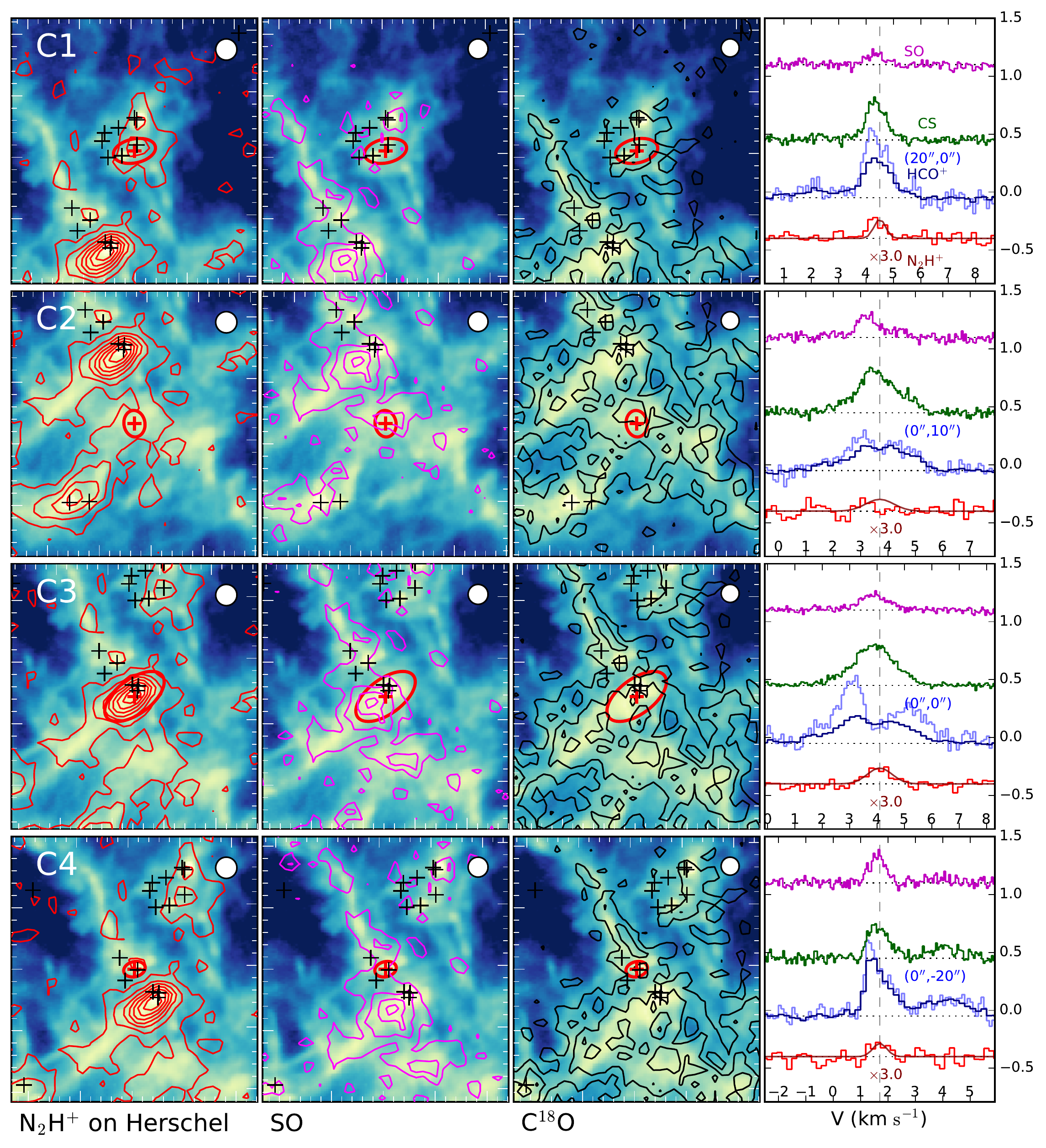}
\caption{\textbf{Images :} Integrated intensity contour maps of $\nthp~(1-0)$ (left), SO~$(3_{2}-2_{1})$ (second column), and $\ceo~(1-0)$ (third column) contour maps on {\em Herschel} 250~$\mu$m image toward the dense cores of C1 to C4. The contour levels are 3$n \times \sigma$ ($n=1, 2, 3, \cdots$). The red cross and ellipse indicate the peak position and the size of the $\nthp$ dense core. The black crosses represent the positions of YSO candidates \citep{harvey2008,poglitsch2010}. The FWHM beam sizes at the $\nthp$, SO, and $\hcop$ frequencies are shown with white circles on the upper right corner. \textbf{Spectra :} The averaged SO~$(3_{2}-2_{1})$, CS~$(2-1)$, $\hcop~(1-0)$, and $\nthp$ isolated component ($1_{0,2}-0_{1,2}$) spectra of each dense cores are presented with magenta, green, navy, and red colors, respectively. $\nthp$ spectra are presented with hyperfine fitting results (maroon lines). The spectrum drawn with the skyblue line is representative of the blue asymmetric $\hcop$ spectrum at the offset position from the core center (in the parenthesis). \label{af:corechem1}}
\end{figure*} \end{center}

\begin{center} \begin{figure*} 
\includegraphics[width=0.9\textwidth,height=1.2\textwidth]{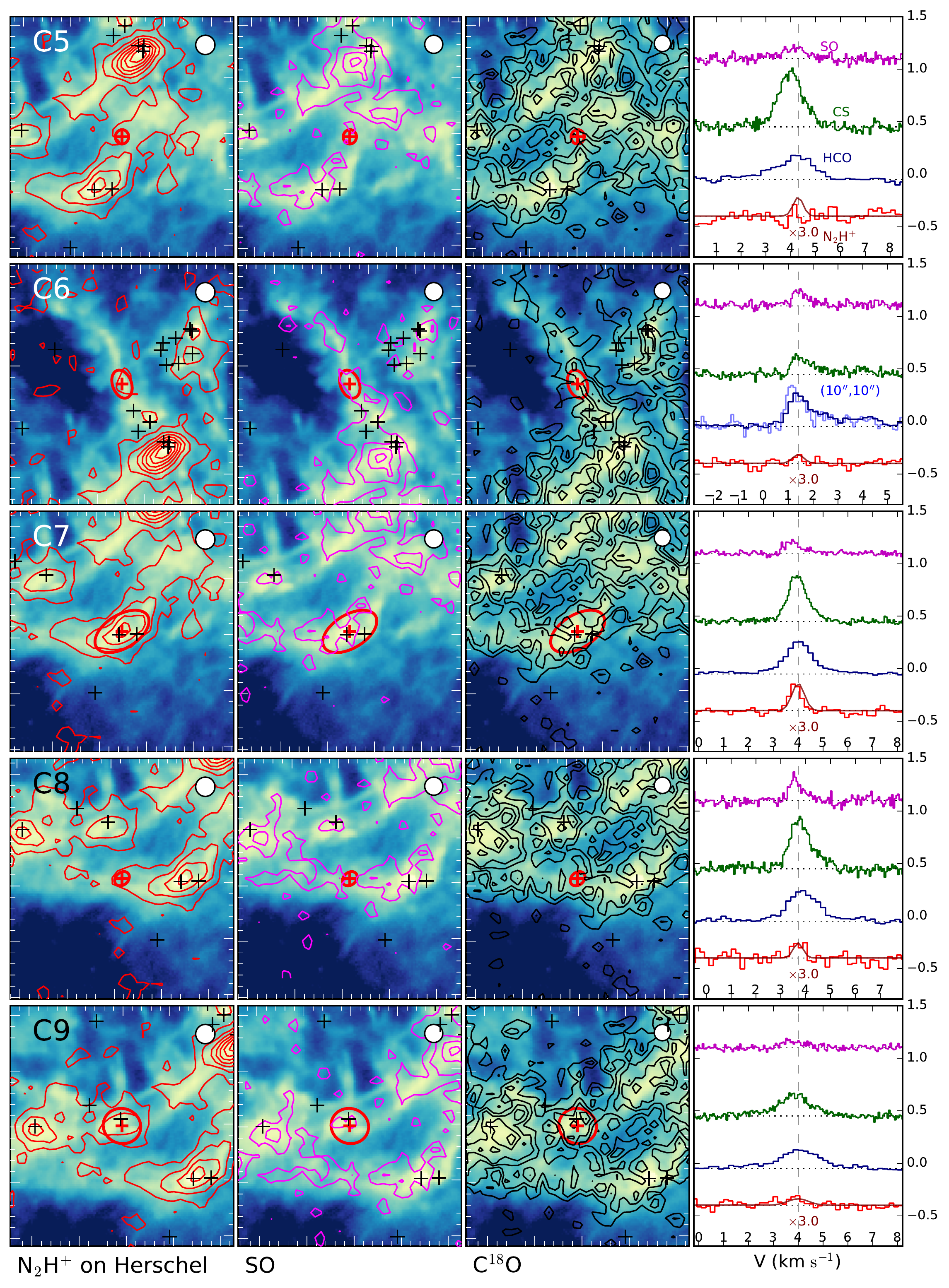}
\caption{Same as Figure~\ref{af:corechem1} for C5 to C9. \label{af:corechem2}}
\end{figure*} \end{center}

\begin{center} \begin{figure*} 
\includegraphics[width=0.9\textwidth,height=1.2\textwidth]{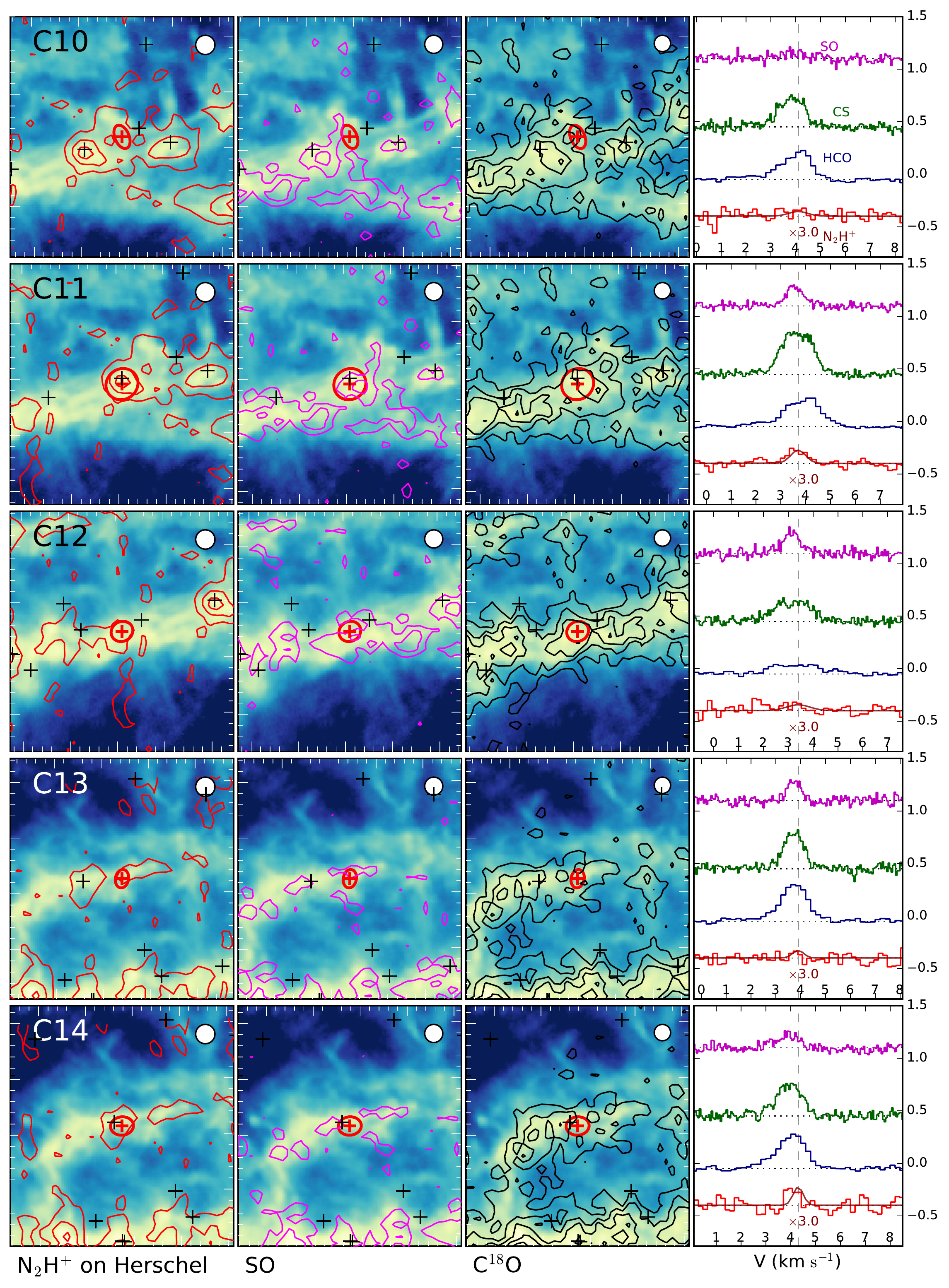}
\caption{Same as Figure~\ref{af:corechem1} for C10 to C14. \label{af:corechem3}}
\end{figure*} \end{center}

\begin{center} \begin{figure*} 
\includegraphics[width=0.9\textwidth,height=1.2\textwidth]{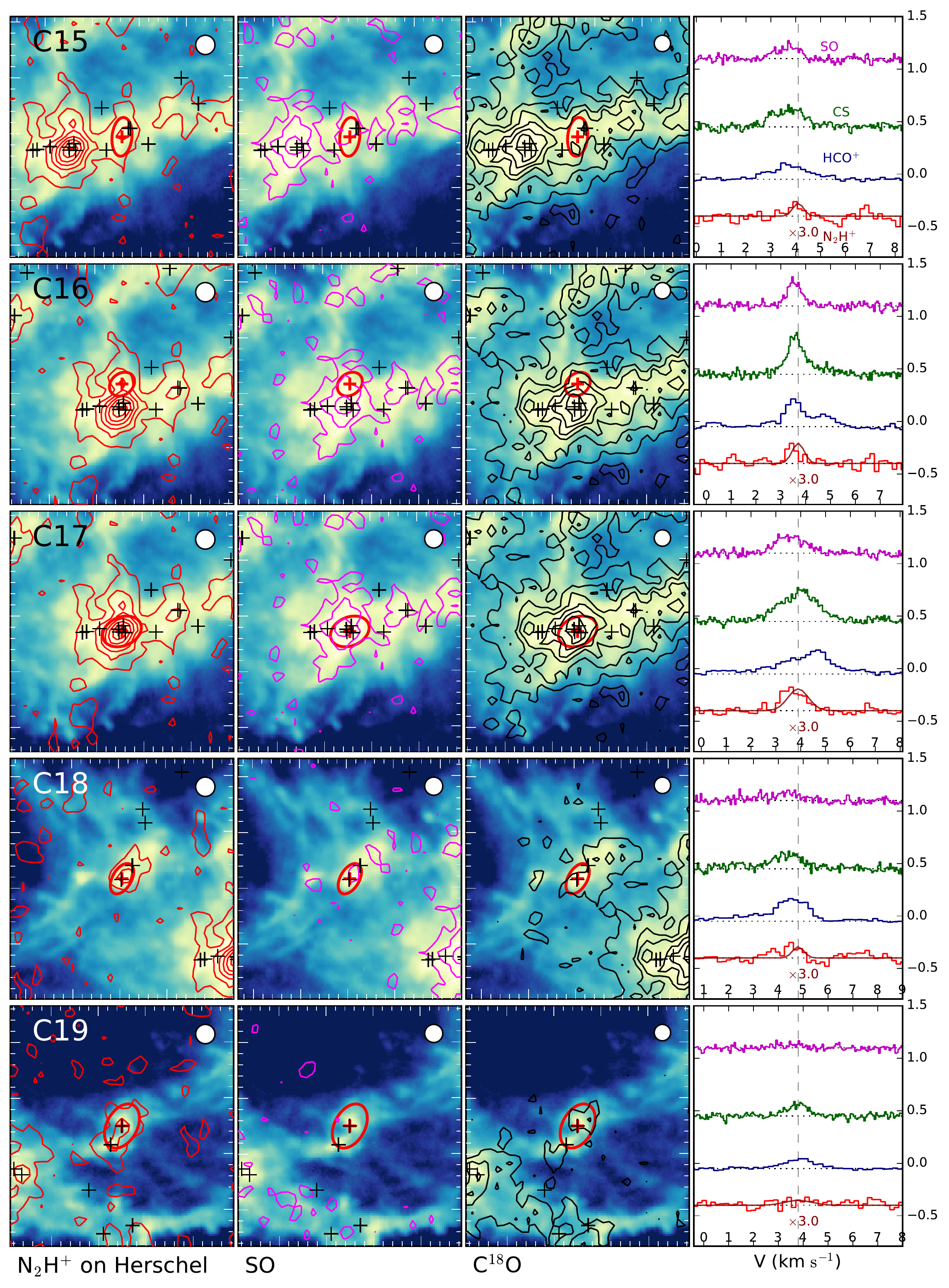}
\caption{Same as Figure~\ref{af:corechem1} for C15 to C19. \label{af:corechem4}}
\end{figure*} \end{center}

\begin{center} \begin{figure*} 
\includegraphics[width=0.9\textwidth,height=0.72\textwidth]{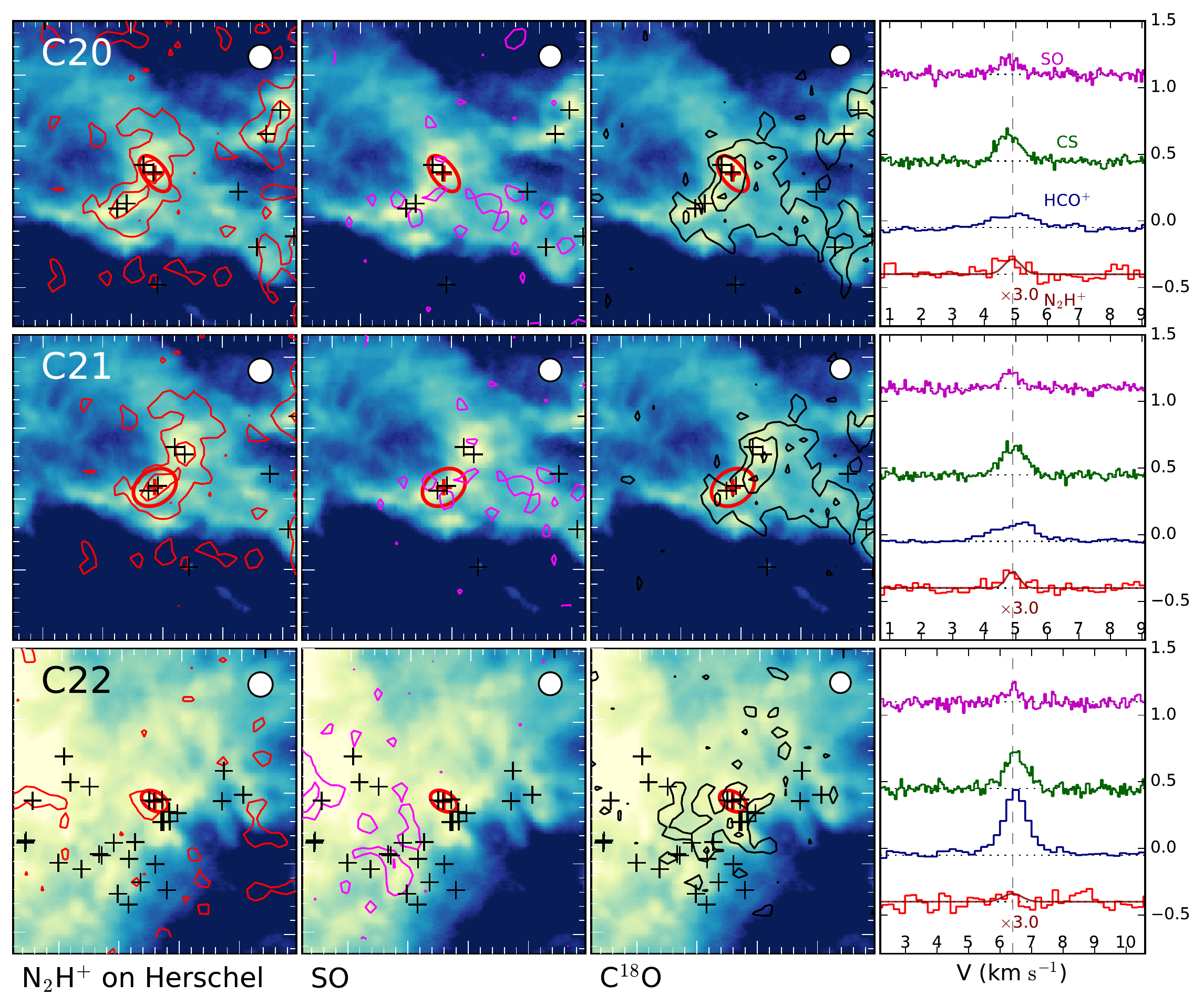}
\caption{Same as Figure~\ref{af:corechem1} for C20 to C22. \label{af:corechem5}}
\end{figure*} \end{center}

\begin{figure*} \epsscale{1.17}
\plotone{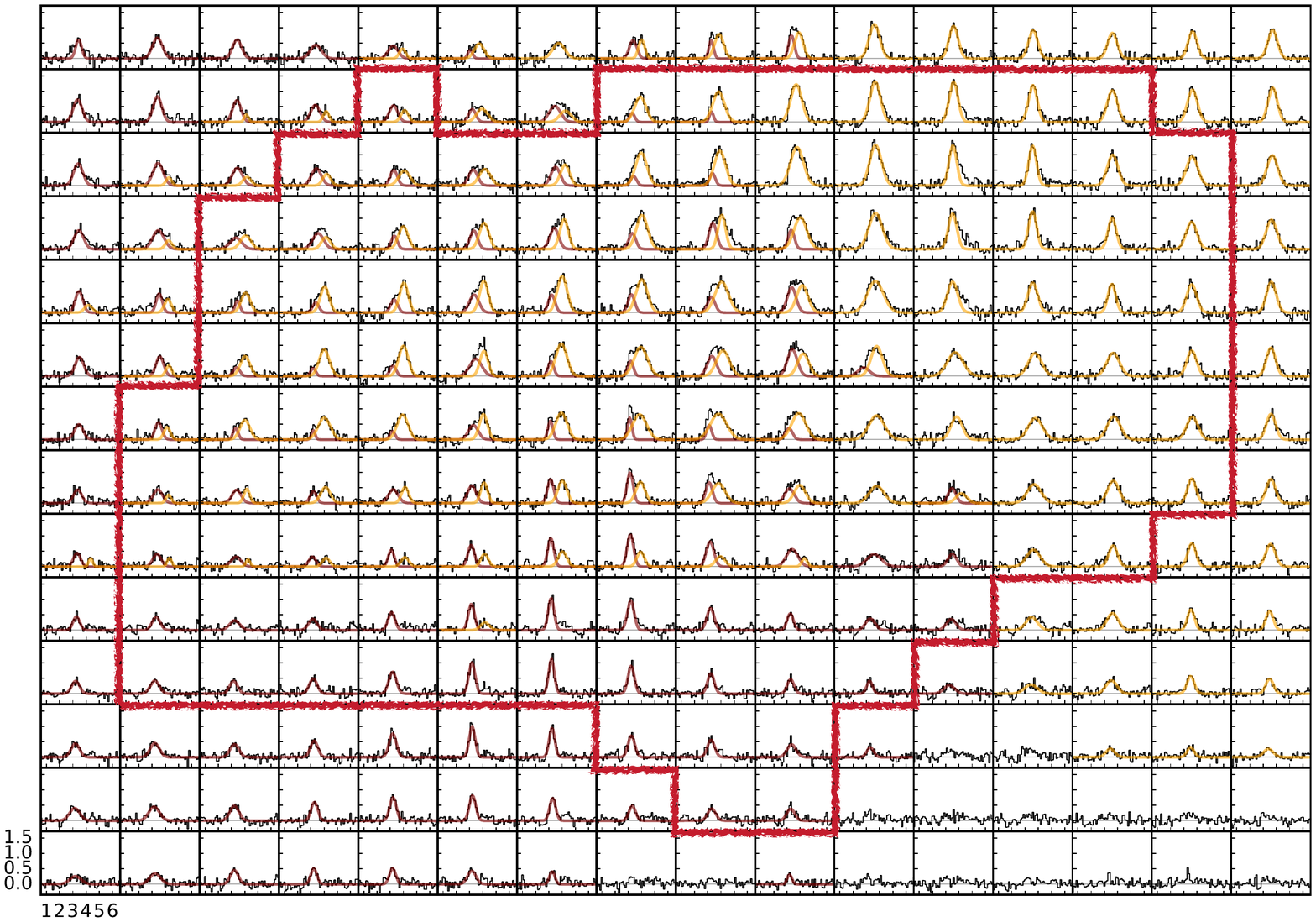}
\caption{$\ceo~(1-0)$ spectra around the C17. The ranges of the velocity and intensity are given to the lower left box in $\kms$ and in K[T$_{\rm A}^{\ast}$], respectively. The red polygon indicates C3. The yellow and red lines are the decomposed Gaussian components of F4 and F6, respectively. \label{af:c3_c18ospec}}
\end{figure*}

\begin{figure*} \epsscale{1.17}
\plotone{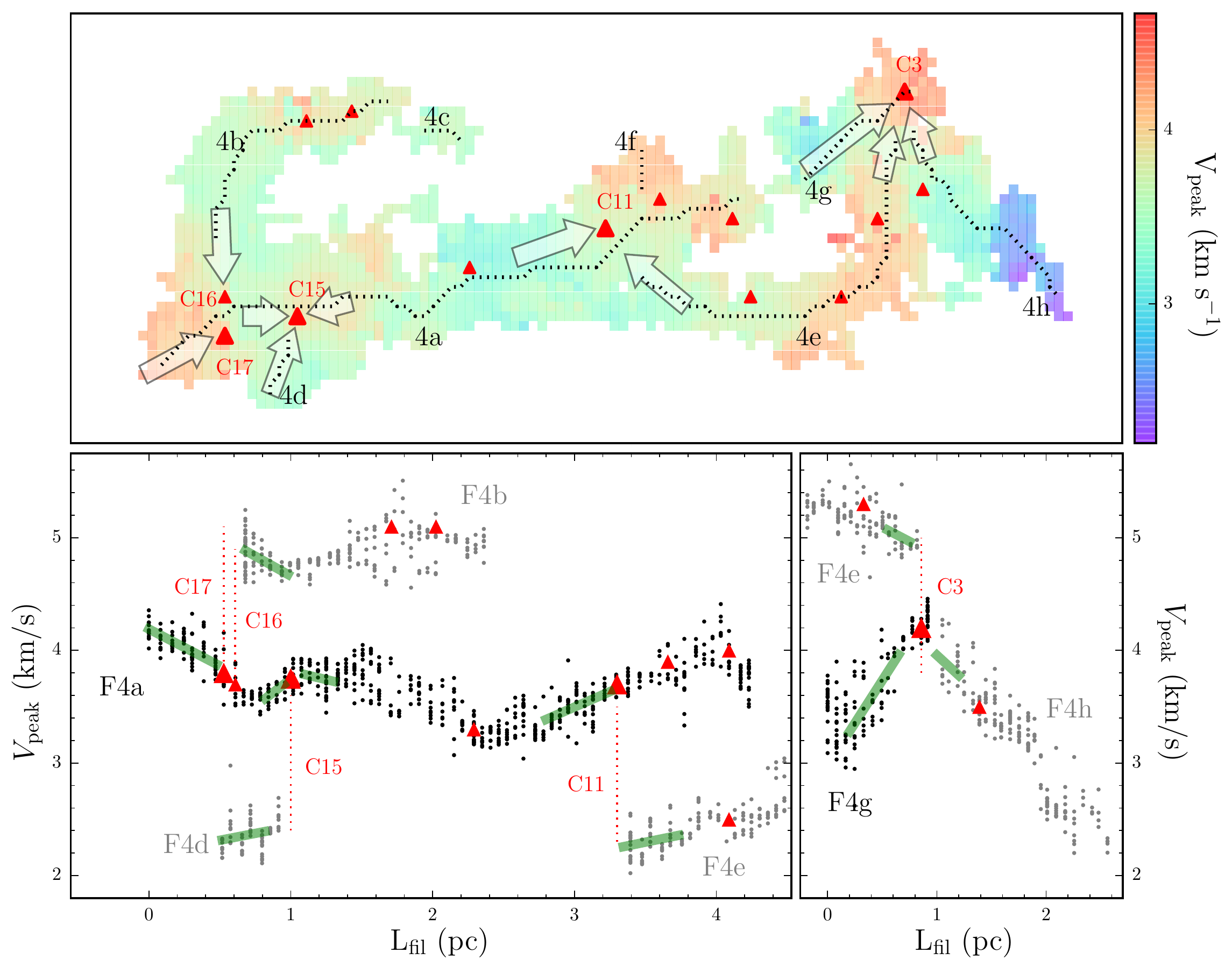}
\caption{{\it Top:} Systemic velocity map of F4 filament. The skeleton is drawn with the dotted line and dense cores with triangles. Large triangle symbols are to indicate dense cores (C3, C11, C15, and C17) which seem to form in a hub-like structure. The possible mass flows near the cores are depicted with the arrows. {\it Bottom:} Systemic velocities along the filament F4. To emphasize the velocity gradient along filaments to cores, velocity distributions of several sub-filament are figured altogether. The offset of $\pm 1.2~\kms$, indicated with the dotted red lines, is given for $V_{\rm sys}$ of F4b, F4d, and F4e to avoid the overlaps of data points. The green lines indicate the linear least squares fit of filament components from which the velocity gradient along the filament to the dense cores is estimated. \label{af:accRfig}}
\end{figure*}

We present the moment 0 maps and position-velocity diagrams of $\tco$ and $\ceo$ (Figure~\ref{af:pvds}), derivation of H$_{2}$ column density from $\ceo$ emission and $\nthp$ core mass (Sections~\ref{as:h2cd} and \ref{as:coremass}, respectively), the integrated intensity maps of $\nthp$, SO, and C18O and the averaged spectra of SO, CS, $\hcop$, and $\nthp$ isolated component of each $\nthp$ core (Figures~\ref{af:corechem1}--\ref{af:corechem5}), $\ceo$ spectra and the Gaussian decomposition results around C17 (Figure~\ref{af:c3_c18ospec}), the systemic velocity map of filament F4 (Figure~\ref{af:accRfig}), and the physical properties of $\ceo$ clumps (Table 6). \\

\subsection{$\rm H_{2}$ Column Density from $\ceo$ Emission with the LTE Condition} \label{as:h2cd}

\noindent $\ceo$ column density is calculated following the formula (Garden et al. 1991; Pattle et al. 2015) 

\begin{equation}
	N = \frac{3 k_{\rm B}}{8 \pi^{\rm 3} B \mu^{\rm 2}} \frac{e^{h B J(J+1)/k_{\rm B} T_{\rm ex}}}{J+1} \frac{T_{\rm ex} + \frac{h B}{3 k_{\rm B}}}{1 - e^{-T_{0} / T_{\rm ex}}} \int \tau \rm d \it v , \label{eq:eqn}
\end{equation}

\noindent where $B$ is the rotational constant, $\mu$ is the permanent dipole moment of the molecule, $J$ is the lower rotational level, and $T_{\rm 0} = h \nu / k_{\rm B}$. $T_{\rm ex}$ is the excitation temperature.  

The rightmost integration of $\tau$, the optical depth of the line, can be calculated following (Pattle et al. 2015): 
\begin{align}
	\int \tau (v) \rm ~d \it v &= \frac{1} {J(T_{\rm ex})-J(T_{\rm bg}) } \int \frac{\tau (v)}{1- e^{- \tau (v)}} ~T_{\rm mb} \rm ~d \it v \nonumber \\
	&\approx \frac{1}{J(T_{\rm ex})-J(T_{\rm bg})} \frac{\tau (v_{\rm 0})}{1 - e^{- \tau (v_{\rm 0})}} \int T_{\rm mb} \rm ~d \it v ,	
\end{align} \label{eq:inttau}

\noindent where $J(T)$ is the equivalent Rayleigh-Jeans temperature function, $J(T) = T_{\rm 0} / ( \rm e^{\it T_{\rm 0} / \it T} -1)$, and $T_{\rm ex}$ and $T_{\rm bg}$ are the excitation temperature, for which we used dust temperature obtained from {\em Herschel} dust continuum emission, and the cosmic microwave background temperature, respectively. $v_{\rm 0}$ is the central velocity, and the optical depth at the central velocity is derived with the abundance ratio of [$\tco / \ceo$] = 5.5 (Frerking et al. 1982) 
and the relation of 
\begin{equation}
	\frac{T_{\rm \ceo, max}}{T_{\rm \tco, max}} = \frac{1 - e^{-\tau_{\ceo}}}{1 - e^{-\tau_{\tco}}} ,
\end{equation} 
where $T_{\rm \ceo, max}$ and $T_{\rm \tco, max}$ are the maximum intensities of $\ceo$ and $\tco~(1-0)$, respectively. $T_{\rm mb}$ is the observed main beam temperature of the line. The area under the fitted Gaussian function is used for $ \int T_{\rm mb} \rm d \it v$, because multiple velocity components in the line of sight are slightly overlapped along the velocity directions.  

H$_{2}$ column density ($\nht$) is derived from $\ceo$ column density ($N_{\ceo}$) with the abundance ratios of $\tco / \ceo = 5.5$ (Frerking et al. 1982) 
and $\rm ^{12}CO / \tco = 69$ (Wilson 1999) 
and the conversion factor of $\nht / N_{\rm ^{12}CO} = 1.1 \times 10^{4}$ (Pineda et al. 2010). 

\begin{figure} \begin{center} 
\includegraphics[width=.47\textwidth]{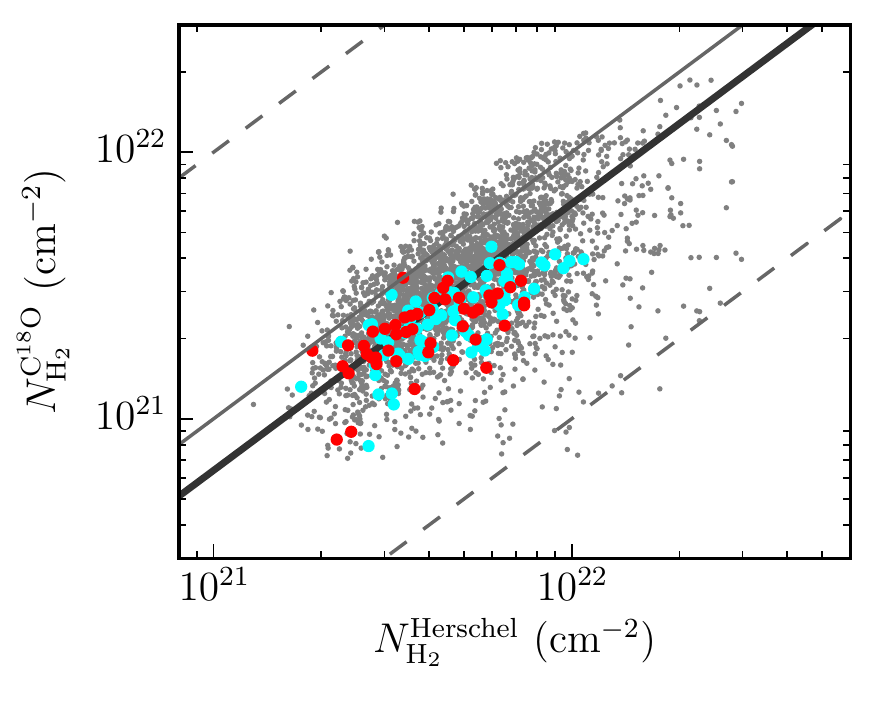} \end{center}
\caption{Comparison of $\nht$ derived from $\ceo~(1-0)$ with that from the {\em Herschel} data \citep{andre2010, arzoumanian2011}. 
The cyan and red dots are for the F2 and F8, respectively. The thin solid gray line shows where $\nht ^{\ceo}$ and $\nht ^{Herschel}$ are identical, and the two dashed lines indicate where the ratio of $\nht ^{\ceo} / \nht ^{Herschel}$ is 10 and 0.1. The least-squares fit result is shown as a thick solid line. \vspace{2mm} } \label{fig:compnh2} \end{figure}

We compare the derived $\nht$ with that of the {\em Herschel} data (Andr\'e et al. 2010; Arzoumanian et al. 2011) 
in Figure~\ref{fig:compnh2}. They are linearly correlated, and $\nht^{\ceo}$ appears slightly smaller than $\nht^{Herschel}$ on average but matches well within the range of uncertainty. The abundance for $\nht^{\ceo}$ to be best matched with $\nht^{Herschel}$ for IC~5146 would be $0.8 \pm 0.2$ for the Cocoon region and $0.7 \pm 0.3$ for the Streamer region. This is mostly due to the use of standard conversion factors of  $\tco / \ceo$, $\rm ^{12}CO / \tco$, and $\nht / N_{\rm ^{12}CO}$. We corrected the H$_{2}$ column density by factors of 0.8 and 0.7 for the Cocoon and the Streamer, respectively, and estimated the masses of filaments and clumps. \\

\setcounter{equation}{0}
\renewcommand{\theequation}{B\arabic{equation}}

\subsection{Mass and Virial Mass of Dense Cores} \label{as:coremass}

\noindent The masses of dense cores are calculated with $\nthp~(1-0)$ molecular line data. First, total column density of $\nthp$ is derived using the equation of Caselli et al. (2002): 
\begin{equation}
N = \frac{8 \pi W}{\lambda^{3} A} \frac{g_{l}}{g_{u}} \frac{1}{J(T_{\rm ex}) - J(T_{\rm bg})} \times  \frac{1}{1- {\rm exp}(-h \nu / k T_{\rm ex})} \frac{Q_{\rm rot}}{g_{l} {\rm exp}(-E_{l} / k T_{\rm ex})},
\end{equation}

\noindent where $W$ is the integrated intensity of $\nthp~(1-0)$ emission, $A$ is the Einstein coefficient, $g_{l}$ and $g_{u}$ are the statistical weights of the lower and upper levels, and $Q_{\rm rot}$ is the partition function. H$_{2}$ column density is estimated from the $\nthp$ column density with the average abundance of $\nthp$ of $6.8(\pm 4.8) \times 10^{-10}$ (Johnstone et al. 2010; Lee \& Myers 2011). 
The most uncertain factors for dense core mass are the excitation temperature and the conversion factor between the column densities of $\nthp$ and H$_{2}$ but the uncertainty caused by these factors is claimed to be less than a factor of 2 (Johnstone et al. 2010). 

The virial mass ($M_{\rm vir}$) for a spherical dense core is estimated as
\begin{equation} \label{eq:mvir}
	M_{\rm vir} = k ~ R~ \bar\sigma_{\rm tot}^{2} / G ,
\end{equation}
\noindent where $R$ is the radius of the core and $\bar \sigma_{\rm tot}$ is the total velocity dispersion of the mean molecular weight ($\mu=2.8$) averaged over the core. Assuming the density profile of $\rho \propto R^{-2}$, $k=1$ is applied. For the virial mass of clumps, the effective radius, the radius of a circle that has the same area as the clump, is applied (Table~\ref{tab:ppclumps}). With the assumption of the density profile of $\rho \propto R^{-2} $ where $R$ is the radius, $k=1$ is used. 


\begin{deluxetable*}{lcccc r@{$\pm$}l r@{$\pm$}l c}
\tablecaption{Physical Properties of $\ceo$ Clumps \label{tab:ppclumps}}
\tablewidth{0pt}
\tablehead{
\colhead{Clump ID} &
\colhead{R.A.} & 
\colhead{Decl.} & 
\colhead{$\bar \sigma$} & 
\colhead{${R_{\rm eff}}$ \tablenotemark{$\ast$}} &
\multicolumn{2}{c}{$\bar N_{\rm H_{2}}$} & 
\multicolumn{2}{c}{$M$} & 
\colhead{$\alpha_{\rm vir}$} \\ 
\colhead{} & 
\colhead{(hh:mm:ss)} &
\colhead{(dd:mm:ss)} & 
\colhead{($\kms$)} & 
\colhead{(pc)} & 
\multicolumn{2}{c}{($10^{20}~ \rm cm^{-2}$)} & 
\multicolumn{2}{c}{($M_{\odot}$)} & 
\colhead{} } 
\startdata 
CL1 & 21:44:30.5 & +47:36:01.1 & 0.39 & 0.16 & 34 & 10 & 6.2 & 0.5 & 0.64 \\ 
CL2 & 21:44:46.0 & +47:39:53.3 & 0.36 & 0.09 & 26 & 9 & 1.6 & 0.3 & 1.04 \\ 
CL3 & 21:44:47.3 & +47:36:53.5 & 0.37 & 0.12 & 31 & 17 & 3.0 & 0.3 & 0.83 \\ 
CL4 & 21:44:47.4 & +47:35:33.5 & 0.30 & 0.08 & 23 & 7 & 1.0 & 0.2 & 0.75 \\ \vspace{2mm}
CL5 & 21:44:53.1 & +47:38:34.2 & 0.30 & 0.09 & 25 & 8 & 1.3 & 0.2 & 0.72 \\ 
CL6 & 21:44:59.1 & +47:36:45.0 & 0.33 & 0.08 & 29 & 9 & 1.3 & 0.2 & 0.87 \\ 
CL7 & 21:45:26.1 & +47:47:58.3 & 0.27 & 0.09 & 21 & 5 & 1.1 & 0.3 & 0.45 \\ 
CL8 & 21:45:28.0 & +47:33:28.5 & 0.39 & 0.11 & 41 & 9 & 3.7 & 0.3 & 0.77 \\ 
CL9 & 21:45:42.8 & +47:34:20.0 & 0.34 & 0.10 & 27 & 10 & 1.8 & 0.3 & 0.84 \\ \vspace{2mm}
CL10 & 21:45:49.8 & +47:33:10.8 & 0.31 & 0.10 & 24 & 7 & 1.8 & 0.3 & 0.68 \\ 
CL11 & 21:46:00.6 & +47:34:21.8 & 0.40 & 0.09 & 36 & 11 & 1.9 & 0.3 & 1.20 \\ 
CL12 & 21:46:01.5 & +47:36:21.9 & 0.30 & 0.09 & 24 & 9 & 1.3 & 0.3 & 0.70 \\ 
CL13 & 21:46:02.3 & +47:38:12.0 & 0.34 & 0.08 & 24 & 9 & 1.1 & 0.2 & 1.09 \\ 
CL14 & 21:46:08.1 & +47:41:32.5 & 0.36 & 0.09 & 31 & 10 & 1.6 & 0.3 & 0.96 \\ \vspace{2mm}
CL15 & 21:46:14.4 & +47:33:43.1 & 0.38 & 0.09 & 48 & 17 & 2.9 & 0.3 & 0.73 \\ 
CL16 & 21:46:18.8 & +47:46:23.5 & 0.36 & 0.14 & 31 & 7 & 4.3 & 0.4 & 0.59 \\ 
CL17 & 21:46:38.1 & +47:34:05.0 & 0.34 & 0.08 & 13 & 12 & 1.7 & 0.2 & 0.81 \\ 
CL18 & 21:46:45.5 & +47:48:35.5 & 0.30 & 0.08 & 21 & 4 & 0.9 & 0.2 & 0.79 \\ 
CL19 & 21:46:57.9 & +47:34:06.3 & 0.33 & 0.10 & 27 & 14 & 1.8 & 0.3 & 0.73 \\ \vspace{2mm}
CL20 & 21:47:00.7 & +47:39:36.5 & 0.28 & 0.08 & 20 & 7 & 0.9 & 0.2 & 0.69 \\ 
CL21 & 21:47:16.5 & +47:39:57.4 & 0.31 & 0.14 & 30 & 9 & 3.2 & 0.4 & 0.48 \\ 
CL22 & 21:47:56.2 & +47:36:39.1 & 0.30 & 0.14 & 24 & 5 & 3.2 & 0.4 & 0.46 \\ 
CL23 & 21:51:43.0 & +47:25:02.2 & 0.28 & 0.12 & 15 & 3 & 1.4 & 0.4 & 0.45 \\ 
CL24 & 21:51:51.1 & +47:29:31.5 & 0.29 & 0.12 & 18 & 4 & 1.7 & 0.4 & 0.46 \\ \vspace{2mm}
CL25 & 21:52:21.9 & +47:17:18.3 & 0.31 & 0.14 & 19 & 4 & 2.6 & 0.5 & 0.46 \\ 
CL26 & 21:52:23.7 & +47:30:28.1 & 0.27 & 0.12 & 15 & 4 & 1.6 & 0.4 & 0.42 \\ 
CL27 & 21:52:52.7 & +47:22:34.5 & 0.32 & 0.17 & 26 & 6 & 5.2 & 0.6 & 0.37 \\ 
CL28 & 21:53:18.1 & +47:20:21.0 & 0.37 & 0.18 & 30 & 11 & 6.9 & 0.7 & 0.35 \\ 
CL29 & 21:53:22.3 & +47:22:40.4 & 0.36 & 0.15 & 36 & 7 & 3.4 & 0.5 & 0.61 \\ 
CL30 & 21:53:28.2 & +47:23:09.6 & 0.32 & 0.12 & 17 & 5 & 1.8 & 0.4 & 0.58 \\ 
\enddata
\tablenotemark{}{} \\
\tablenotemark{$\ast$}{Effective radius is the radius of a circle that has the same area as the clump. }
\vspace{5mm}

\end{deluxetable*}
\vspace{5mm}

\end{document}